\PassOptionsToPackage{numbers,sort&compress}{natbib}
\documentclass[preprint,12pt]{elsarticle}
\usepackage[framemethod=TikZ]{mdframed}
\usepackage{xcolor}
\usepackage{amsthm}
\usepackage{pgf}
\usepackage{pgfpages}
\usepackage{amsmath}
\usepackage{amsfonts}
\usepackage{amssymb}
\usepackage{graphicx}
\usepackage{bm}
\usepackage{lipsum}
\usepackage{array}
\usepackage{booktabs}
\usepackage{tikz}
\usepackage[hyphens]{url}
\usepackage{hyperref}
\hypersetup{
    breaklinks=true
}
\usepackage{amsmath}
\usepackage{empheq}
\usepackage{mdframed}
\usepackage{lipsum}
\usepackage{PGF}
\usepackage{pgfkeys}
\usepackage{pgffor}
\usepackage{pgfcalendar}
\usepackage{pgfpages}
\usepackage{array}
\usepackage{booktabs}
\usepackage{multirow}
\usepackage{longtable}
\usepackage{caption}
\usepackage{afterpage}
\usepackage{lscape}
\usepackage{colortbl}
\usepackage{xcolor}
\usepackage{makecell}
\usepackage{latexsym}
\usepackage{float}
\usepackage[english]{babel}
\usepackage{tikz,tikz-3dplot}
\usepackage{tabularx}
\usepackage{makecell}
\usetikzlibrary{arrows.meta}
\tdplotsetmaincoords{80}{45}
\tdplotsetrotatedcoords{-90}{180}{-90}
\RequirePackage{subfig}
\allowdisplaybreaks
\usepackage[skins,theorems]{tcolorbox}
\tcbset{highlight math style={enhanced,
  colframe=red,colback=white,arc=0pt,boxrule=1pt}}
\newcounter{theo}[section] 
\renewcommand{\thetheo}{\arabic{section}.\arabic{theo}}

\newcounter{prf}[section]
\renewcommand{\theprf}{\arabic{section}.\arabic{prf}}

\newcounter{lem}[section]
\renewcommand{\thelem}{\arabic{section}.\arabic{lem}}

\theoremstyle{theoremdd}
\newtheorem{thmd}{Theorem}[section]
\allowdisplaybreaks

\begin{document}

\begin{frontmatter}

\title{\textcolor[rgb]{0.00,0.40,0.80}{\bfseries{\boldmath
Analysis of Phantom Cosmology With Different Potentials: Inflation and Isotropization
}}}

\author[inst1]{Behzad Tajahmad\corref{cor1}}
\ead{behzadtajahmad@yahoo.com}

\author[inst1]{Hossein Kheiri}
\ead{h-kheiri@tabrizu.ac.ir}

\cortext[cor1]{Corresponding author}

\address[inst1]{Faculty of Mathematics, Statistics and Computer Science, University of Tabriz, Tabriz, Iran}

\begin{abstract}
We investigate inflationary dynamics and isotropization in a phantom-field model with five families of potentials on Bianchi I, Bianchi III, and Kantowski–Sachs backgrounds. First, the general properties of the model are established, and Heusler's proposition is extended. To address the limitations of this proposition, a powerful dynamical-systems approach is then adopted to provide a comprehensive analysis of the cosmological evolution.

In total, 87 critical points are identified and their properties are analyzed. Throughout the analysis, we emphasize the physical interpretation of the mathematical results and aim to provide a thorough explanation of the rationale behind the choice of each mathematical tool and variable. Within this framework, the conclusions of Collins and Hawking and of Burd and Barrow are revisited, and several challenging questions with deep conceptual implications are posed and discussed.
\end{abstract}

\end{frontmatter}
\newpage
\tableofcontents
\newpage
\section{Introduction\label{sec:intro}}
Inflation was introduced into the standard cosmological model by Guth as a mechanism
aimed at addressing the problems of homogeneity, isotropy, and the cosmological horizon
\cite{ref1}. Its effectiveness in this context follows from the fact that inflation corresponds
to either power-law or exponential expansion of the universe, while the Hubble horizon
remains approximately constant during this phase.

Nevertheless, because the Friedmann--Lema\^{i}tre--Robertson--Walker (FLRW) metric is assumed from the outset, the
origins of homogeneity and isotropy are not fully explained. A more fundamental treatment
would begin with a general metric, establish the occurrence of inflation, and subsequently
demonstrate that the cosmological evolution approaches a FLRW geometry.
Numerical investigations of cosmological models containing spherically inhomogeneous
configurations have been carried out to determine the conditions for the onset of inflation
\cite{ref2,ref3}. In addition, a semi-numerical analysis of inhomogeneous,
quasi-isotropic universes was developed using the long-wavelength iteration scheme
\cite{ref4,ref5,ref6}. These analyses showed that sufficiently large initial inhomogeneities
can inhibit inflation.

Given the complexity of the problem, a possible first step toward addressing the isotropy
issue is to consider a homogeneous but anisotropic metric. This strategy was originally
adopted by Collins and Hawking \cite{ref7}. They showed that the isotropy problem can be
resolved only for Bianchi types I, V, $\mathrm{VII}_{0}$, and $\mathrm{VII}_{h}$ when the
matter content satisfies the dominant energy condition. They further demonstrated that
the subset of spatially homogeneous cosmological models that become isotropic at infinite
times has measure zero within the space of all spatially homogeneous models.

It was subsequently anticipated that an inflationary phase, during which the dominant
energy condition is violated, might allow the cosmic no-hair theorem to be established.
For Bianchi-type universes containing a real scalar field with a convex positive potential
and a vanishing local minimum, however, Heusler showed that a no-hair theorem does not
hold \cite{ref8}. In fact, the attainment of isotropy requires the corresponding
FLRW model to be compatible with the Lie group associated with the
underlying Bianchi-type metric.

Besides Bianchi-type geometries, the Kantowski--Sachs model also provides a description of
spatially homogeneous universes. This model has been investigated by several authors both
in the presence and in the absence of a cosmological constant, with matter represented by
a perfect fluid behavior\cite{ref9,ref10,ref11, ref12}. The study of anisotropic inflation remains an active and challenging
area of research; see, for example, \cite{ref13}.

In 1998, two independent observational studies reported that the expansion of the universe
is accelerating \cite{ref14,ref15}. These results provided evidence for the existence of
dark energy, whose present-day equation-of-state parameter $\omega$ is smaller than
$-1/3$. Analyses based solely on Type Ia supernovae indicate that $\omega$ was larger
than $-1$ in the recent past, whereas its present value is smaller than $-1$. This
conclusion holds both when spatial flatness is assumed
\cite{ref16,ref17,ref18,ref19,ref20} and when it is not
\cite{ref21}.

The phantom dark energy model represents a distinct class of dynamical dark energy scenarios characterized by an equation of state parameter $w < -1$, which leads to a super-accelerated expansion and a future finite-time singularity known as the ``Big Rip'' \cite{phantom1,phantom2}.
Unlike quintessence or the cosmological constant, phantom fields possess a negative kinetic term in the Lagrangian, violating the null energy condition. Although this property can explain current observational hints of $w < -1$ from combined cosmological probes \cite{ref17}, phantom models suffer from quantum instabilities at high energy scales, such as ghost and vacuum decay problems \cite{phantom5}. Ghost condensation and effective field theories with higher-derivative
operators have been investigated as frameworks for studying
phantom-like behavior and its stability properties
\cite{phantom6,phantom7}. In single-field k-essence models, however,
crossing the phantom divide encounters theoretical obstructions
\cite{ref29}. Interacting phantom dark energy models have also
been studied as a means of avoiding the Big Rip singularity
\cite{ref30}.

Recent data from DESI has revived interest in phantom-like behavior, as some analyses show mild tension with $w = -1$ in favor of $w < -1$ at low redshifts \cite{phantom8}. Consequently, phantom dark energy remains a viable albeit exotic candidate for late-time cosmic acceleration, provided that its theoretical pathologies can be consistently tamed within a UV-complete framework \cite{phantom9}.

Despite its well-known pathologies,
the phantom model has proved successful in describing late-time cosmic acceleration, making
it natural to ask whether the same model can also accommodate early-time inflation together
with the isotropization of the Universe. If it can, one must determine whether such a
simultaneous behavior is generic or whether it requires a specific potential or a particular
background geometry. More refined questions also arise: can inflation occur without
isotropization, or isotropization without inflation? And if one of these phenomena is
realized, does it necessarily entail the other? The latter questions are directly motivated
by the results of Collins and Hawking~\cite{ref7} and by the framework of Burd and
Barrow~\cite{aniso14}.
To answer these questions comprehensively, it is insufficient to restrict attention to a
single background or a single potential such as the exponential or power-law forms, which are
among the most frequently studied. A systematic treatment requires considering a family of
potentials across a class of anisotropic geometries, so that the conditions for inflation and
isotropization can be delineated precisely.

In the present work, we conduct a comprehensive investigation of a phantom scalar field
characterized by five classes of convex positive potentials---none of which is required to
possess a local minimum---within the frameworks of Bianchi type I, Bianchi type III, and
Kantowski--Sachs cosmological models. The potentials selected are well-established in the
literature and have each been shown, in various contexts, to account for cosmic inflation,
late-time accelerated expansion, or related cosmological phenomena. 
Our aim is to determine
which of these potentials, and under which anisotropic background geometry, can simultaneously
produce an inflationary regime and drive the Universe toward isotropy within the phantom
framework.
Our investigation proceeds in two stages: we first derive and apply a generalization of the
Heusler theorem to obtain partial results, and then, because that approach
yields incomplete information, we employ a full dynamical systems analysis to characterize
the critical structure of the phase space and determine the cosmological behavior.

\section{
The Model, Geometry of Background, and Equations
}
We begin by considering the phantom gravitational action expressed in the form:
\begin{align}\label{action}
S=\int d^{4}x \, \sqrt{-g} \left[
\frac{R}{16 \pi G} +\frac{1}{2} \partial_{\mu}\varphi \partial^{\mu}\varphi -V(\varphi)
\right].
\end{align}
Here, $g$ denotes the determinant of the metric tensor, $R$ represents 
the Ricci scalar curvature, $\varphi$ designates the phantom scalar field, 
and $V(\varphi)$ corresponds to its associated potential.\\
As previously mentioned, the three well-known metrics, namely Bianchi type I, 
Bianchi type III, and Kantowski--Sachs, are examined in the present study. 
These three metrics can be written collectively in the following unified and 
compact form:
\begin{align}\label{metric}
ds^{2}= -\mathrm{d}t^{2}+a(t)^{2}\mathrm{d}r^{2}
+b(t)^{2} \left( \mathrm{d}\theta^{2}+\mathcal{F}(k,\theta)^{2}\mathrm{d}\Phi^{2} \right),
\end{align}
where
\begin{align}\label{F_definition}
\mathcal{F}(k, \theta)=
\left\{
\begin{array}{lll}
\theta, &\quad k=0, &\quad \text{Bianchi type I (BI);} \\[4mm]
\dfrac{\sinh \left(\sqrt{-k\,}\,\theta\right)}
      {\sqrt{-k\,}}, &\quad k<0, &\quad \text{Bianchi type III (BIII);} \\[4mm]
\dfrac{\sin \left(\sqrt{k\,}\,\theta\right)}
      {\sqrt{k\,}}, &\quad k>0, &\quad \text{Kantowski--Sachs (KS).}
\end{array}
\right.
\end{align}
The two-dimensional reference metric
$\mathrm{d}\theta^{2}
+\mathcal{F}^{2}(k,\theta)\,\mathrm{d}\Phi^{2}$
has constant Gaussian curvature $k$, as follows from
\begin{align}\label{F_curvature}
\frac{\partial^{2}\mathcal{F}}{\partial\theta^{2}}
+k \,\mathcal{F}=0.
\end{align}
Equivalently, wherever $\mathcal{F}\neq0$,
\begin{align}
k = -\frac{1}{\mathcal{F}}
\frac{\partial^{2}\mathcal{F}}{\partial\theta^{2}}.
\end{align}
The definition in Eq.~(\ref{F_definition}) also has the continuous
flat limit
\begin{align}
\lim_{k\to 0}\mathcal{F}(k,\theta)=\theta.
\end{align}
For $k\neq0$, its magnitude can be absorbed into the
redefinitions
\begin{align}
\bar{\theta}=\sqrt{|k|}\,\theta,
\qquad
\bar{b}(t)=\frac{b(t)}{\sqrt{|k|}}.
\end{align}
The metric then takes the same form with the normalized curvature
parameter $\bar{k}=\operatorname{sgn}(k)$ (i.e. $\bar{k}=(-1, 0, +1) \equiv (\text{BIII, BI, KS})$).
Thus, allowing arbitrary nonzero values of $k$ does not
introduce additional local geometries within this family.
The physical Gaussian curvature of the transverse two-surfaces
is $k/b^{2}(t)$ and remains unchanged under these
redefinitions.

\begin{figure}[!h]
	\centering
	
	\tikzset{surface/.style={draw=blue!70!black, fill=blue!40!white, fill opacity=.6}, >={Stealth[scale=1.5]}}
	
	\newcommand{\coneback}[4][]{
		\draw[canvas is xy plane at z=#2, #1] (45-#4:#3) arc (45-#4:225+#4:#3) -- (O) --cycle;
	}
	\newcommand{\conefront}[4][]{
		\draw[canvas is xy plane at z=#2, #1] (45-#4:#3) arc (45-#4:-135+#4:#3) -- (O) --cycle;
	}
	\begin{tikzpicture}[tdplot_main_coords, grid/.style={help lines,blue!40!white,opacity=0.2},scale=1]
		\coordinate (O) at (0,0,0);
		\coneback[surface]{-3}{2}{-12}
		\conefront[surface]{-3}{2}{-12}
		\draw[->, thick] (-4.2,0,0) -- (4.2,0,0) {};
		\draw[->, thick] (0,4.5,0) -- (0,-4.5,0) {};
		\coneback[surface]{3}{2}{12}
		\draw[-,thick] (0,0,-4) -- (0,0,-3.15) node[above] {};
		\draw[->, thick] (0,0,3.00) -- (0,0,4.2) node[above] {$\Theta$};
		\conefront[surface]{3}{2}{12}
		\fill (4,0,2.5) node[below right] {$k>0 ; \; \mathrm{KS}$};
		\node at (0,1.4,1.7) [above left] {$k <0 \;\; \mathrm{BIII}$};
		\fill (1.3,0.5,2) node[above left] {};
		\fill (5,-0.25,4) node[above] {$k=0; \; \mathrm{BI}$};
		\draw[->,red] (1.3,0.5,2) -- (5,-0.25,4) node[below, pos=0.65, rotate=55.1459,scale=0.70,black] {};
		\node[black] at (0,0,2.955) {$\text{Expanding Universe}$};
		\node[black] at (0,0,-3.03) {$\text{Contracting Universe}$};
		\node[black] at (0,0.17,0.35) [scale=0.6] {$O$};
		\node[black] at (0,-4.8,0) [scale=1.2] {$\sigma$};
		\node[black] at (5.2,-0.05,0) {$\mathrm{\;\;}\sqrt{V} \mathrm{\; or\; } i \psi$};
\draw[-,dashed, thick, draw=black!90!blue, opacity=0.8] (0,0,-3) -- (0,0,2.85) node[above] {};
	\end{tikzpicture}
	\caption{
Segregation of BI, BIII, and KS phase spaces based on the coordinate system.
	}
	\label{fig1}
\end{figure}
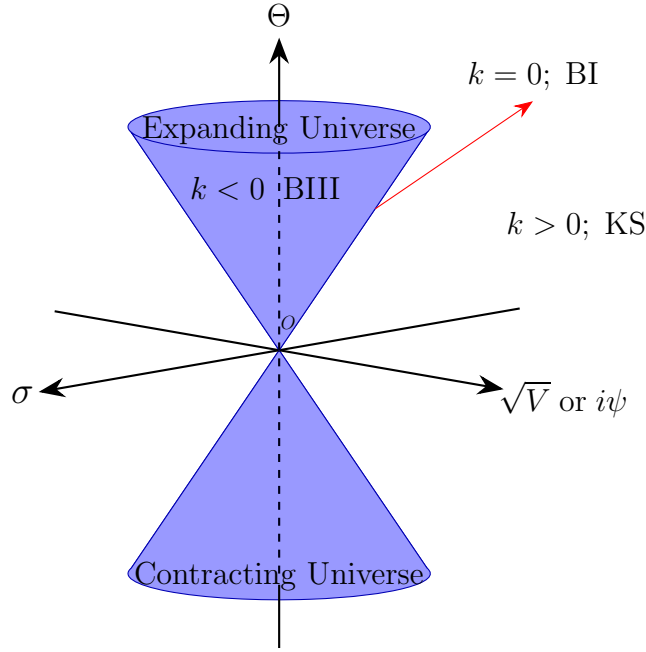

Upon varying the action with respect to the metric, 
the Einstein field equations are obtained as follows:
\begin{align}
2H_{a}H_{b}+H_{b}^{2}+\frac{k}{b^{2}} &=8\pi G \left( -\, \frac{1}{2}\dot{\varphi}^{2}+V \right), \label{eq3}\\
2\dot{H}_{b}+3H_{b}^{2}+\frac{k}{b^{2}} &=8\pi G \left( \frac{1}{2}\dot{\varphi}^{2}+V \right), \label{eq4}\\
\dot{H}_{a}+\dot{H}_{b}+H_{a}^{2}+H_{b}^{2}+H_{a}H_{b} &=8\pi G \left( \frac{1}{2}\dot{\varphi}^{2}+V \right), \label{eq5}
\end{align}
where an overdot denotes a differentiation with respect to cosmic time 
(Newton's dot notation), and $H_{a}$ and $H_{b}$ represent the Hubble 
parameters along the radial and angular directions, respectively, defined as:
\begin{align*}
H_{a} = \frac{\dot{a}}{a}, \qquad H_{b} = \frac{\dot{b}}{b}.
\end{align*}
As is evident from the structure of the equations, the distinctions 
among the three metrics manifest exclusively through the parameter $k$.\\
Varying the action with respect to the scalar field yields 
the Klein--Gordon equation:
\begin{align}\label{eq6}
\ddot{\varphi}+(H_{a}+2H_{b})\dot{\varphi}-\frac{\mathrm{d}V}{\mathrm{d}\varphi}=0.
\end{align}
Henceforth, all physical quantities are expressed in units of the Planck 
mass, $m_{\mathrm{P}}=1/\sqrt{8\pi G} \equiv 1$.
The system is fully characterized 
by the three independent equations (\ref{eq3}), (\ref{eq4}), and (\ref{eq6}). 
It can be straightforwardly demonstrated that equation (\ref{eq5}) is not 
independent, but rather derivable from the remaining equations---a 
consequence that reflects the Bianchi identity. Following a series of 
algebraic manipulations, the governing equations can be recast as a set of 
four first-order ordinary differential equations, which are independent of 
$k$, together with the constraint equation (\ref{eq3}), which is preserved 
throughout the dynamical evolution. This reformulation is carried out in 
terms of the expansion rate $\Theta=H_{a}+2H_{b}$ and the shear scalar 
$\sigma=(H_{a}-H_{b})/\sqrt{3}$, as follows:
\begin{align}
&\dot{\varphi}=\psi, \label{eq7}\\
&\dot{\Theta}=-\, \frac{1}{3}\Theta^{2}-2\sigma^{2}+\psi^{2}+V, \label{eq8}\\
&\dot{\sigma}=-\, \frac{1}{3\sqrt{3}}\Theta^{2}+\frac{1}{\sqrt{3}}\sigma^{2}-\Theta \sigma
+\frac{1}{\sqrt{3}} \left(-\, \frac{1}{2}\psi^{2}+V\right), \label{eq9}\\
&\dot{\psi}=-\Theta \psi +\frac{\mathrm{d}V}{\mathrm{d}\varphi}, \label{eq10},\\
&\frac{1}{3}\Theta^{2}+\frac{k}{b^{2}}= \sigma^{2}-\frac{1}{2}\psi^{2}+V.\label{eq11}
\end{align}
Equation (\ref{eq11}) constitutes the constraint equation.\\
As is apparent, all newly introduced variables carry well-defined 
physical significance within the first-order system: $\Theta$ 
identifies the expansion ($\Theta>0$)/contraction ($\Theta<0$) of the universe, $\sigma$ 
measures the anisotropy, and $\psi$ serves the purpose 
of reducing the order of the equations. Notably, equations 
(\ref{eq7})--(\ref{eq10}) are completely independent of the particular 
homogeneous model under consideration. Given that equation (\ref{eq11}) 
is conserved, the specific homogeneous model must be prescribed through 
the initial conditions. Although the equations reflect a four-dimensional 
phase space with coordinates $(\Theta, \sigma, z=\sqrt{V}, \psi)$, 
conceptual insights can nonetheless be extracted by adopting either the 
three-dimensional coordinate system\footnote{Here, $i=\sqrt{-1}$.} 
$(\Theta, \sigma, z=\sqrt{V})$ or $(\Theta, \sigma, i\psi)$. 
In both aforementioned three-dimensional coordinate systems, 
the BI solutions reside on the light cone, 
BIII solutions are located in the interior of the cone, 
and the KS solutions lie in the exterior region 
(see Fig.~\ref{fig1}).
Although $\sigma$ quantifies the degree of anisotropy---and its 
vanishing can be interpreted as an indicator of isotropy---a more 
stringent criterion, originally proposed by Hawking~\cite{ref7, aniso14}, 
is adopted in the present work:
\begin{align}
\frac{\sigma}{\Theta} \to 0  \qquad \text{as} \qquad t \to \infty.
\end{align}
We refer to $\sigma\to 0$ as \emph{weak condition of isotropization} and to $(\sigma/\Theta)\to 0$ as \emph{strong condition of isotropization}, although the terminology may be reversed in some literature. Here, “weak” and “strong” are meant in the sense of physical plausibility and cosmological relevance. Hawking’s criterion is more natural, since isotropization is not about the absolute vanishing of shear, but about the shear becoming negligible \emph{relative to the expansion rate}. Physically, $\sigma$ measures anisotropic distortion of fluid elements, whereas $\Theta$ measures the overall volume expansion; thus $(\sigma/\Theta)$ quantifies the fractional importance of anisotropy in the cosmic dynamics. 
In this sense, $\sigma=0$ is an exact and very restrictive condition, essentially corresponding to a perfectly isotropic geometry such as FLRW, while $\sigma\to 0$ alone is still insufficient if $\Theta$ tends to zero at a comparable rate. For example, if $\sigma\sim t^{-1}$ and $\Theta\sim t^{-1}$, then both quantities vanish asymptotically, but $\sigma/\Theta$ remains finite, so anisotropy still contributes at a fixed fraction of the expansion. This is why the dimensionless ratio $(\sigma/\Theta)$ is the natural cosmological measure of isotropization. It is also the more relevant quantity observationally, because anisotropic signatures depend not on the absolute size of shear alone, but on its strength relative to the Hubble expansion scale.

The scope of this study is limited to expanding phantom cosmological models (with \textit{real} scalar field and also \textit{real} potentials); 
accordingly, the analysis is restricted to strictly positive values of 
$\Theta$, corresponding to the upper half of the cone (expanding universe). Contraction is not acceptable at all, since we know that the universe has been in an expanding phase (``decelerated'' and ``accelerated'' \emph{expansion}) from the beginning until now.

In the following section, we begin with a generalization of the Heusler theorem. We will observe that this theorem alone does not provide a complete and transparent answer to the questions under consideration. We therefore continue our analysis within the framework of dynamical systems, which constitutes one of the most powerful and precise mathematical approaches for investigating the qualitative and asymptotic behavior of cosmological systems. This approach has been extensively used in the literature, for example, see refs.~\cite{refBeh,dy1,dy2,dy3,dy4,dy5,dy6,dy7,dy8,dy9,dy10,dy11}. 
The dynamics of the system can be investigated using either the conventional dynamical-systems approach or the $\mathfrak{B}\text{oundary}$-function ($\mathfrak{B}$-function) method proposed in Ref.~\cite{BehAnnals}. The latter provides a unified framework for reconstruction, phase-space analysis, and the study of exact solutions in terms of a single dimensionless function. Specifically, the $\mathfrak{B}$-function is defined as the Lie derivative of $\ln \mathcal{G}(x)$ along the vector field $X=(x+c)\mathrm{d}/\mathrm{d}x$, on a singleton $Q=\{x\}$, as
\begin{equation}
    \mathfrak{B}[x,c;\mathcal{G}]
    \equiv \mathcal{L}_{X}\ln \mathcal{G}(x),
\end{equation}
where $c$ is a constant with the same dimensions as $x$. By relating reconstruction to the behaviour of the scalar field(s), this method can be used to constrain model parameters and admissible initial and boundary conditions, thereby helping to identify physically viable models and parameter domains for subsequent data analysis. In this work, however, we adopt the conventional dynamical-systems approach, which is more familiar to most readers.

\section{Intrinsic Properties of the Dynamical System and 
Generalization of the Heusler Theorem to the Phantom Model}
In this section, the Heusler theorem is extended to the phantom 
model, first for an arbitrary potential of general form, and subsequently 
applied to the following five specific classes of well-known potentials:
\begin{align*}
	\text{Potential A:}&\quad V_{\mathrm{A}}(\varphi)=V_{0a} \varphi^{n};\\
	\text{Potential B:}&\quad V_{\mathrm{B}}(\varphi)=V_{0b} \exp (-\zeta \varphi)+V_{1b};\\
	\text{Potential C:}&\quad V_{\mathrm{C}}(\varphi)=\cosh \left(\xi \varphi \right)-1;\\
	\text{Potential D:}&\quad V_{\mathrm{D}}(\varphi)=V_{0d}	\sinh^{-\alpha} (\beta \varphi);\\
	\text{Potential E:}&\quad V_{\mathrm{E}}(\varphi)=2M^{2} \cos^{2} \left( \frac{\varphi}{2l} \right).
\end{align*}
Here,
$V_{0a}$,
$V_{0b}$,
$V_{1b}$,
$V_{0d}$,
$M$,
$n$,
$\xi$,
$\alpha$,
$\beta$,
and
$l$
are constant. The potentials $A$, $B$, $C$, $D$, and $E$ have been adopted from 
refs.~\cite{pota, potb, potc, potd, pote}, respectively. The selected potential functions are well-established in the literature and are known to exhibit remarkably rich phenomenological behavior.
We shall assume these constants to be real, since our focus is on real potentials and real phantom scalar field. Special cases involving imaginary values are discussed in some parts where the result is so important; however, after suitable redefinitions, the potential and the field remain real in those cases as well.

The reason for excluding a purely imaginary phantom field, as well as a genuinely complex one, is that in the purely imaginary case the phantom scalar field effectively reduces to a canonical field, while in the complex case, with both nonzero real and imaginary parts, it becomes a quintom field. Since the present work is devoted exclusively to the phantom model, which is a technically demanding and time-consuming framework, we do not pursue these extensions here.

For BI ($k=0$) and BIII ($k<0$), the asymptotic 
behavior of the solutions can be extracted directly by examining equations 
(\ref{eq7})--(\ref{eq11}). Since our focus is confined to expanding 
cosmological models, it is assumed that $\Theta$ is strictly positive, 
or vanishes at some prescribed time $t_{0}$. As will become apparent 
in the subsequent analysis, generalizing the theorem necessitates 
restricting attention to non-negative potentials. In this regard, various 
sufficient conditions may be formulated; however, the most natural and 
physically well-motivated choices are the following\footnote{Note that 
$\cosh(\xi\varphi) \geq 1$ holds identically.}: the quantities 
$V_{0a}$, $V_{0b}$, $V_{1b}$, $V_{0d}$, and $\beta$ are required to be 
non-negative constants, and furthermore, the phantom scalar field is 
constrained to remain non-negative, i.e., $\varphi \geq 0$. It should 
be emphasized that the theorem is formulated exclusively for $k \leq 0$,
the rationale for which will become clear below.

In light of equation (\ref{eq8}), it follows that:
\begin{equation}
\dot{\Theta}=V -\frac{1}{3}\Theta^{2}-2\sigma^{2}+\psi^{2}.
\end{equation}
Now, imposing the condition $|\psi| \leq \sqrt{2}|\sigma|$, 
the following expression is obtained:
\begin{align}\label{con1}
\dot{\Theta} \leq V -\frac{1}{3} \Theta^{2}.
\end{align}
It is worth emphasizing that, although the above relation is automatically 
satisfied under the assumption of a purely imaginary phantom scalar field 
---thereby rendering the condition 
\linebreak
$|\psi| \leq \sqrt{2}|\sigma|$ 
superfluous---the phantom scalar field is taken to be real throughout 
this work. Indeed, were the phantom field purely imaginary, the action 
would reduce to that of a canonical (conventional) scalar field, whereas 
the present investigation is exclusively concerned with the phantom model.\\
In view of equation (\ref{eq11}), and given that $k \leq 0$, 
the condition $|\psi| \leq \sqrt{2}|\sigma|$ leads to the following 
expression:
\begin{align}\label{con2}
V -\frac{1}{3} \Theta^{2} \leq 0.
\end{align}
Consequently, by virtue of equations (\ref{con1}) and (\ref{con2}), 
it follows that:
\begin{align}\label{eq14}
\dot{\Theta} \leq V -\frac{1}{3} \Theta^{2} \leq 0.
\end{align}
It is noteworthy that the derivation of this relation is entirely 
independent of any assumption regarding the sign of the potential; 
the relation holds universally, irrespective of both the sign and 
the functional form of the potential.
Employing equation (\ref{eq8}) together with equation 
(\ref{eq11}) multiplied by $2$, and invoking the non-negativity of the potential 
along with the condition $k \leq 0$, one arrives at the following 
inequality:
\begin{align}\label{eq15}
\dot{\Theta} +\Theta^{2}=3V-\frac{2k}{b^{2}} \geq 0.
\end{align}
It is precisely at this juncture that the non-negativity condition on 
the potential finds its origin.\\
Introducing the definition
\begin{align}\label{eq16}
\Theta =3\frac{\dot{a}_{\mathrm{ave.}}}{a_{\mathrm{ave.}}}=
\frac{\dot{A}}{A},
\end{align}
where $a_{\mathrm{ave.}}=\sqrt[3]{ab^{2}}$ denotes the mean scale factor, 
and invoking the positivity of the scale factors, the following arguments 
can be put forth:\\
Substituting (\ref{eq16}) into equation (\ref{eq11}) yields 
$(\ddot{A}/A) \geq 0$. Consequently, $\dot{A}$ is monotonically 
non-decreasing, provided that $\dot{A}|_{t_{0}}>0$. That is, without 
loss of generality, it is assumed that $\dot{A}$ is positive at some 
prescribed time $t_{0}$, which guarantees that $\Theta$ remains positive 
for all $t > t_{0}$, corresponding to a perpetually expanding universe. 
Accordingly, $\Theta$ converges monotonically to some positive constant, 
denoted $\Theta_{\infty}$, as $t \to \infty$.

It has thus been established that $\Theta$ is a positive, monotonically 
decreasing function, and therefore the following asymptotic behavior 
must hold:
\begin{align}\label{eq17}
\dot{\Theta} \to 0 \qquad \text{and} \qquad \Theta \to \Theta_{\infty} \geq 0
\qquad \text{as} \qquad t \to \infty.
\end{align}
The expansion rate $\Theta$ converges to the critical value $\Theta_{\infty}$.

Subtracting equation (\ref{eq11}) from equation (\ref{eq8}), and 
applying the condition $|\psi| \leq \sqrt{2}|\sigma|$, one obtains:
\begin{equation}\label{eq18}
\dot{\Theta} = \frac{k}{b^{2}}-3\sigma^{2} +\frac{3}{2}\psi^{2} \leq 0,
\end{equation}
which is consistent with equation (\ref{eq14}). Since $\dot{\Theta} \to 0$ 
as $t \to \infty$, and given the previously imposed conditions 
$|\psi| \leq \sqrt{2}\sigma$ and $k \leq 0$, both terms $(k/b^{2})$ 
and $(-3\sigma^{2}+(3/2)\psi^{2})$ are individually non-positive; 
hence, each must vanish separately. The vanishing of $(k/b^{2})$ is 
immediate, as the expanding universe implies that the scale factor $b$ 
grows monotonically with time.\\
The vanishing of $\sigma$ is the central objective, and can be 
inferred from the condition $\psi \to 0$ as $t \to \infty$. Since 
$\psi = \dot{\varphi}$, the phantom scalar field must be strictly 
decreasing. Given that the field was constrained to be non-negative, 
it is bounded below by zero; the strict monotonic decrease therefore 
ensures its convergence to zero at late times, which in turn drives 
$\sigma \to 0$, yielding the desired result.\\
Therefore, the phantom scalar field $\varphi$ must be both non-negative 
and strictly decreasing\footnote{
In general, a scalar field that decreases with time can be motivated on strong physical grounds. However, there is no direct observational evidence for such a field. Nevertheless, it is often treated as a dynamical cosmological constant. As is well known, the cosmological constant has a clear physical interpretation: it corresponds to the vacuum energy density, i.e.\ the sum of the zero-point energies of quantum fields. Its value is estimated to be of order $10^{74}\,\mathrm{GeV^4}$. By contrast, if it exists at the present epoch, it is of order of the square of the current Hubble parameter,
\[
\rho_\Lambda = \frac{3}{8\pi G} H_0^2 \sim 10^{-47}\,\mathrm{GeV^4}.
\]
Thus, $\rho_\Lambda$ is smaller than the vacuum energy density by about $10^{-121}$ orders of magnitude. This can be understood if one begins with a large scalar-field energy density,
\[
\rho_\varphi = \frac{1}{2}\dot{\varphi}^{\,2} + V(\varphi),
\]
which has evolved to $\rho_\varphi \sim 10^{-47}\,\mathrm{GeV^4}$ at the present time. Moreover, to achieve nearly 60 e-folds of inflation, the field $\varphi$ must have decayed substantially. Then, the transition from the cold Big-Bang to the hot Big-Bang requires further decay of the scalar field through particle production. If the field decays monotonically toward the end of inflation, the cosmological constant problem can be avoided.

}.

In light of the assumptions delineated above and the constraint equation 
(\ref{eq11}), the following asymptotic relation can be established:
\begin{equation}
\Theta \to \sqrt{3V} \qquad \text{as} \qquad t \to \infty.
\end{equation}
Equation (\ref{eq10}) governs the dynamics of a damped harmonic 
oscillator; consequently, the system must ultimately settle at the 
minimum of the potential, henceforth denoted $V_{\mathrm{min.}}$. 
Thus, as $t \to \infty$, one has $\Theta_{\infty}=\sqrt{3V_{\mathrm{min.}}}$. 
For all potentials examined in the present work, the non-negativity and 
strict monotonic decrease imply that the unique minimum is located at 
zero. The simultaneous vanishing of both $\Theta$ and $\sigma$ in the 
limit $t \to \infty$ precludes any definitive conclusion regarding the 
isotropization of the universe in this particular regime solely on the 
basis of the present theorem; a more powerful framework is therefore 
required, which will be addressed through the ``dynamical systems approach'' in the subsequent analysis.

It should be noted that, had we not imposed the condition $k \le 0$, we would not have been able to establish the theorem. 
The interested reader 
may verify that relaxing this condition to $k > 0$ renders the 
final combination unattainable.

The foregoing results are now encapsulated in the following theorem.
\begin{thmd}[\textbf{Generalization of the Heusler Theorem to the Phantom Model}]
Let $k \le 0$ (this condition covers BI and BIII metrics) and suppose 
that the ``phantom scalar field $\varphi$ is a real non-negative and strictly decreasing function of time'' such that $|\psi|=|\dot{\varphi}| \le \sqrt{2}\, |\sigma|$, 
and that ``the potential function is non-negative.'' In particular, for the 
five potential functions considered herein, the non-negativity condition 
is satisfied provided that the quantities $V_{0a}$, $V_{0b}$, $V_{1b}$, 
$V_{0d}$, and $\beta$ are non-negative constants. 
If the universe is expanding at a given time $t_{0}$, i.e. if $\Theta(t_{0}) \geq 0$, then the following results hold:
\begin{itemize}
    \item[\maltese]
    For all subsequent times $t \geq t_{0}$, one has $\Theta(t) \geq 0$ 
    and $\dot{\Theta}(t) \leq 0$; that is, the expansion rate is 
    non-negative and monotonically decreasing as we expect from the observational point of view.
    \item[\maltese]
    In the asymptotic regime $t \to \infty$, it follows that 
    $\sigma \to 0$ and $(k/b^{2}) \to 0$.
    \item[\maltese]
    In the asymptotic regime $t \to \infty$, the expansion rate 
    satisfies $\Theta \to \sqrt{3V_{\mathrm{min.}}}$, where 
    $V_{\mathrm{min.}}$ denotes the global minimum of the potential 
    function. For the five specific potentials examined in the present 
    study, this reduces to $\Theta \to 0$, since each of these potentials 
    attains its minimum at zero.
\end{itemize}
\end{thmd}
As demonstrated above, the foregoing theorem applies only to phantom-field gravitational models whose potentials possess nonzero minima. Consequently, it yields no conclusion regarding the isotropization of no conclusion regarding the isotropization of BI and BIII models considered here for the five potentials under investigation. Indeed, all five potentials have vanishing minima, $V_{\mathrm{min.}}=0$,  in which case both $\Theta$ and $\sigma$ vanish simultaneously as $t \to \infty$. The ratio $(\sigma/\Theta)$ therefore becomes indeterminate, and the isotropization criterion cannot be established on the basis of the theorem.

However, this should not be regarded as a deficiency of the theorem itself. Rather, it appears to reflect a generic limitation shared by a broad class of gravitational models considered in the literature: the theorem does not apply to potentials whose minimum vanishes. This restriction is common to a variety of gravitational frameworks.

Nevertheless, a complementary analytical approach is warranted for 
the examination of such potentials.
The dynamical-systems approach offers a powerful framework for this purpose. 
In the following section, we systematically apply this methodology to investigate the relevant aspects of the problem and address the remaining open questions.

\section{Analysis of Inflation and Isotropization via the Dynamical 
Systems Approach\label{sectdysy}}
To cast the system within the dynamical systems framework, the 
governing equations must first be reformulated as an autonomous 
system. To this end, the following variables are 
introduced:
\begin{align}
&S=\frac{\sigma}{\Theta}, \quad U=\frac{\sqrt{V}}{\Theta},
\quad P=\frac{\psi}{\Theta}, \quad \lambda=-\frac{\mathrm{d}V/\mathrm{d}\varphi}{V},
\nonumber \\
&\Gamma =V\frac{\mathrm{d}^{2}V/\mathrm{d}\varphi^{2}}{\left( \mathrm{d}V/\mathrm{d\varphi} \right)^2},
\quad
f=\lambda^{2} \left(\Gamma -1\right).
\end{align}
Before proceeding, let us clarify the meaning of the defined variables.

The variable $S$ is required to tend to zero, which is precisely Hawking’s isotropization condition, namely strong isotropization. Naturally, if at a critical point, $S$ takes a nonzero constant value, or if it is a time-dependent function whose limit as $t\to\infty$ is nonzero, then the strong isotropization condition is not satisfied. Both positive and negative signs of this quantity are physically admissible. Since we focus on an expanding universe, according to the definition of the shear scalar 
$\sigma=(H_{a}-H_{b})/\sqrt{3}$, the positive and negative values merely indicate differences in the expansion rates along different directions.

The variable $U$ must be real and positive. Since it is defined through the square root of the potential, and the potential itself is real and positive, a negative $U$ would imply $\Theta<0$, corresponding to a contracting universe. Therefore, to ensure cosmic expansion, $U$ must be real-valued and non-negative at a physically admissible critical point.
This variable also allows us to determine whether the potential contributes to the dynamics responsible for inflation and isotropization. In particular, if it vanishes at a critical point, the potential does not play an active role in driving these processes at that point.
The motivation for defining this quantity in terms of the square root of the potential is that, in the equation governing $U$, it appears with the odd powers one and three, whereas it enters the remaining equations quadratically. For convenience, and to avoid introducing $\sqrt{U}$ explicitly into the equations, we define this quantity as
$\sqrt{V}/\Theta$,
rather than as
$V/\Theta^{2}$.

The quantity $P$ measures the rate of change of the phantom field relative to the expansion rate. Since $\psi=\dot{\varphi}$ and $\Theta>0$ for an expanding universe, a negative value of $P$ at a critical point indicates a phantom field that decreases monotonically with time. Although this is not a mandatory requirement, such behavior is physically more desirable, as discussed earlier. The value of $P$ allows us to assess whether the kinetic energy of the phantom field contributes to the dynamics responsible for inflation and isotropization at a given critical point. If $P$ vanishes at that point, the kinetic contribution is absent.

The parameter $\lambda$ is the logarithmic slope of the potential; in fact, $|\lambda|$ measures how rapidly the potential changes relative to its own magnitude. One may regard $|\lambda|$ as a kind of potential-Hubble parameter, since it has a clear interpretation analogous to the familiar Hubble parameter associated with the scale factor. The reason we often use logarithmic derivatives is that this eliminates the effect of the absolute size of the quantity and retains only its relative rate of change.

The quantities $\Gamma$ and $f$ do not have independent equations of motion of their own in our system; rather, they are introduced solely as physically meaningful and well-motivated auxiliary variables.\\
The parameter $\Gamma$ is a well-known quantity in the inflationary literature. It introduced in~\cite{gammaref} to characterize the cosmological dynamics of the potential. The properties of $\Gamma$ determine whether tracking solutions exist. For a given potential, three possibilities arise:
\begin{align}\label{eqgamma1}
\left\{
      \begin{array}{ll}
	\Gamma<1 &\Longrightarrow \hbox{Thawing}, \\
	\Gamma=1 &\Longrightarrow \hbox{Scaling}, \\
	\Gamma>1 &\Longrightarrow \hbox{Tracker}.
 \end{array}
       \right.
\end{align}
Nevertheless, this definition is intended primarily to characterize the global behavior of the system, rather than specific critical points, which we will examine individually. We therefore refrain from discussing the conventional classification to which each potential belongs. Moreover, in the reference cited above, this definition was formulated for a canonical scalar field rather than for a phantom field. Owing to the negative sign of its kinetic term, phantom dynamics differ from those of a canonical scalar field; consequently, definitions introduced for canonical fields should, in principle, be applied with caution to such models.\\
Although the function $f$ does not appear directly among the coordinates of the critical points, it nevertheless helps determine the existence of critical branches and related structures. Through $f$, the dynamical equations for all five potential families can be written in closed form. If two potentials share the same $f$, this implies that their relative-slope evolution is identical.\\
For real-valued potentials of a canonical scalar field, the following classification may likewise be introduced on the basis of Eq.~(\ref{eqgamma1}):
\begin{align}
\left\{
      \begin{array}{ll}
	f<0 &\Longrightarrow \hbox{Thawing}, \\
	f=0 &\Longrightarrow \hbox{Scaling}, \\
	f>0 &\Longrightarrow \hbox{Tracker}.
 \end{array}
       \right.
\end{align}

By virtue of equations (\ref{eq7})--(\ref{eq10}), the governing 
system can be recast in terms of the newly introduced variables 
as the following autonomous system:
\begin{align}
\Theta^{\prime}=&\; \Theta \left( -\frac{1}{3}-2S^{2}+U^{2}+P^{2} \right),\\
S^{\prime}=&-\frac{\sqrt{3}}{9}-\frac{2S}{3}+\frac{\sqrt{3}}{3}S^{2}+2S^{3}\nonumber \\
&+P^{2}\left(\frac{\sqrt{3}}{6}+S\right)+
U^{2}\left(\frac{\sqrt{3}}{3}-S\right),\\
U^{\prime}=&\; U\left( \frac{1}{3}+2S^{2}-U^{2}+P^{2}-\frac{\lambda}{2}P \right),\\
P^{\prime}=&\; P\left( -\frac{2}{3}+2S^{2}+P^{2} \right)+U^{2}\left(\lambda -P\right),\\
\lambda^{\prime}=&-Pf,
\end{align}
where $f$ admits the following general structure:
\begin{align*}
f(\lambda)=\alpha_{1}\lambda^{2}+\alpha_{2}\lambda+\alpha_{3},
\end{align*}
and a prime denotes differentiation with respect to the rescaled time 
variable $\tau$, defined via $\prime = \mathrm{d}/\mathrm{d}\tau = 
\Theta^{-1}\mathrm{d}/\mathrm{d}t$. 
Had we used overdots, denoting derivatives with respect to cosmic time $t$, instead of primes, the right-hand side of the dynamical equations would have acquired an additional factor of $\Theta$. This would introduce the trivial critical point
$\Theta = 0$,
corresponding to a non-expanding universe\footnote{We emphasize, however, that $\Theta=0$ does not necessarily represent
a critical point or a non-expanding universe; it may instead
correspond to a turning point, such as a bounce or the onset of
recollapse, or to a static configuration. The investigation of
such phenomena may be better suited to alternative choices of
variables or parametrizations. In the present formulation, the
decoupling of $\Theta$ gives it a specific role within the scope
and objectives of this work. Accordingly, within the parametrization
adopted here, the condition $\Theta=0$ carries no further implications
for the reduced system beyond the interpretation stated above.}. Such a solution is not relevant to the cosmological branch considered here, since observations indicate that the Universe has undergone expansion throughout its known history.

In the autonomous system introduced above, $\Theta$ does not appear in any of the remaining evolution equations. Therefore, its evolution equation is not coupled to the other equations in the dynamical sense, and $\Theta$ can be treated separately from the rest of the system. Once the solutions for $S$, $P$, and $U$ have been determined, $\Theta$ can be reconstructed by a straightforward integration. Consequently, the dynamical analysis may be performed in the reduced four-dimensional phase space
$
\{ S, P, U, \lambda \}
$,
rather than in the full five-dimensional space
$
\{ \Theta, S, P, U, \lambda \}
$.
This reduction does not entail any loss of generality. In particular, the variables $\{S,P,U,\lambda\}$ form a closed subsystem.

This type of reduction is standard in dynamical-systems analyses of cosmological models. More generally, when the expansion variable---the Hubble parameter in homogeneous and isotropic cosmologies, or the expansion scalar $\Theta$ in anisotropic settings---can be reconstructed independently once the closed subsystem has been solved, including its evolution equation in the full autonomous system may introduce redundant critical points and lead to potentially misleading physical interpretations. The degree of freedom associated with $\Theta$ parametrises a one-dimensional expanding manifold along which the cosmological evolution proceeds.

For the five potential functions examined in the present study, the 
corresponding simplified forms of $f$ are listed in Table 
\ref{table31}.
\begin{table}[H]
	\centering
	\caption{
		List of potentials and corresponding function $f$\cite{Roy2018}.
	}\label{table31}
\renewcommand{\arraystretch}{1.3}
	\setlength{\extrarowheight}{3pt}
	\begin{tabular}{c c c c c c}
		\hline \hline
		\textbf{
			Name of Potential
		 }&
		\textbf{
			Potential Form $V(\varphi)$
		 }&
			$f$
		 &
			$\alpha_1$
		&
			$\alpha_2$
		 &
			$\alpha_3$
		 \\
		\hline
			Potential $\mathrm{A}$
		 &
			$V_{0a} \varphi^n$
		 &
			$-\frac{\lambda^2}{n}$
		 &
			$-\frac{1}{n}$
		 &
			$0$
		 &
			$0$
		 \\
		Potential $\mathrm{B}$
		 &
			$V_{0b}\, e^{-\zeta\varphi} + V_{1b}$
		 &
			$-\lambda^2 + \zeta\lambda$
		 &
			$-1$
		 &
			$\zeta$
		 &
			$0$
		 \\
		Potential $\mathrm{C}$
		 &
			$\cosh(\xi\varphi)-1$
		 &
			$-\frac{1}{2}\lambda^2 + \frac{1}{2}\xi^2$
		 &
			$-\frac{1}{2}$
		 &
			$0$
		 &
			$\frac{1}{2}\xi^2$
		 \\
		Potential $\mathrm{D}$
		 &
			$V_{0d}\sinh^{-\alpha}(\beta\varphi)$
		 &
			$\frac{\lambda^2}{\alpha} - \alpha\beta^2$
		 &
			$\frac{1}{\alpha}$
		 &
			$0$
		 &
			$-\alpha\beta^2$
		 \\
		Potential $\mathrm{E}$
		 &
			$2M^{2} \cos^2\left(\frac{\varphi}{2l}\right)$
		 &
			$-\frac{1}{2}\lambda^2 - \frac{1}{2l^2}$
		 &
			$-\frac{1}{2}$
		 &
			$0$
		 &
			$-\frac{1}{2l^2}$
 \\
		\hline \hline
	\end{tabular}
\end{table}

For the five potentials considered here, we identified $87$ critical points, including continuous families of critical points, in the general case. A substantial fraction of these points possess a non-hyperbolic structure and therefore require a center-manifold analysis. Since a systematic analytical treatment of the center manifolds for such a large set of critical points is highly cumbersome, we employed  a perturbative approach as an alternative means of determining their local stability properties. The conclusions reported below are based on an extensive numerical analysis. As a consistency check, we also applied the perturbative procedure to those critical points for which a center-manifold analysis is not required. In all such cases, the perturbative results agree exactly with the stability properties inferred analytically.

Our perturbative analysis was carried out systematically by considering a comprehensive set of all possible perturbation modes for a wide range of choices of the constants. Clearly, numerical values also had to be assigned to the constant parameters appearing in the potentials in order to generate the plots. To this end, we tested numerous parameter choices, guided by the available analytical expressions and by variations in the qualitative behavior of the solutions, particularly with respect to inflation, isotropization, and stability. This procedure enabled us to obtain a comprehensive characterization of the different stability properties.

A further issue concerned the perturbation amplitude, which in the present context corresponds to the displacement from the coordinates of a critical point. An appropriate choice of this amplitude can be made only after determining the sensitivity of the system to perturbations in the different cases. We tested several values and found no appreciable qualitative differences in the resulting behavior. We therefore chose a displacement of magnitude $10^{-4}$ from each critical point. More precisely, if the value of $S$ at a critical point is $S_{0}$, we initiated the trajectory integrations from
\begin{equation*}
S_{0}-2\times 10^{-4}, \qquad
S_{0}-1\times 10^{-4}, \qquad
S_{0}, \qquad
S_{0}+1\times 10^{-4}, \qquad
S_{0}+2\times 10^{-4}.
\end{equation*}
Using a substantially larger number of initial points instead of the five points specified above, or adopting different separations between them, did not provide any additional useful information. We nevertheless explored these alternatives and ultimately selected the above configuration as the most efficient and computationally economical choice.

The interval over which the evolution was monitored, namely the range shown on the horizontal axis, was not the same in all cases. For some perturbations, the change in behavior and the corresponding asymptotic outcome became apparent only after a relatively long evolution, whereas for others these features emerged rapidly.

More than $550$ plots were taken into account in determining the behavior and dynamical character of the critical points. It is evident that including all figures generated by the perturbative analysis, together with a detailed discussion of all possible behaviors of each of the $87$ critical points, would be impractical within the scope of this paper. We therefore summarize the outcome of this extensive and time-consuming work as follows:
\begin{enumerate}
\item We sum up the results in tables in the~\ref{appendix_tables}. For each potential, we provide two tables. The first contains the coordinates of all critical points and their corresponding eigenvalues. The second provides all relevant information concerning the possible behaviors associated with the critical points in the general case, including the exact solution, the general stability character, the corresponding cosmological solution, and related properties.

    \item We present a point-by-point analysis of the critical points, together with the plots obtained from the perturbative analysis, only for potential $A$, which serves as a representative example. For the remaining potentials, we provide only a summary of the results directly relevant to the questions and objectives of this paper, since analogous analyses can be readily performed by following the same procedure.
\end{enumerate}
\subsection{The Target Behavior in Our Dynamical-Systems Analysis\label{ourgoal}}
First and foremost, it should be acknowledged that, from a physical point of view, the critical points that can be regarded as admissible are those whose stability is of ``saddle type''. The critical point of interest in the present analysis is neither the point corresponding to the beginning of creation, namely the ``Big Bang'', for which a ``repeller'' would be appropriate, nor the ``final point of the universe'', for which an ``attractor'' would be appropriate. ``All intermediate points'', that is, the points lying between the Big Bang and the final state of the universe, should be of ``saddle'' type. Therefore, for the purpose of achieving our objective, we regard only saddle points as physically relevant and exclude all the remaining points, although all of them have been listed in the tables provided in the supplementary material.

For the sake of clarity, we divided isotropization into two types, namely weak and strong isotropization. 
Our goal is the \emph{simultaneous} occurrence of inflationary expansion and
strong isotropization in a single background and for a single potential; we
wish to determine in which of the $15$ possible situations ($5$ potentials and $3$
types of background) this can take place. Note that the ``simultaneity'' to
which we refer may manifest itself either through two distinct critical
points, each of which serves one of our objectives, or through a single
critical point. Note that we have identified inflationary
expansion exhaustively in the tables, whereas strong isotropization has been
flagged without regard to whether the universe is expanding—that is, it has
been flagged merely on the ground that it occurs. A point that carries only
the feature of strong isotropization is nevertheless unacceptable to us if it
lacks the expansionary property of the universe (i.e., a positive expansion
rate).
In summary, our objective is finding one of the following packages:
\begin{center}
\tcbhighmath[boxrule=2pt,arc=1pt,colback=blue!10!white,colframe=blue,
]{
\begin{gathered}
\textbf{\textcolor[rgb]{1.00,0.00,0.00}{Package 1}-- A CP subject to the following conditions:}\\
\bigg\{
\textbf{``\textcolor[rgb]{0.50,0.00,0.50}{Inflationary Expansion}''}
\, \bm{+} \, \textbf{``\textcolor[rgb]{0.50,0.00,0.50}{Strong Isotropization}''}
 \, \bm{+} \, \textbf{``\textcolor[rgb]{0.50,0.00,0.50}{Saddle Nature}''}
\bigg\}\\[6pt]
\hline \\
\textbf{\textcolor[rgb]{1.00,0.00,0.00}{Package 2}-- Two CPs of the form:}\\
\text{I- A CP satisfying the following conditions:}\\ \bigg\{
\textbf{``\textcolor[rgb]{0.50,0.00,0.50}{Inflationary Expansion}''}
 \, \bm{+} \, \textbf{``\textcolor[rgb]{0.50,0.00,0.50}{Saddle Nature}''}
\bigg\}\\
\text{II- A CP characterized by the following properties:}\\
\bigg\{
\textbf{``\textcolor[rgb]{0.50,0.00,0.50}{Expanding Universe}''}
\, \bm{+} \, \textbf{``\textcolor[rgb]{0.50,0.00,0.50}{Strong Isotropization}''}
 \, \bm{+} \, \textbf{``\textcolor[rgb]{0.50,0.00,0.50}{Saddle Nature}''}
\bigg\}
\end{gathered}
}
\end{center}
If this condition is not realized, inflation accompanied by weak isotropization may be considered a less favorable alternative; nevertheless, it would not constitute the ideal or most desirable scenario. If we are forced to choose between inflation and isotropization, inflation takes priority over isotropization.
\subsection{Investigation of Potential $\mathrm{A}$:
$V_{0a}\, \varphi^{n}$
}
In this subsection, a point-by-point analysis of potential $A$ is presented. The figures produced by perturbation method for potential $\mathrm{A}$ have been illustrated in the~\ref{appendix_figures}.\\
For this potential, we examined the standard powers commonly used in the literature, in particular in Planck-related phenomenological analyses, i.e.
\begin{align*}
	n=1,\, 2,\, \frac{2}{3},\, \frac{4}{3},\, 3,\, 4.
\end{align*}
No substantial change was observed in the qualitative dynamics or in the stability character of the system. This is because $n$ does not enter the coordinates of the critical points, the eigenvalues, or the exact analytic solutions. Therefore, in reporting the center-manifold diagrams we restrict the presentation to the representative case $n=2$, which is also among the most frequently used choices.

This potential contains nine critical points.
The critical-point coordinates and eigenvalues have been listed in Table~\ref{tableVA} (in the~\ref{appendix_tables1}); the corresponding stability type, exact solutions, associated cosmological model, Inflationary expansion behavior, weak and strong isotropization, and related physical properties have been summarized in Table~\ref{tableSVA}.

\paragraph{$\blacklozenge$ Preliminary physical summary of the critical points for potential $\mathrm{A}$}
As shown in the tables, a large subset of the critical points of this potential satisfies
\begin{align}
	U=0.
\end{align}
At these points, the potential contribution to the normalized dynamical variable is switched off. Consequently, most of them represent shear-, kinetic-, or curvature-driven regimes rather than a genuinely potential-dominated phantom phase. Only
$\mathrm{CP}_{A,6-1}$ and $\mathrm{CP}_{A,6-2}$
have
$
U=\pm 1/\sqrt{3}.
$
Given
$
U=\sqrt{V}/\Theta,
$
the physically preferred branch is the one compatible with an expanding universe (i.e. $\Theta>0$ or consequently $U \ge 0$.

\subsubsection{Analysis of $\mathrm{CP}_{A,1}$\label{CPA1}}
This critical point has the coordinates
\begin{align}
	S=-\frac{\sqrt{3}}{6},
	\qquad
	U=0,
	\qquad
	P=0,
	\qquad
	\lambda=\lambda ,
\end{align}
and the eigenvalues of the Jacobian evaluated at it are
\begin{align}
	-\frac{1}{2},\,\frac{1}{2},\, -\frac{1}{2},\,0.
\end{align}
The simultaneous presence of positive and negative nonzero eigenvalues is sufficient to identify the point as a saddle, independently of the center-manifold analysis. The zero eigenvalue gives a one-dimensional center manifold, but the saddle character is already fixed by the linear sector. The perturbative center-manifold analysis confirms this result, as illustrated in Fig.~\ref{FigPotA1}.

The exact solution gives
\begin{align}
	\Theta=\frac{2}{t},
	\qquad
	\sigma=-\frac{1}{\sqrt{3}\,t},
	\qquad
	\frac{\sigma}{\Theta}=-\frac{\sqrt{3}}{6},
	\qquad
	a_{\mathrm{ave.}}(t)=c_{0}t^{2/3}.
\end{align}
For $t>0$, the universe is expanding since $\Theta>0$\footnote{
The conditions of an expanding universe are: $\{\Theta >0 \;\; \& \;\; U \ge 0\}$.}. However,
\begin{align}
	a_{\mathrm{ave.}}(t)\propto t^{2/3}
\end{align}
corresponds to a non-accelerating power-law expansion, so inflation does not occur. Since the scale factor grows too slowly to address the horizon and flatness problems, this point cannot by itself provide a viable early-universe inflationary phase.\\
For a canonical scalar field, this scaling, i.e. $t^{2/3}$,
arises both during the post-inflationary epoch, when the inflaton oscillates coherently and reheating takes place in an approximately quadratic potential, and during the standard matter-dominated era. For a phantom scalar field with a positive monomial potential, however, this scaling cannot generally be interpreted as the consequence of coherent field oscillations. It can occur only when the dominant cosmic component has an effective matter-like equation of state, $w_{\mathrm{eff}}\simeq 0$, for example, during an ordinary matter-dominated epoch. A detailed justification of this statement is beyond the scope of the present paper.

Regarding isotropization,
\begin{align}
	\sigma\propto \frac{1}{t}\to 0 \qquad (t\to\infty),
\end{align}
so weak isotropization is achieved. However,
\begin{align}
	\frac{\sigma}{\Theta}=-\frac{\sqrt{3}}{6}
\end{align}
is constant and nonzero; hence strong isotropization is unattainable. This point therefore describes a regime in which shear decays in absolute magnitude but does not become negligible relative to the expansion rate.

Since
\begin{align}
	\psi=0,
	\qquad
	\varphi(t)=c_{1},
	\qquad
	V=0,
\end{align}
the scalar field is frozen and the phantom dynamics is inactive. Indeed, in such cases, the phantom field does not contribute to the kinetic term. Furthermore, the phantom-field energy density vanishes at this critical point. The point is therefore better interpreted as a geometric/shear-dominated regime than as an active phantom phase.

Geometrically, this point belongs to the $\mathrm{KS}$ model.\\
The absence of inflation together with the absence of strong isotropization is not in tension with the result of Burd and Barrow\footnote{
Burd and Barrow concluded that, for canonical scalar field, if inflation occurs then isotropy is always reached.
}, since no inflationary phase is present.

This result is also qualitatively consistent with the Collins--Hawking perspective\footnote{
Collins and Hawking demonstrated that ordinary matter can achieve isotropy without inflation within the BI universe.}:
in the absence of inflation, weak isotropization may occur, although complete or strong isotropization is not achieved. Nevertheless, the present critical point belongs to the $\mathrm{KS}$ rather than the $\mathrm{BI}$ class, and the scalar field is phantom rather than canonical.

\subsubsection{Analysis of $\mathrm{CP}_{A,2-1}$\label{CPA21}}
The coordinates of this critical point are given by
\begin{align}
	S=\frac{1}{\sqrt{3}},
	\qquad
	U=0,
	\qquad
	P=0,
	\qquad
	\lambda=\lambda,
\end{align}
with the corresponding eigenvalues being
\begin{align}
	2,\,1,\,0,\,0.
\end{align}
There are two positive and two zero eigenvalues, so the center manifold is two-dimensional. The perturbative center-manifold analysis shows that the point is a saddle; the neutral directions therefore do not turn the point into a pure source. This is confirmed in Fig.~\ref{FigPotA21}.

The exact solutions turn out to be
\begin{align}
	\Theta=\frac{1}{t},
	\qquad
	\sigma=\frac{1}{\sqrt{3}\,t},
	\qquad
	\frac{\sigma}{\Theta}=\frac{1}{\sqrt{3}},
	\qquad
	a_{\mathrm{ave.}}(t)=c_{0}t^{1/3}.
\end{align}
For $t>0$, the universe expands, but the expansion is much too slow to be inflationary.\\ Since $\sigma\to0$, weak isotropization is realized, whereas
\begin{align}
	\frac{\sigma}{\Theta}=\frac{1}{\sqrt{3}}
\end{align}
remains constant and nonzero; strong isotropization is therefore absent.

Moreover
\begin{align}
	\psi=0,
	\qquad
	\varphi(t)=c_{1},
	\qquad
	V=0.
\end{align}
Thus, the phantom scalar field remains constant, and hence this branch does not correspond to a dynamically active phantom phase. Again, like $\mathrm{CP}_{A,1}$, the phantom field does not contribute to the kinetic term and the energy density associated with the phantom field is zero at this critical point.

This point corresponds to a $\mathrm{BI}$ universe and represents a non-inflationary anisotropic regime with weak isotropization only. This behavior is compatible with the Collins--Hawking result for BI in weak sense, where shear decay can occur without inflation, although complete isotropization is not reached. Nevertheless, in the present case, the scalar field is phantom rather than canonical, providing an additional distinction from the Collins--Hawking framework.

\subsubsection{Analysis of $\mathrm{CP}_{A,2-2}$\label{CPA22}}
This critical point, which belongs to the $\mathrm{BI}$ universe, is located at
\begin{align}
	S=-\frac{1}{\sqrt{3}},
	\qquad
	U=0,
	\qquad
	P=0,
	\qquad
	\lambda=\lambda,
\end{align}
with corresponding eigenvalues
\begin{align}
	\frac{2}{3},\,1,\,0,\,0.
\end{align}
The system has two positive and two zero eigenvalues, resulting in a two-dimensional center manifold. Similar to $\mathrm{CP}_{A,2-1}$, the perturbative center-manifold analysis identifies this point as a saddle, as illustrated in Fig.~\ref{FigPotA22}.

The exact relations are found to be
\begin{align}
	\Theta=\frac{1}{t},
	\qquad
	\sigma=-\frac{1}{\sqrt{3}\,t},
	\qquad
	\frac{\sigma}{\Theta}=-\frac{1}{\sqrt{3}},
	\qquad
	a_{\mathrm{ave.}}(t)=c_{0}t^{1/3}.
\end{align}
Thus, for $t>0$, the universe is expanding but non-inflationary. As in the previous point, the shear decays and weak isotropization is achieved, while the constant nonzero value of $\sigma/\Theta$ excludes strong isotropization.

The scalar field is again constant:
\begin{align}
	\psi=0,
	\qquad
	\varphi(t)=c_{1},
	\qquad
	V=0.
\end{align}
Hence, this point also does not describe an active phantom regime. Its only difference from $\mathrm{CP}_{A,2-1}$ is the sign of the shear, which reverses the anisotropic orientation; that is, in the present case, $H_{a}$ remains smaller than $H_{b}$, in contrast to the previous case, without altering the physical conclusion. Therefore, the two points are physically equivalent with respect to inflationary behavior and isotropization. Hence, we do not analyze this case further.
\subsubsection{Analysis of $\mathrm{CP}_{A,3-1}$\label{CPA31}}
This critical point is characterized by the coordinates
\begin{align}
	S=\sqrt{\frac{2-3P^{2}}{6}},
	\qquad
	U=0,
	\qquad
	P=P,
	\qquad
	\lambda=0.
\end{align}
Reality of $S$ requires\footnote{The requirement that the Hubble parameter be real in different directions implies that the parameter $S$ must also be real.}
\begin{align}
	2-3P^{2}\ge 0
	\qquad \Longrightarrow \qquad
	P^{2}\le \frac{2}{3},
\end{align}
so the allowed range is
\begin{align}\label{arp1}
	-\sqrt{\frac{2}{3}}\le P\le \sqrt{\frac{2}{3}}.
\end{align}
The corresponding eigenvalues are as follows:
\begin{align}
	\frac{4-\sqrt{4-6P^{2}}}{3},\,
	1,\,
	0,\,
	0 .
\end{align}
In the allowed domain, the first eigenvalue is non-negative and the second is positive, while the remaining two eigenvalues vanish. The stability must therefore be determined through a center-manifold analysis.
For $P=\pm\sqrt{2/3}$, the center manifold is three-dimensional, with the perturbative analysis revealing saddle behavior, whereas for the remaining allowed values of $P$, it is two-dimensional and the point acts as a repeller (source), as illustrated in Figs.~\ref{FigPotA31-inf} and \ref{FigPotA31-iso}.

The exact solutions are found to be
\begin{equation}
	\begin{split}
		&\Theta=\frac{1}{(1-2P^{2})t},
		\qquad
		\sigma=-\frac{\sqrt{12-18P^{2}}}{6(2P^{2}-1)t},\\
		&\frac{\sigma}{\Theta}=\frac{\sqrt{12-18P^{2}}}{6},
	\end{split}
\end{equation}
and
\begin{equation}
	\begin{split}
		&a_{\mathrm{ave.}}(t)=c_{0}\, t^{1/(3-6P^{2})},
		\qquad
		\psi=\frac{P}{(1-2P^{2})t},\\
		&\varphi(t)=\frac{P\ln t}{1-2P^{2}}+c_{1},
		\qquad
		V=0.
	\end{split}
\end{equation}
For $t>0$, the sign of $\Theta$ is controlled by $1-2P^{2}$. The universe expands for $P^{2}<1/2$, contracts for $P^{2}>1/2$, and the critical solution becomes singular at $P^{2}=1/2$. The range over which expansion occurs coincides with the previously determined allowed range of $P$, since this range constitutes a subinterval of Eq.~(\ref{arp1}). Accordingly, the physically admissible range is now given by
$
-\sqrt{1/2}<P<\sqrt{1/2}
$.

Since
\begin{align}
	a_{\mathrm{ave.}}(t)\propto t^{m};
	\qquad
	m=\frac{1}{3-6P^{2}},
\end{align}
power-law inflation in the expanding branch requires\footnote{
It should be emphasized that the sign of $c_{0}$, which determines the sign of the scale factor, has no bearing on inflation, expansion, contraction, or related cosmological behavior. This is because our criterion for expansion is solely the positivity of the Hubble parameter, i.e., the expansion rate. In the expression for the Hubble parameter, any overall sign of the scale factor cancels out and therefore does not affect the classification of the cosmological evolution. More generally, the sign of the scale factor itself is not a physical criterion for distinguishing between expansion and contraction, nor for identifying inflation. Rather, it merely specifies the orientation of the corresponding expansion/contraction relative to the chosen coordinate direction. In particular, a negative scale factor along a given spatial direction indicates that the expansion/contraction in that direction proceeds opposite to the conventionally assigned positive direction of the coordinate axis.
For this reason, we do not discuss the signs of the individual scale-factor coefficients, as they do not affect the physical characterization of the cosmological evolution.
}
\begin{align}
	m>1
	\qquad \Longleftrightarrow \qquad
	P^{2}>\frac{1}{3}.
\end{align}
Thus,
\begin{align}
	\frac{1}{3}<P^{2}<\frac{1}{2} \quad \Longrightarrow
\left\{-\sqrt{\frac{1}{2}}<P<-\sqrt{\frac{1}{3}} \quad \text{or} \quad
\sqrt{\frac{1}{3}}<P<\sqrt{\frac{1}{2}} \right\}
\end{align}
gives power-law inflation in the expanding branch. In these intervals the scale factor grows fast enough to address the horizon and flatness problems. Note that, as mentioned earlier, within these intervals, the critical point becomes a repeller and is therefore not physically admissible in the present context. Although a repelling critical point may appropriately describe an initial state, it is incompatible with a scenario in which the cosmological trajectory must pass through the point. As is clear, the power-law accelerated expansion is determined by the complete field equations rather than by the sign of $\dot{\varphi}$, because the sign of $\dot{\varphi}$, which is determined by the sign of $P$, does not affect the average scale factor, since $P$ enters the exponent of the scale factor quadratically.

Weak isotropization is realized with
\begin{align}
	\sigma\propto \frac{1}{t}.
\end{align}
However,
\begin{align}
	S=\frac{\sigma}{\Theta}=\frac{\sqrt{12-18P^{2}}}{6}
\end{align}
vanishes only when
\begin{align}
	12-18P^{2}=0
	\qquad \Longrightarrow \qquad
	P^{2}=\frac{2}{3},
\end{align}
and the spacetime becomes shear-isotropic with the saddle nature for the corresponding CP, but it falls outside the expanding regime. Thus, despite its saddle nature, it is not physically admissible.

Therefore, for $\mathrm{CP}_{A,3-1}$, the precise classification is as follows:
\begin{itemize}
	\item for $P=\pm\sqrt{2/3}$, strong isotropization holds (i.e. $S=0$), but in contracting phase, and the point is a saddle;
	\item for $1/3<P^{2}<1/2$, the expanding universe undergoes power-law inflation and the point is a repeller;
	\item for $P^{2}<1/3$, the expansion is non-inflationary and the point is a repeller;
	\item for $1/2<P^{2}\le2/3$, the branch is contracting with an inflationary power-law repeller behavior.
\end{itemize}
It is worth noting that weak isotropization occurs in all cases.

The scalar field evolves as
\begin{align}
	\varphi(t)=\frac{P\ln t}{1-2P^{2}}+c_{1}.
\end{align}
Hence, in the expanding branch $P^{2}<1/2$, the field increases for $P>0$ and decreases for $P<0$. If one requires an expanding, inflationary universe with a decreasing scalar field, the interval
\begin{align}
	-\frac{1}{\sqrt{2}}<P<-\frac{1}{\sqrt{3}}
\end{align}
is physically more appealing, however, since the point is repeller under this condition, it cannot be an appropriate candidate for inflation. Strong isotropization, however, is still absent in this branch, except at $P=\pm\sqrt{6}/3$, which lies outside the expanding interval.

Although this point can generate inflation, the conditions $U=0$ and $V=0$ imply that the inflationary behavior is not a standard potential-dominated phantom phase. It is better interpreted as an effective kinetic/geometric inflationary regime, and its physical meaning should therefore be treated with care.\\
Within the inflationary expansion regime, i.e.
$(1/3)<P^{2}<(1/2)$,
the following interpretations can be deduced according to the sign of $P$:
\begin{itemize}
  \item \textbf{For $\mathbf{P>0}$:}\\
For positive values of $P$, we have:
\[
\varphi(t)=r_{1} \ln (t)+c_{1};
\qquad r_{1}=\frac{P}{1-2P^{2}}>0,
\]
for which
\[
\psi=\dot{\varphi}=\frac{r_{1}}{t}.
\]
Although the scalar potential vanishes, $V=0$, the scalar field retains a nontrivial kinetic contribution. The resulting cosmological solution exhibits power-law accelerated expansion. Thus, the accelerated expansion in this branch is realized through the full dynamical structure of the model rather than through a constant potential energy. In particular, the increasing logarithmic evolution of the phantom field provides a time-dependent scalar contribution that scales as $t^{-2}$.
\item \textbf{For $\mathbf{P<0}$:}\\
A similar interpretation can be applied to the branch with a decreasing logarithmic scalar field,
\[
\varphi(t)=-r_{2} \ln (t)+c_{1};
\qquad r_{2}=\frac{|P|}{1-2P^{2}}>0,
\]
for which
\[
\psi=\dot{\varphi}=-\frac{r_{2}}{t}.
\]
Despite the opposite direction of the scalar-field evolution, the kinetic contribution retains the same time dependence, since it depends on $\dot{\varphi}^{2}\propto t^{-2}$. 
\end{itemize}

Irrespective of the specific parameter values, this critical point belongs to the $\mathrm{BI}$ universe. For a particular choice of $P$, its cosmological behavior is consistent with the Collins--Hawking result, although their analysis was formulated for a canonical scalar field rather than a phantom one. Likewise, over an appropriate range of $P$, the Burd--Barrow result is recovered when the isotropization is interpreted in the weak sense. Nevertheless, the stability properties of the critical point impose an important restriction on its physical admissibility and, consequently, on the set of critical points that can be accepted in the present cosmological scenario.

\subsubsection{Analysis of $\mathrm{CP}_{A,3-2}$\label{CPA32}}
This critical point has a structure analogous to that of $\mathrm{CP}_{A,3-1}$, with all corresponding features being identical except for two differences, neither of which alters the overall physical picture. First, the sign of the shear-related parameter $S$ is reversed. This merely reverses the anisotropic orientation; in the present case, $H_{a}$ remains smaller than $H_{b}$, in contrast to the previous case, without affecting the physical conclusion concerning the admissible range. In fact, the admissible range of $P$ is identical for both critical points.
Second, the eigenvalue
$(4-\sqrt{4-6P^{2}})/3$
is replaced by
$(4+\sqrt{4-6P^{2}})/3$.
This modification does not lead to any qualitative change in the stability behavior, since both eigenvalues remain positive throughout the admissible range. Nevertheless, we do not rely solely on this structural similarity and have explicitly examined the perturbative behavior of both critical points. The qualitative perturbative dynamics of $\mathrm{CP}_{A,3-1}$ and $\mathrm{CP}_{A,3-2}$ are found to be identical. Since all other physical and dynamical properties coincide, apart from the two differences noted above, we refrain from repeating the same analysis and omit a redundant discussion of this critical point.

\subsubsection{Analysis of $\mathrm{CP}_{A,4}$\label{CPA4}}
At this point,
\begin{align}
	S=\frac{1}{\sqrt{3}},
	\qquad
	U=+1,
	\qquad
	P=0,
	\qquad
	\lambda=0,
\end{align}
and the eigenvalues are
\begin{align}
	1,\,-2,\,-1,\,0.
\end{align}
The coexistence of positive and negative eigenvalues immediately identifies the point as a saddle. The zero eigenvalue only indicates a one-dimensional center manifold and does not alter this conclusion. The perturbative center-manifold analysis confirms the result (See Fig.~\ref{FigPotA4}).

The exact solutions are given by
\begin{align}
	\Theta=\theta_{0},
	\qquad
	\sigma=\frac{\theta_{0}}{\sqrt{3}},
	\qquad
	\frac{\sigma}{\Theta}=\frac{1}{\sqrt{3}},
	\qquad
	a_{\mathrm{ave.}}(t)=e^{\theta_{0}t/3},
\end{align}
\begin{align}\label{eq4.61}
	\psi=0,
	\qquad
	\varphi(t)=c_{1},
	\qquad
	V=\theta_{0}^{2}.
\end{align}
For $\theta_{0}>0$, the universe is expanding exponentially, and inflation is realized. The scale factor then grows faster than the horizon scale, allowing the model to address the horizon and flatness problems. Since such a solution effectively violates the dominant energy condition, accelerated expansion is expected.

However,
\begin{align}
	\sigma=\frac{\theta_{0}}{\sqrt{3}}=\text{constant},
	\qquad
	\frac{\sigma}{\Theta}=\frac{1}{\sqrt{3}}=\text{constant and nonzero}.
\end{align}
Thus neither weak nor strong isotropization occurs. This point, which belongs to $\mathrm{BIII}$ universe, is therefore inflationary but not isotropizing. This is important in comparison with the Burd and Barrow result: here inflation alone is not sufficient to isotropize the spacetime.

According to Eq.~(\ref{eq4.61}), this critical point corresponds to a frozen-field configuration in which the scalar field remains constant at a nonzero value, while the scalar potential takes a constant positive value. The resulting persistent potential energy, through the corresponding cosmological constraint, yields a constant positive expansion rate, $\Theta$, and consequently an exponentially expanding average scale factor. Thus, the critical point represents an exact inflationary solution with de Sitter-like exponential expansion, despite the absence of dynamical evolution of the scalar field at the fixed point. In particular, the slow-roll parameter, $\epsilon_{\Theta}$, and the deceleration parameter, $q$, demonstrate
$$
\epsilon_{\Theta}=-\; \frac{\dot{\Theta}}{\;\; \Theta^2}=0,
\qquad
q=-1,
$$
confirming the inflationary character of this solution.

It is important to note that the same parameter $\theta_{0}$ that governs the exponential growth of the scale factor also controls the degree of anisotropy. In fact, the expansion rate $\Theta$ is exactly equal to $\theta_{0}$ and therefore directly determines the relative growth rate of the average scale factor: larger $\theta_{0}$ corresponds to faster expansion and, consequently, to a higher inflationary rate. At the same time, increasing $\theta_{0}$ enhances the anisotropy. Conversely, reducing $\theta_{0}$ to suppress the anisotropy necessarily slows down the expansion. Hence, within this solution, an arbitrarily rapid inflationary expansion cannot be achieved simultaneously with an arbitrarily small degree of anisotropy; increasing the inflationary rate inevitably enhances the anisotropic character of the solution.

\subsubsection{Analysis of $\mathrm{CP}_{A,5}$}
This point is spatially analogous to $\mathrm{CP}_{A,4}$, except that
\begin{align}
	U=-1.
\end{align}
Given
\begin{align}
	U=\frac{\sqrt{V}}{\Theta},
\end{align}
the branch $U<0$ is not physically preferred for an expanding universe under the standard interpretation $\sqrt{V}\ge0$, unless $\Theta<0$. It is therefore naturally interpreted as a contracting branch.\\
More fundamentally, within the underlying cosmological gravitational model, and assuming the model itself is physically viable, the potential $U$ is intrinsically required to be positive. However, when the model is formulated mathematically, $U$ enters the resulting equations quadratically. Consequently, solving the reduced system in the reverse direction—i.e., inferring $U$ from the mathematical solutions of the cosmological equations—naturally produces both positive and negative roots. While both branches are mathematically admissible solutions of the reduced equations, they are not physically equivalent: the negative-$U$ branch is an algebraic artifact of the squared dependence and does not represent a physically realizable cosmological state.

The eigenvalues are again
\begin{align}
	1,\,-2,\,-1,\,0,
\end{align}
so the point is a saddle with a one-dimensional center manifold. The perturbative center-manifold analysis confirms this conclusion, as shown in Fig.~\ref{FigPotA5}.

The exact solution gives
\begin{equation}
	\begin{split}
		&\Theta=-|\theta_{0}|,
		\qquad
		\sigma=-\frac{|\theta_{0}|}{\sqrt{3}},\\
		&\frac{\sigma}{\Theta}=\frac{1}{\sqrt{3}},
		\qquad
		a_{\mathrm{ave.}}(t)=e^{-|\theta_{0}|t/3}.
	\end{split}
\end{equation}
The universe is contracting, so inflation in the desired expanding early-universe sense does not occur. Moreover, neither weak nor strong isotropization is present, since $\sigma$ is constant and nonzero and $\sigma/\Theta$ remains constant and nonzero.

The scalar field is again constant:
\begin{align}
	\psi=0,
	\qquad
	\varphi(t)=c_{1},
	\qquad
	V=\theta_{0}^{2}.
\end{align}The positive constant energy density supports a contracting de Sitter
solution characterized by a constant negative

The critical point corresponds to a frozen-field configuration in which the scalar field remains constant at a nonzero value, while the scalar potential is also constant and nonzero. The latter leads to a persistent positive energy density, which gives rise to a constant negative Hubble parameter (expansion rate) and consequently to exponential contraction of the average scale factor. Thus, the critical point represents an exact contracting de Sitter-like state, despite the absence of dynamical evolution of the scalar field at the fixed point.

Thus, from the viewpoint of a viable expanding early-universe scenario, this point has no physical relevance and mainly serves as the mathematical counterpart of the contracting branch.

The emergence of such points does not indicate a deficiency of the model; rather, it results from the presence of quadratic terms, which allow both signs of $U$ as mathematically admissible solutions of the system.

\subsubsection{Analysis of $\mathrm{CP}_{A,6-1}$}
This is one of the ``challenging'' critical points of potential $\mathrm{A}$ and it is characterized by the coordinates
\begin{align}
	S=0,
	\qquad
	U=\frac{1}{\sqrt{3}},
	\qquad
	P=0,
	\qquad
	\lambda=0,
\end{align}
with the following eigenvalues:
\begin{align}
-1,\,-\frac{2}{3},\,-1,\,0.
\end{align}
All nonzero eigenvalues are negative and one eigenvalue is zero, giving a one-dimensional center manifold. The perturbative center-manifold analysis shows that the point is an attractor (See Fig.~\ref{FigPotA61}). Because of the attractor nature of this point, we used the term ``challenging'', as some researchers regard the inflationary point as an attractor-type critical point. In our analysis, however, the cosmological evolution must pass through the inflationary point. Therefore, except at the beginning and end of cosmic evolution, the relevant critical points should be of saddle type, rather than of any other type.

The exact solutions turn out to be
\begin{align}
	\Theta=\theta_{0},
	\qquad
	\sigma=0,
	\qquad
	\frac{\sigma}{\Theta}=0,
	\qquad
	a_{\mathrm{ave.}}(t)=e^{\theta_{0}t/3},
\end{align}
\begin{align}\label{eq4.72}
	\psi=0,
	\qquad
	\varphi(t)=c_{1},
	\qquad
	V=\frac{\theta_{0}^{2}}{3}.
\end{align}
For $\theta_{0}>0$, the universe expands exponentially, and a complete inflationary phase is realized. Moreover,
\begin{align}
	\sigma=0
	\qquad \text{and} \qquad
	\frac{\sigma}{\Theta}=0,
\end{align}
so both weak and strong isotropization hold exactly. This point therefore represents an isotropic de Sitter branch within the $\mathrm{BI}$ sector.

Physically, this is the most favorable point of potential $\mathrm{A}$ for describing the isotropic endpoint of an inflationary phase. Since it is an attractor, however, it is better interpreted as a stable final state than as a natural saddle-type transition channel in the early-universe phase portrait.

According to Eq.~(\ref{eq4.72}), similarly to the preceding argument of $\mathrm{CP}_{A,4}$, the critical point considered here corresponds to a frozen-field configuration in which the scalar field remains fixed at a nonzero value, while the scalar potential also attains a constant positive value. This persistent potential energy leads to a constant positive expansion rate and, consequently, to exponential expansion of the average scale factor. Hence, despite the absence of any dynamical evolution of the scalar field at the fixed point, the critical point describes an exact inflationary, de Sitter-like state. Fortunately, in contrast to point $\mathrm{CP}_{A,4}$, the isotropization parameter at this critical point is no longer coupled to the parameter governing both the expansion and inflationary rates. However, the crucial distinction lies in the stability character of the critical point. At $\mathrm{CP}_{A,4}$, the point was of saddle type, as physically expected for a transient critical point. Here, by contrast, it is an attractor, which prevents this solution from satisfying all of our physical requirements simultaneously. Consequently, this critical point cannot be regarded as an \textit{ideal candidate} fulfilling the three essential criteria we seek: ``strong isotropization'', ``inflationary expansion'', and ``saddle-type stability''.
It is also worth noting that $\mathrm{CP}_{A,4}$ belongs to the $\mathrm{BIII}$ universe, whereas the present critical point belongs to the $\mathrm{BI}$ universe. This difference suggests that the underlying spacetime metric may play a more significant role in determining the stability and cosmological behavior than might ordinarily be anticipated. Nevertheless, no general conclusion or definitive statement can be drawn from this comparison alone; at this stage, such an interpretation should be regarded only as a plausible possibility rather than an established result.

From the Burd--Barrow perspective, this point provides a clean example: inflation occurs and complete isotropization is obtained. It therefore directly realizes the classical intuition that accelerated expansion can dynamically suppress anisotropy. However, this behavior is in tension with the Collins--Hawking result, since isotropization without inflation does not occur in the $\mathrm{BI}$ case considered here. It should be emphasized, however, that the scalar field in the present model is phantom rather than canonical.\\
These comparisons are intended to determine whether the qualitative conclusions established for canonical scalar fields continue to hold in phantom-field cosmologies.

\subsubsection{Analysis of $\mathrm{CP}_{A,6-2}$}
This point is the time-reversed counterpart of $\mathrm{CP}_{A,6-1}$:
\begin{align}
	S=0,
	\qquad
	U=-\frac{1}{\sqrt{3}},
	\qquad
	P=0,
	\qquad
	\lambda=0,
\end{align}
with the same eigenvalues
\begin{align}
-1,\,-\frac{2}{3},\,-1,\,0.
\end{align}
The center manifold is one-dimensional, and the perturbative center-manifold analysis identifies the point as an attractor, as shown in Fig.~\ref{FigPotA62}.

At this critical point, as in the case of $\mathrm{CP}_{A,5}$, the potential $U$ becomes negative, which leads to a negative expansion rate, $\Theta<0$, and hence to a contracting phase. Such a behavior is incompatible with the observationally established expansion history of the Universe, which has remained globally expanding throughout its post-Big-Bang evolution. The negative expansion rate is also evident from the exact solution, for which
\begin{align}
	\Theta=-|\theta_{0}|,
	\qquad
	\sigma=0,
	\qquad
	\frac{\sigma}{\Theta}=0,
	\qquad
	a_{\mathrm{ave.}}(t)=e^{-|\theta_{0}|t/3}.
\end{align}
\begin{align}
	\psi=0,
	\qquad
	\varphi(t)=c_{1},
	\qquad
	V=\frac{\theta_{0}^{2}}{3}.
\end{align}
Therefore, further analysis of the properties of this critical point is not physically warranted. As demonstrated and discussed for the analogous case of $\mathrm{CP}_{A,5}$, a critical point characterized by $U<0$ does not correspond to a physically admissible cosmological solution and may instead be regarded as a spurious mathematical branch arising from the quadratic dependence of $U$ in the governing equations.

\subsection{Investigation of All Potentials \label{sectionotherpots}}
The detailed analysis of potential $\mathrm{A}$ was completed in the previous subsection. It is evident that continuing with a point-by-point analysis of the critical points of all the potentials would take the paper beyond the scope of a normal and reasonably focused article, owing to the large number of critical points, namely $87$, and the different configurations that they may assume. The analysis of all the points can readily be performed by following the procedure used for potential $\mathrm{A}$ and by referring to the tables provided; it is, however, merely time-consuming. We have carried out this analysis and present the resulting findings below in a condensed form, while keeping the objectives of the present paper in mind.

According to subsection~\ref{ourgoal}, we sum up the results for ``all'' potentials in the following four categories:
\begin{enumerate}
    \item Occurrence of inflation alone, without any type of isotropization, either weak or strong;
    \item Occurrence of isotropization alone, either weak or strong;
    \item Occurrence of inflation accompanied by weak isotropization;
    \item Occurrence of inflation accompanied by strong isotropization.
\end{enumerate}

It should also be noted that, as is apparent from the tables, some purely imaginary points were occasionally obtained. However, our analysis is focused on real-valued potentials and on the corresponding real phantom field, since a purely imaginary phantom field is transformed into the canonical field. Hence, we ignore them.

The results of our cumbersome analysis are as follows\footnote{
It is clear that the critical points introduced in the following classification may belong to the corresponding category only under particular algebraic conditions and not necessarily in all cases.
}:

\begin{enumerate}

    \item \textbf{Occurrence of inflationary expansion alone, without isotropization}

    \begin{itemize}

        \item \textbf{BI universe:}

        For this universe, no saddle point was found for any of the potentials that generates inflation alone, without any form of isotropization, whether weak or strong. 

        \item \textbf{BIII universe:}

        For this type of universe, all the potentials except potential $\mathrm{D}$ possess saddle points that generate inflation alone. The points satisfying the required condition are:
        \[
        \mathrm{CP}_{A,4}, \qquad
        \mathrm{CP}_{B,8}, \qquad
        \mathrm{CP}_{C,5}, \qquad
        \mathrm{CP}_{E,5}.
        \]
The exact solutions associated with these critical points have
the same functional form as those associated with $\mathrm{CP}_{A,4}$.
Consequently, the same analysis applies
(see Subsubsection~\ref{CPA4}).

These demonstrate that the theory of Burd and Barrow---which states that whenever inflation occurs, isotropization necessarily occurs as well---does not hold for the phantom model considered here.

        \item \textbf{KS universe:}

        For this type of universe, as in the BI universe, none of the potentials contains a point at which inflation occurs alone without any accompanying isotropization.

    \end{itemize}

    \item \textbf{Occurrence of isotropization alone, either weak or strong}

    \begin{itemize}

        \item \textbf{BI universe:}

In this universe, all the potentials possess the relevant saddle-type critical points and can produce just ``weak'' isotropization without the occurrence of inflation. It should be noted, however, that some points seem to be capable of producing strong isotropization, such as 
  \[
        \mathrm{CP}_{A,3-1}, \quad
        \mathrm{CP}_{A,3-2}, \quad
        \mathrm{CP}_{B,1-1}, \quad
        \mathrm{CP}_{B,1-2}, \quad
        \mathrm{CP}_{B,2-1}, \quad
        \mathrm{CP}_{B,2-2}, \quad
        \mathrm{CP}_{B,5-1},
        \]
        \[
        \mathrm{CP}_{C,8-1}, \quad
        \mathrm{CP}_{C,9-1}, \quad
        \mathrm{CP}_{D,5-1}, \quad
        \mathrm{CP}_{D,6-1},
        \]
but if one applies the strong isotropization condition to the parameters, then quickly finds that all points turn out to describe a contracting universe, due to $\Theta$ becoming negative.

Therefore, the conclusion of Collins and Hawking that isotropization without the occurrence of inflation can be achieved in a BI universe also holds in ``weak form'' for the phantom field considered here.

The saddle-type critical points that can produce weak isotropization in the BI universe are
        \[
        \mathrm{CP}_{A,2-1}, \quad
        \mathrm{CP}_{A,2-2}, \quad
        \mathrm{CP}_{B,4-1}, \quad
        \mathrm{CP}_{B,4-2}, \quad
        \mathrm{CP}_{C,2-1}, \quad
        \mathrm{CP}_{C,2-2}, \quad
        \]
        \[
        \mathrm{CP}_{D,4-1}, \quad
        \mathrm{CP}_{D,4-2}, \quad
        \mathrm{CP}_{E,2-1}, \quad
        \mathrm{CP}_{E,2-2}.
        \]
Since the exact solutions corresponding to these critical points
share the same functional form as those of $\mathrm{CP}_{A,2-1}$ and $\mathrm{CP}_{A,2-2}$,
we do not repeat the detailed analysis here and refer the reader
to Subsubsections~\ref{CPA21} and~\ref{CPA22}.

        \item \textbf{BIII universe:}

        In this universe, none of the studied potentials generates a saddle-type critical point capable of producing isotropization alone, whether weak or strong.

        \item \textbf{KS universe:}

For this metric, each potential gives rise to weak—but not strong—isotropization, without inflation, provided that its corresponding parameters satisfy certain conditions.
The corresponding points are:
        \[
        \mathrm{CP}_{A,1}, \quad
        \mathrm{CP}_{B,3}, \quad
        \mathrm{CP}_{C,1}, \quad
        \mathrm{CP}_{D,3}, \quad
        \mathrm{CP}_{E,1}.
        \]
    \end{itemize}
These critical points yield exact solutions of the same
functional form as those obtained for $\mathrm{CP}_{A,1}$.
Their properties can therefore be discussed along the same
lines as in Subsubsection~\ref{CPA1}.

As follows from the above analysis, a weak-form version of the Collins--Hawking theorem may be formulated for the KS universe in the phantom-field setting.

\item \textbf{Occurrence of inflation accompanied by weak isotropization}

For this case, we obtain the following outcomes:
    \begin{itemize}

        \item \textbf{BI universe:}

        Only three potentials, namely $\mathrm{B}$, $\mathrm{C}$, and $\mathrm{D}$, are generative of saddle-type critical points at which inflation and weak isotropization occur simultaneously. The corresponding critical points are as follows:
        \[
        \mathrm{CP}_{B,2-1}, \quad
        \mathrm{CP}_{B,2-2}, \quad
        \mathrm{CP}_{C,3-1}, \quad
        \mathrm{CP}_{C,3-2}, \quad
        \mathrm{CP}_{C,4-1},
        \]
        \[
        \mathrm{CP}_{C,4-2}, \quad
        \mathrm{CP}_{D,1-1}, \quad
        \mathrm{CP}_{D,1-2}, \quad
        \mathrm{CP}_{D,2-1}, \quad
        \mathrm{CP}_{D,2-2}.
        \]
        Potentials $\mathrm{A}$ and $\mathrm{E}$ do not exhibit this
capability, even in the other backgrounds considered.\\
  The above-mentioned critical points have similar phase-space
  coordinates, given by
        \begin{align*}
        S=\pm &\sqrt{\frac{2-3P^{2}}{6}}, \qquad U=0, \qquad P=P, \\
        \lambda \propto &\text{ the constant coefficient of the scalar field and the power of the}\\
         &\text{scalar-field function appearing in the potential function, viz.,}\\
         & (\lambda_{\text{Potential B}} \, ; \lambda_{\text{Potential C}}\, ; \lambda_{\text{Potential D}}) = (\zeta \, ; \pm \xi \, ; \pm \alpha \beta),
        \end{align*}
        and the same functional form of exact solutions as:
        \begin{align}\label{eq101}
        &\Theta = \frac{1}{(1-2P^{2})}\frac{1}{t}, \quad
        \sigma = \pm \frac{\sqrt{12-18P^{2}}}{6(2P^{2}-1)} \frac{1}{t}, \quad
        \frac{\sigma}{\Theta}= \frac{\sqrt{12-18P^{2}}}{\pm 6}\\
        & a_{\mathrm{ave.}}=c_{0}t^{1/(3-6P^{2})},\\
        & \varphi = \frac{P}{1-2P^{2}} \ln (t) +c_{1}, \quad
        \psi= \frac{P}{1-2P^{2}} \frac{1}{t}, \quad
        V=0.
        \end{align}
The value of $\lambda$ plays a crucial role in determining whether
a critical point is a saddle. This explains why potential
$\mathrm{A}$ is absent from this class: in that case, $\lambda=0$,
and the corresponding critical points are repellers (See $\mathrm{CP}_{A,3-1}$ and $\mathrm{CP}_{A,3-2}$). In particular,
the sign of $\lambda$ affects the saddle character of these points.

To clarify the allowed values of $\lambda$, we consider its
evolution equation, $\lambda^{\prime}=-Pf$. At a critical point,
$\lambda^{\prime}=0$, which requires $Pf=0$. For a family of critical
points in which $P$ is arbitrary,
this condition requires $f=0$ yielding
        \begin{align}\label{eqlam02}
        \lambda = \frac{-\alpha_{2} \pm \sqrt{\alpha^{2}_{2}-4\alpha_{1}\alpha_{3}}}{2\, \alpha_{1}}.
        \end{align}

According to Table~\ref{table31}, the radicand is negative for
potential $\mathrm{E}$, yielding two purely imaginary values of
$\lambda$, whereas potentials $\mathrm{B}$, $\mathrm{C}$, and
$\mathrm{D}$ admit two real values. For potential $\mathrm{A}$,
$\lambda$ vanishes.

The entries for potential $\mathrm{E}$ reveal a similar structure
among the critical points
$\mathrm{CP}_{E,3-1}$--$\mathrm{CP}_{E,3-4}$. However, the
corresponding formal values of $\lambda$ are purely imaginary,
so these points cannot be interpreted directly as critical points
of the real phase space in the present variables. The stability
analysis classifies the corresponding branches as repellers,
apart from the special case $P^{2}=2/3$, which belongs to the
contracting branch and is of saddle nature. We therefore exclude these points from
the subsequent discussion.

A qualification is necessary at $U=0$. In the present setting,
this condition implies $V=0$, and the expression defining
$\lambda$ takes the indeterminate form $0/0$. Thus, $\lambda$
is not defined there by its original expression, $\lambda = -(\mathrm{d}V/\mathrm{d}\varphi)/V$; its behaviour at
that point must instead be inferred from Eq.~(\ref{eqlam02}). This signals
a limitation of the chosen phase-space variables rather than
a physical contradiction.

For the aforementioned critical points (i.e. $\mathrm{CP}_{B,2-1}$, $\mathrm{CP}_{B,2-2}$, ...), the associated exact
solutions have the same functional form as those for
$\mathrm{CP}_{A,3-1}$ and $\mathrm{CP}_{A,3-2}$, although
$\lambda$ is nonzero and the eigenvalue spectra differ.
The nonzero values of $\lambda$ allow saddle behaviour
for the relevant parameter ranges. Accordingly, the discussion
of the exact solutions in Subsubsections~\ref{CPA31}
and~\ref{CPA32} also applies here, whereas the stability
properties must be assessed separately: these points can
be saddles rather than repellers.

        \item \textbf{BIII universe:}

        None of the potentials provides a saddle critical point containing inflationary expansion along with isotropization in BIII background.

        \item \textbf{KS universe:}

        In this universe, the situation is the same as in the BI universe: only potentials $\mathrm{B}$, $\mathrm{C}$, and $\mathrm{D}$ produce such critical points, whereas potentials $\mathrm{A}$ and $\mathrm{E}$ do not.

        The saddle-type critical points rendering this situation are:
        \[
        \mathrm{CP}_{B,5-1}, \quad
        \mathrm{CP}_{C,8-1}, \quad
        \mathrm{CP}_{C,9-1}, \quad
        \mathrm{CP}_{D,5-1}, \quad
        \mathrm{CP}_{D,6-1}.
        \]
    \end{itemize}
For these critical points, the associated exact solutions share
the same functional form. By appropriately tuning the parameters,
one obtains power-law inflation on the saddle branch. Since the
shear scalar scales as $\sigma \propto t^{-1}$, the resulting
expansion is only weakly isotropic. However, the expansion scalar
has the same time dependence, $\Theta \propto t^{-1}$, and hence
the shear-to-expansion ratio, $\sigma / \Theta$,
remains constant. This ratio can be set to zero only for special
parameter values. In that limit, the solution becomes exactly
isotropic, but it is no longer compatible with the inflationary
saddle branch. Moreover, the corresponding solution belongs to
the contracting, rather than the expanding, branch, and the
associated critical points become repellers. Thus, this limit
does not provide a physically viable realization of expanding,
isotropic phantom inflation. The physically relevant solution
associated with these critical points is therefore an expanding
power-law inflationary solution with weak isotropization.

An interesting feature of these solutions is that both the scalar
field and the potential are nonvanishing and time-dependent. They
scale, respectively, as
\[
\varphi(t)\propto \ln t,
\qquad
V(t)\propto t^{-2}.
\]
These critical points may therefore be interpreted as defining
a regime in which both the phantom kinetic sector and the
potential contribute nontrivially to the dynamics. In this
sense, both the scalar field and the potential participate in
the realization of transient power-law inflation with weak
isotropization, and the resulting solution can be regarded as a
genuinely phantom-active regime.

Therefore, these solutions satisfy the weak-form version of the
Burd--Barrow theorem in the BI and KS
backgrounds within the phantom model.

    \item \textbf{Occurrence of inflation accompanied by strong isotropization}

For this case, which is one of our main objectives, the status is as follows:
    \begin{itemize}

        \item \textbf{BI, BIII, and KS universes:}

        None of these three universes are generative of inflationary expansion and strong isotropization simultaneously. Indeed, the Burd--Barrow theorem does not hold in its strong form for the phantom model under the assumptions considered.\\
        This result does not mean that the phantom model is incapable of producing such points. Rather, such points---which constitute our primary desired outcome---can be obtained by changing the type of potential/background.

For the \textit{canonical} scalar field, this can be achieved in our case by performing
the transformation
\[
\varphi \to i\widetilde{\varphi},
\qquad \widetilde{\varphi}\in\mathbb{R},
\] 
or by imposing suitable conditions on the model parameters. For example, the constant coefficients may be tuned so that the potential, scalar field, and isotropization factor become real-valued without leading to unphysical behavior. This may result in saddle-type critical points that satisfy both inflation and strong isotropization simultaneously for canonical/ordinary scalar field. We identify two such points as representative examples. However, the study of the situations corresponding canonical-scalar-field cases are beyond the scope of this paper because, as mentioned earlier, we focus exclusively on the phantom model.\\
\textbf{Two Examples:}\\
For the potentials $\mathrm{C}$ and $\mathrm{D}$, one may easily found that, in BI metric, the  purpose can be achieved by shifting toward particular oscillatory potentials, namely a sine or cosine potentials. More precisely, we obtain the following results:
        \begin{enumerate}

            \item For potential $\mathrm{C}$, which is given by
            \[
            V_{\mathrm{C}}(\varphi)
            =
            \cosh(\xi\varphi)-1,
            \]
            redefining
            \[
            \xi=i\, \widetilde{\xi};
            \qquad
            \widetilde{\xi} \in \mathbb{R},
            \]
            changes the form of the potential to
            \[
            \widetilde{V}_{\mathrm{C}}(\varphi)
            =
            \cos(\widetilde{\xi}\varphi)-1.
            \]
            Nevertheless, upon imposing suitable conditions, one obtains a critical point that satisfies all the required criteria. For example, the critical point $\mathrm{CP}_{C,10-1}$ is of saddle type of canonical scalar field in the BI universe under the condition
            \[
            0<\widetilde{\xi}<6.
            \]
            Inflation occurs at this point, and the condition for strong isotropization is also satisfied.

            It should be noted that potential $\mathrm{E}$ also contains a cosine function; however, the crucial difference is that the cosine is squared in that potential which is solvable. \\
            Note that this expression can be recast in the following form:
\begin{equation}
    \label{eq:recast}
V_{\mathrm{E}}=2M^{2}\cos^{2}\left( \frac{\varphi}{2l} \right)
=M^{2}\left[ \cos \left( \frac{\varphi}{l} \right)+1 \right].
\end{equation}
It follows that an equivalent structure---up to minor discrepancies that we shall overlook---must be present in that context. This expectation is indeed borne out: the anticipated critical point is $\mathrm{CP}_{E,2}$.

It is worth remarking that, in all the cases discussed above, $\lambda$ takes imaginary values. The origin of this behavior becomes transparent upon rewriting $\lambda$ as
\begin{equation}
    \label{eq:lambda_rewrite}
\lambda= - \, \frac{\mathrm{d}V/\mathrm{d}\varphi}{V}
=-\, \frac{1}{\psi} \frac{\dot{V}}{V}
=-\, \frac{1}{P}\frac{V^{\prime}}{V}.
\end{equation}
Although the potential itself is real, the overall expression acquires an imaginary character due to the presence of $P$—a quantity that is itself rendered imaginary by the field redefinition to the canonical field. The simultaneous and purely imaginary nature of both $P$ and $\lambda$ is readily understood by inspection of the field equations: $P$ and $\lambda$ enter the equations in precisely such a combination that the factors of $i=\sqrt{-1}$ cancel on both sides, leaving the equations manifestly real.

Here, we discuss how $\lambda$ and $P$ cancel each other in
the equations, whereas this aspect was not examined in the
previous case. This difference in approach reflects the fact
that, previously, we focused on the critical points, while
here we have changed the potential and the structure of the
model. We therefore examine the equations to determine their
form for the canonical field.

            \item For potential $\mathrm{D}$, which is of the form
            \[
            V_{\mathrm{D}}(\varphi)
            =
            V_{0d}\sinh^{-\alpha}(\beta\varphi),
            \]
            redefining
            \[
            \beta=i\widetilde{\beta},
            \qquad
            V_{0d}=i\widetilde{V}_{0d},
            \]
            where
            \[
            \widetilde{\beta},\widetilde{V}_{0d}\in\mathbb{R},
            \]
            transforms the potential into
            \[
            \widetilde{V}_{\mathrm{D}}(\varphi)
            =
            \widetilde{V}_{0d}
            \sin^{-\alpha}(\widetilde{\beta}\varphi).
            \]

            By imposing suitable conditions, a critical point satisfying all the required criteria can then be obtained. In particular, the critical points $\mathrm{CP}_{D,7-1}$ and $\mathrm{CP}_{D,8-1}$ are of saddle type in the BI universe, inflation occurs at these points, and the condition for strong isotropization is satisfied under the ``cautious'' conditions
            \[
            0<\alpha^{2}\widetilde{\beta}^{2}<6
            \]
            and with $\alpha$ being a negative integer.

            The term ``cautious'' was used because other conditions could also have been imposed, although doing so would have reduced the available freedom. For example, one could restrict the interval from $(0,6)$ to $(2,6)$ for $\alpha^{2}\widetilde{\beta}^{2}$ and remove the requirement that $\alpha$ be negative in order to obtain a saddle point.
        \end{enumerate}
    \end{itemize}
\end{enumerate}
Therefore, as is observed, neither of the two packages of solutions
discussed in Subsection~\ref{ourgoal} can be obtained in the phantom model for any of the five potentials across the three backgrounds considered. It seems that the canonical scalar field admits a richer set of critical points meeting our target criteria than the phantom field does.

\subsection{A challenging question: must inflation and strong isotropization
manifest themselves through a critical point?\label{sectchall}}

This question stems from a more fundamental, philosophical one: ``in general, when can a system manifest the occurrence of a given state through a critical point?''

First of all, it should be noted that the number of critical points depends entirely on the dimensionless variables that one defines for the system.
Indeed, the point is that a system which has been rewritten in terms of two sets of different variables usually share equivallent points, but the one with judicious choices, may reveal a greater number of critical points. Sometimes in the literature, authors introduce extra variables when rewriting the system even though there is no need for them, in the sense that the new variable the author has in mind possesses neither an independent existence nor an independent equation; rather, it is added merely in order to extract more information from the system. However, in this approach a feature usually appears that may at first sight look paradoxical and obscure to interpret. For example, a value is obtained for the newly added variable that appears contradictory, and only with care can one understand what this seemingly contradictory datum means. For instance, the Hubble constant takes an infinite value and an apparent contradiction arises, whereas this is not the case: we have merely added a one-dimensional manifold that encodes the expansion of the universe, but in the system it acquires a value that appears contradictory. This is not really a contradiction---had we not added it, the issue would not have arisen at all, and now that we have added it, it demands careful interpretation free of superficial judgment (for further reading we refer the reader to \cite{behmot}). The point is that neither an injudicious and careless choice is appropriate, since it does not reveal all the dimensions of the system, nor is the introduction of a variable without any basis appropriate, since it creates false contradictions that do not correspond to reality; and even if it can reveal some dimensions, this occurs by virtue of the elementary physics we already know about the system, not because the value of the variable itself tells us anything. In fact, knowing the physics of the problem, we would be forced to justify the problematic and apparently contradictory value in a professional manner, which means it is of no use to us and only causes trouble.

Apparently, our general rule should be the following: our choices should in principle be focused on the variables that are the principal carriers of the data we seek to extract; for instance, in our case, the expansion rate, the isotropization, and so forth were of primary interest. The fact that we also included the potential among the variables, in addition to completing the set, was because we wanted to obtain its value in order to determine whether the agent driving the appearance of this point is the potential or whether the kinetic term comes into play, and other important questions of this kind.

Let us now turn to the main question. As to whether inflation and strong
isotropization must manifest themselves given this choice of variables, the
answer is that inflation must indeed manifest itself, whereas demanding strong
isotropization to do so may be overly optimistic. Moreover, there is no
guarantee that, after passing through a critical point satisfying this
condition, the isotropization state will remain the same. The reason is that
primordial inflation was a highly critical behavior at the early times of
creation, indeed, inflation is a very sudden and rapid phenomenon that occurs within a
very short time interval, of the order of $10^{-32}\,(\mathrm{s})$, its
manifestation in the form of a critical point is reasonable and expected, by
virtue of the nature of its behavior and the variables we have chosen; given that a saddle critical point is attractive from one side and repulsive from the other, the cosmic system was in fact attracted toward the
saddle critical point and then powerfully repelled by its repulsive force, and
this violent repulsion is precisely primordial inflation. The time scale of inflation is so small that the entire period of evolution can occur in a
very close neighborhood of the critical point, governed by the
approximate/exact behavior of the exact solutions, leaving no room for dispute
on this matter.
All of this processes occurs on an expansion manifold of the universe; we observed that the
inflationary regimes we obtained were sometimes contracting and sometimes
expanding, the correct case being the expanding one, since the universe has
been expanding from the beginning and this expansion is developing; therefore
the manifold that supports inflation must be of the expanding type. The
repulsive strength of the saddle point originates from the vector fields; in
fact, the length of the vector-field arrows, as well as their accumulation in the repulsive region, is greater than at the other points, which indicates the strength of the field. Such a situation has occurred in inflation.

As to whether the evolutionary trajectory of the universe will necessarily pass through a particular critical point satisfying our desired conditions, the answer is that it depends entirely on the initial conditions. This is the root of phenomena such as fine-tuning: our system exhibits a particular behavior and possesses many possibilities; hence it is necessarily far more sensitive to the initial conditions than it appears.

Isotropization, whether weak or strong, is not a phenomenon like inflation; isotropization occurs over time, not at a particular point. In fact, unlike inflation, it is not a sudden, impulsive behavior. The first thing that comes to mind is that, at a critical point, if, for
example, we arrive at a solution of the form $S=(\sigma/\Theta)\propto f(t)$,
in which $f(t)$ is a decaying function or zero, the condition is met and our
objective is thereby attained; however, this is not the appropriate and correct
answer.
Indeed, even if the point in question produces strong isotropization and the cosmic evolutionary trajectory passes through it, there is no guarantee that this property will be preserved as the model evolves further and moves away from that critical point. Nevertheless, if it does occur, it means that the model has this capability, provided that it can preserve the property along the trajectory, which in turn depends on the initial conditions, on the subsequent evolution, and on the examination of all later stages, rather than merely on the short initial interval of the creation and formation of the universe on which we have focused in this paper. On the other hand, the non-satisfaction of the strong isotropization condition cannot be taken as evidence that the model is incapable of producing strong isotropization due to the following reasons:

\begin{enumerate}
\item First, in the later stages of evolution and beyond these critical points, isotropization may happen.
The important point is that the critical point we adopt for inflation should
not exhibit a contradictory behavior for $S$; that is, the behavior should be
either decreasing in time or constant. If it is increasing, then, even though
it is transient and may last only moments, it is not acceptable, since it
disrupts the orderly course of the evolution and contradicts the physics that
existence has revealed to us. 
    Indeed, during inflation, $S$ should in principle decrease rather than increase, or at least remain constant.
    The reason why $S$ should decrease in time during inflation seems clear and reasonable. The question then arises: why is a constant
value for $S$ justifiable? First, note that $S$ can be constant either because the shear scalar and the
expansion rate are both constant, or because their time-dependent parts
coincide—for example, both being proportional to $t^{-1}$.
A constant value is justifiable because, first, it is transient, and we saw that it is usually accompanied by a decreasing function for $\sigma$ itself; second, we may even offer the following philosophical justification: this inflation was like a very large shock (an impulse) to the system of existence, which the system itself perceived perhaps with a delay---albeit an extremely small, transient one---and, over time, brought itself to equilibrium and adapted through isotropization, rather than instantly and momentarily. This testifies to the enormous power of the primordial inflation that occurred. Therefore, these repeated constant values of $S$ do not constitute definitive evidence of a defect in the model.

In the models we studied, we did not observe such a case that would contradict our expectations. This, then, provides the first reason why no defect can be attributed to the model.

\item Second, perhaps within the package we examined---that is, this model together with this kind of potentials and these three types of backgrounds---a small change in any of the components or in the whole, for example a change in the shape of the potential/model/background, could enforce the strong isotropization condition together with inflation at a saddle critical point. Note that we are not obliged to require inflation and strong isotropization
to occur simultaneously at a single critical point; each feature can instead
take place at a critical point of its own. For instance, inflation may occur
at one point, and strong isotropization subsequently at another critical
point. If the evolution trajectory passes through both, the universe will
experience both inflation and isotropization. Of course, the subsequent
stages of evolution must still be traced in order to determine whether the
system preserves the property of strong isotropization on large scales up to
the present epoch.
\end{enumerate}

Apart from the above discussions, note that isotropization at sufficiently large cosmological scales is approximate, not exactly precise. The cosmological principle states that on large scales the universe is homogeneous and isotropic and is described by the FLRW metric. But these are not exact mathematical statements about the actual Universe; in standard cosmology, these are properties of the background spacetime, while the observed structures are placed on top of it as perturbations. The universe is more than what is detectable and observable by us. To clarify, suppose the homogeneous and isotropic universe is like a sphere and the anisotropic one like an ellipsoid. If your observational sphere lies inside and concentric with this ellipsoid, the non-observation of anisotropy does not imply that it is not anisotropic; in fact, this claim pertains only to your observational horizon, not to the entire existing universe. You cannot see it because you do not have sufficient detection range.

Therefore, in view of what has been discussed, we can deduce that the phantom model, despite the non-realization of strong isotropization together with inflation at the critical points found for all the potentials, possesses the richness necessary to describe the early universe. However, the canonical field seems to perform better than the phantom in the description of the early inflation and isotropization.

\section{Conclusion}
In this paper, we aimed to investigate inflation and isotropization in the early Universe within the phantom model. The phantom model is one of the successful models for describing the late-time acceleration of the Universe. In principle, from the perspective of unifying gravitational models, cosmological models should provide an adequate description of all stages of cosmic evolution, including the earliest moments of its origin and phenomena such as inflation and isotropization. This was one of the motivations for the present investigation.

We examined three backgrounds---BI, BIII, and KS---with five different well-known forms of the potential in the phantom model, corresponding to fifteen different cases.

To this end, we first extended the well-known cosmic no-hair theorem (Heusler theorem) to phantom model containing any form of convex potentials. We found that, irrespective of the restrictions that this theorem imposes on the constant parameters of the potential and on the behaviour of the phantom field, it ultimately cannot provide an adequate answer because the shear scalar and the expansion rate vanish simultaneously. We therefore continued the analysis using the dynamical-systems approach.

We have sought to keep the analysis as general as possible. In total, we identified $87$ critical points and analysed their stability and cosmological properties. Most of the critical points were non-hyperbolic, and we consequently analysed their centre manifolds. Our approach to the centre-manifold analysis was based on the perturbation method, with a precision of $10^{-4}$. More than $550$ figures were generated during the stability analysis using this method.

In the dynamical-systems analysis, we first specified that, for the fifteen combinations (models) under consideration, we sought to achieve one of the following two objectives:
\begin{enumerate}
    \item a saddle critical point that simultaneously describes inflationary expansion and strong isotropization;
    \item two saddle critical points in a fixed background and potential, one describing inflationary expansion and the other strong isotropization on an expanding manifold.
\end{enumerate}

It should be noted that, unlike studies that allow attractor points, we regarded only saddle points as physical. Our reasoning is that the Universe passes through these points, and points along its evolutionary trajectory are of saddle type, apart from the initial and final points of the Universe, which may be repeller and attractor, respectively.

For convenience and to summarize the main results, we considered five important classes:
\begin{itemize}
    \item \textbf{Class I: Inflation alone, without either weak or strong isotropization.}\\
    In this case, only the BIII Universe produced physically acceptable points, for all potentials except potential D.

    \item \textbf{Class II: Weak isotropization alone in an expanding Universe, without inflation.}\\
    With the exception of the BIII Universe, both of the remaining backgrounds, i.e. BI and KS, produced acceptable critical points for all the potentials. In this case, the two backgrounds exhibited essentially similar behaviour: in both, the phantom field and the corresponding potential were inactive at the critical points and were instead frozen. The constant parameters of the potential functions did not appear in the eigenvalues either, and therefore could not be used to adjust the stability, although the saddle points themselves existed and required no such adjustment.

    \item \textbf{Class III: Strong isotropization alone in an expanding Universe, without inflation.}\\
    None of the backgrounds produced this behaviour for any of the potentials considered.
    This cancel the existence of the second option mentioned above.

    \item \textbf{Class IV: Inflationary expansion accompanied by weak isotropization.}\\
    With the exception of the BIII Universe, both of the remaining backgrounds, i.e. BI and KS, produced physically acceptable critical points for potentials B, C, and D. This was the best and most physically desirable outcome that we obtained. In other words, it was the optimal outcome among the fifteen cases considered.

    Unlike in Class II, the fact that both the BI and KS backgrounds produced critical points satisfying the conditions of this class does not imply that these backgrounds have approximately equivalent effects and roles. In both backgrounds, the constant parameters of the potential functions appeared in the eigenvalues and helped make it possible to adjust the stability of the points to saddle type. In this sense, they played a role. However, a pronounced and significant difference between the two backgrounds in producing points in this class was that, in the BI Universe, the contributions of the potential and the phantom field to the kinetic contribution were zero, and the phantom field was effectively inactive. In the KS Universe, by contrast, both the potentials and the phantom field were active and had non-negligible time dependence.

    Combining the findings of this class with those of Class II, and eliminating by symmetry the role of the constant parameters in the potential functions in the eigenvalues of both the BI and KS backgrounds, one may infer that the role of the potential and the scalar field of the phantom model in producing desirable critical points vanishes and is frozen in the BI Universe. In the KS Universe, however, they may or may not play a role, depending on what is required.

    \item \textbf{Class V: Inflationary expansion accompanied by strong isotropization.}\\
    No physically acceptable critical point was found for any case in the phantom model.
\end{itemize}

A summary of our findings, without referring to the names of the points and focusing only on the ability of each potential to produce the corresponding critical points in the different backgrounds studied in the phantom cosmological model, is presented in Table~\ref{tablenutshell}\footnote{
    In this table, the terms ``Univ.'', ``Inf. Alone'', ``W. Iso. Alone'', ``S. Iso. Alone'', ``Inf. + W. Iso.'', ``Inf. + S. Iso.'', and ``All'' are abbreviations for ``The corresponding Universe'', ``Occurrence of Inflationary expansion Alone without any form of isotropization'', ``Occurrence of Weak Isotropization Alone in an expanding Universe without inflation'', ``Occurrence of Strong Isotropization Alone in an expanding Universe without inflation'', ``Inflationary expansion along With Weak Isotropization'', ``Inflationary expansion along with Strong Isotropization'', and ``All of the Potentials'', respectively.
    }.
\begin{table}[H]
    \centering
    \caption{Summary of our findings, without reference to the names of the points and focusing only on the ability of each potential to produce the corresponding critical points in the different backgrounds studied in the phantom cosmological model.}
    \label{tablenutshell}
    \small
    \begin{tabular}{!{\vrule width 1.2pt} c||c||c||c||c||c!{\vrule width 1.2pt}}
        \Xhline{1.2pt}
        \cellcolor{pink!100} \textbf{Univ.} &\cellcolor{orange!50} \textbf{Inf. Alone} &\cellcolor{red!50!orange!45} \textbf{W. Iso. Alone} &\cellcolor{red!50} \textbf{S. Iso. Alone} &\cellcolor{green!45} \textbf{Inf. + W. Iso.} & \cellcolor{blue!25} \textbf{Inf. + S. Iso.} \\ \Xhline{1.2pt}
        \cellcolor{pink!100}\textcolor[rgb]{0.00,0.00,0.50}{BI}   &\cellcolor{orange!50} $\textcolor[rgb]{0.00,0.00,0.50}{\times}$ &\cellcolor{red!50!orange!45} \textcolor[rgb]{0.00,0.00,0.50}{All} &\cellcolor{red!50} $\textcolor[rgb]{0.00,0.00,0.50}{\times}$ &\cellcolor{green!45} \textcolor[rgb]{0.00,0.00,0.50}{B, C, D} & \cellcolor{blue!25}  $\textcolor[rgb]{0.00,0.00,0.50}{\times}$ \\ \hline
        \cellcolor{pink!100}BIII &\cellcolor{orange!50} A, B, C, E &\cellcolor{red!50!orange!45} $\times$ &\cellcolor{red!50} $\times$ &\cellcolor{green!45} $\times$ & \cellcolor{blue!25} $\times$ \\ \hline
        \cellcolor{pink!100}\textcolor[rgb]{0.00,0.00,0.50}{KS}   &\cellcolor{orange!50} $\textcolor[rgb]{0.00,0.00,0.50}{\times}$ &\cellcolor{red!50!orange!45} \textcolor[rgb]{0.00,0.00,0.50}{All} &\cellcolor{red!50} $\textcolor[rgb]{0.00,0.00,0.50}{\times}$ &\cellcolor{green!45} \textcolor[rgb]{0.00,0.00,0.50}{B, C, D} & \cellcolor{blue!25} $\textcolor[rgb]{0.00,0.00,0.50}{\times}$ \\
        \Xhline{1.2pt}
    \end{tabular}
\end{table}

Thus, neither of the two objectives we pursued was achieved. Among the fifteen possible combinations leading to $87$ critical points, we found neither a saddle critical point that simultaneously describes inflationary expansion and strong isotropization, nor at least two saddle critical points belonging to a fixed background and potential, one describing inflationary expansion and the other strong isotropization in an expanding Universe.

\textit{The best and optimal outcome was inflationary expansion accompanied by weak isotropization. This occurred only in the BI and KS backgrounds, and only for potentials B, C, and D.}

\textit{The BIII Universe played a role only in producing inflation without any isotropization, whether weak or strong, and then only for all potentials except potential D.}
\textit{Therefore, it appears that, unlike the BIII Universe, the BI and KS Universes are more reasonable choices for describing the Universe at its earliest moments. Of these two, the KS background is more appealing because it gives a more prominent role to the active phantom field and potential.}

One important finding was that potential A, which is a power-law potential, did not have its parameter---the power of the scalar field---appear in the eigenvalues of any critical point. We were therefore unable to change the nature of the critical points using the parameter associated with this potential; instead, we could influence it only through the degrees of freedom in the coordinates of the points. For example, one subtle finding was that when strong isotropization is imposed, so that $S=0$, the dimension associated with ``$S$'' becomes attracting. This was very helpful in analysing the critical points. For the other potentials, by contrast, the constant parameters of the potentials played the primary role in determining the stability of most critical points. This may have contributed to their greater success, although the existence of potential E changes this proposition from a sufficient condition to a necessary one.
An interesting feature of potential E was that most of its points corresponded to a canonical field rather than a phantom field\footnote{
The sixteen critical points listed at the end of Table~\ref{tableSVE}, namely
$\mathrm{CP}_{7-1}$--$\mathrm{CP}_{8-8}$, have purely imaginary phantom
scalar fields, which can be mapped to canonical scalar fields.
}; that is, it automatically produced points associated with canonical, rather than phantom, fields. For this reason, and in view of the discussion in Subsection~\ref{sectionotherpots}, it may be argued that potential E is more suitable for a canonical field than for a phantom field.

We argued that failure to achieve the main objective of the paper---the occurrence of inflation and strong isotropization in the phantom model for a fixed background and potential---does not imply a deficiency in the phantom model. We discussed this issue in detail in Section~\ref{sectchall}.
If all potentials and backgrounds considered in the literature were tested during inflation and post-inflation stages, and strong isotropization were not confirmed, one could then suspect a deficiency in the model. Otherwise, despite the breadth of the present investigation, no general conclusion can be drawn.

The theorems of Burd and Barrow, as well as those of Collins and Hawking, do not hold universally for the phantom model with different potentials: they are sometimes correct and sometimes not. This makes it necessary to avoid theorems that issue general conclusions in these cases. At best, such theorems can be formulated for case studies.

Finally, at the end of the paper, we discussed the transformation of a phantom field into a canonical field and argued that, unlike the phantom field, the canonical field can achieve our ideal objective in these same backgrounds with oscillatory potentials, namely sine and cosine potentials.

\textit{Overall, on the basis of our detailed analysis, one may argue that potentials of hyperbolic-sine or hyperbolic-cosine form, and consequently potentials of the form
$\exp(\pm \, c \, \varphi)$,
where $c$ is a constant parameter, are suitable for phantom scalar-field models, whereas oscillatory potentials such as sine and cosine appear to be more suitable for canonical fields than for phantom fields.\\
It appears that a canonical field with oscillatory potentials performs better than a phantom model with hyperbolic-type potentials in describing the early Universe, although the phantom model performs better in describing late-time accelerated expansion.}

\appendix
\section{Tables of Critical Points and Their Corresponding Properties\label{appendix_tables}}
This appendix is divided into two subsections. The first subsection presents the tables containing the coordinates of the critical points and their corresponding eigenvalues. The tables in the second subsection provide supplementary information, including the exact solutions, stability types, and other relevant characteristics.
\subsection{Tables of Coordinates of Critical Points and Their Corresponding Eigenvalues\label{appendix_tables1}}
In this subsection, we present the tables containing the coordinates of the critical points together with their corresponding eigenvalues, which are denoted by $\varepsilon$.

In labeling the critical points, we have adopted a notation in which the name of the corresponding potential is given first, followed by an index identifying the critical point. For critical points that can be expressed in a closed form, we distinguish the resulting distinct points using a hyphenated index, thereby indicating their correspondence to the same closed-form solution.
For example, Critical Point (CP) ``$3$'' admits a closed-form expression which, upon explicit expansion, yields two distinct critical points. We therefore denote these two points by ``$3-1$'' and ``$3-2$'', respectively. By this, we intend to highlight a degeneracy-like feature: these critical points have different phase-space coordinates but correspond to exact solutions of the same functional form. 
\begin{table}[H]
\small
    \centering
    \caption{
        List of critical points and their corresponding eigenvalues for the potential $\mathrm{A}$.
    }
    \label{tableVA}
    \begin{tabular}{|c||c||c||c||c||c|}
        \hline \hline
        $\mathrm{CP}$
        &
        $S$
        &
        $U$
        &
        $P$
        &
        $\lambda$
        &
        $\varepsilon_{S}, \varepsilon_{U}, \varepsilon_{P}, \varepsilon_{\lambda}$
        \\
        \hline
        $\mathrm{CP}_{A,1}$
        &
        $-\frac{\sqrt{3}}{6}$
        &
        $0$
        &
        $0$
        &
        $\lambda$
        &
        $-\frac{1}{2}, \, \frac{1}{2}, \, -\frac{1}{2}, \, 0$
        \\
        \hline
        $\mathrm{CP}_{A,2-1}$
        &
        $+\frac{1}{\sqrt{3}}$
        &
        $0$
        &
        $0$
        &
        $\lambda$
        &
        $2, \, 1, \, 0, \, 0$
        \\
        \hline
        $\mathrm{CP}_{A,2-2}$
        &
        $-\frac{1}{\sqrt{3}}$
        &
        $0$
        &
        $0$
        &
        $\lambda$
        &
        $\frac{2}{3}, \, 1, \, 0, \, 0$
        \\
        \hline
        $\mathrm{CP}_{A,3-1}$
        &
        $+\sqrt{\frac{2-3P^2}{6}}$
        &
        $0$
        &
        $P$
        &
        $0$
        &
        $\frac{4-\sqrt{4-6P^{2}}}{3}, \, 1, \, 0, \, 0$
        \\
        \hline
        $\mathrm{CP}_{A,3-2}$
        &
        $-\sqrt{\frac{2-3P^2}{6}}$
        &
        $0$
        &
        $P$
        &
        $0$
        &
        $\frac{4+\sqrt{4-6P^{2}}}{3}, \, 1, \, 0, \, 0$
        \\
        \hline
        $\mathrm{CP}_{A,4}$
        &
        $\frac{1}{\sqrt{3}}$
        &
        $+1$
        &
        $0$
        &
        $0$
        &
        $1, \, -2, \, -1, \, 0$
        \\
        \hline
        $\mathrm{CP}_{A,5}$
        &
        $\frac{1}{\sqrt{3}}$
        &
        $-1$
        &
        $0$
        &
        $0$
        &
        $1, \, -2, \, -1, \, 0$
        \\
        \hline
        $\mathrm{CP}_{A,6-1}$
        &
        $0$
        &
        $+\frac{1}{\sqrt{3}}$
        &
        $0$
        &
        $0$
        &
        $-1, \, -\frac{2}{3}, \, -1, \, 0$
        \\
        \hline
        $\mathrm{CP}_{A,6-2}$
        &
        $0$
        &
        $-\frac{1}{\sqrt{3}}$
        &
        $0$
        &
        $0$
        &
        $-1, \, -\frac{2}{3}, \, -1, \, 0$
        \\
        \hline \hline
    \end{tabular}
\end{table}

\begin{table}[H]
\small
    \centering
    \caption{
        List of critical points and their corresponding eigenvalues for the potential $\mathrm{B}$.
    }
    \label{tableVB}
    \begin{tabular}{|c||c||c||c||c||c|}
        \hline \hline
        $\mathrm{CP}$
        &
        $S$
        &
        $U$
        &
        $P$
        &
        $\lambda$
        &
        $\varepsilon_{S}, \varepsilon_{U}, \varepsilon_{P}, \varepsilon_{\lambda}$
        \\
        \hline
$\mathrm{CP}_{B,1-1}$
&
$+\sqrt{\frac{2-3P^{2}}{6}}$
&
$0$
&
$P$
&
$0$
&
$\frac{4+\sqrt{4-6P^{2}}}{3}, \, 1, \, 0, \, -P\zeta$
\\ \hline
$\mathrm{CP}_{B,1-2}$	
&
$-\sqrt{\frac{2-3P^{2}}{6}}$
&
$0$
&
$P$
&
$0$
&
$\frac{4-\sqrt{4-6P^{2}}}{3}, \, 1, \, 0, \, -P\zeta$
\\
\hline		
$\mathrm{CP}_{B,2-1}$
&
$+\sqrt{\frac{2-3P^2}{6}}$
&
$0$
&
$P$
&
$\zeta$
&
$\frac{4+\sqrt{4-6P^{2}}}{3},\, 1-\frac{P\zeta}{2}, \, P\zeta, \, 0$
\\ \hline
$\mathrm{CP}_{B,2-2}$
&
$-\sqrt{\frac{2-3P^2}{6}}$
&
$0$
&
$P$
&
$\zeta$
&
$\frac{4-\sqrt{4-6P^{2}}}{3},\, 1-\frac{P\zeta}{2}, \, P\zeta, \, 0$
\\
\hline		
$\mathrm{CP}_{B,3}$	
&
$-\frac{\sqrt{3}}{6}$	
&
$0$	
&
$0$	
&
$\lambda$
&
$-\frac{1}{2}, \, \frac{1}{2}, \, -\frac{1}{2}, \, 0$
\\
\hline		
$\mathrm{CP}_{B,4-1}$
&
$+\frac{1}{\sqrt{3}}$
&
$0$
&
$0$
&
$\lambda$
&
$2, \, 1, \, 0, \, 0$
\\ \hline
$\mathrm{CP}_{B,4-2}$	
&
$- \frac{1}{\sqrt{3}}$
&
$0$
&
$0$
&
$\lambda$
&
$\frac{2}{3}, \, 1, \, 0, \, 0$
\\
\hline		
$\mathrm{CP}_{B,5-1}$
&
$\frac{\sqrt{3}(2-\zeta^{2})}{6(\zeta^{2}+1)}$
&
$+\frac{\sqrt{\zeta^{2}+2}}{\sqrt{2}(\zeta^{2}+1)}$
&
$\frac{\zeta}{\zeta^{2}+1}$
&
$\zeta$
&
\makecell{
$\frac{-\zeta^{2}-2}{2 \zeta^{2}+2}, \, \frac{-\zeta^{2}-2+\sqrt{-7 \zeta^{4}+4 \zeta^{2}+36}}{4 \zeta^{2}+4},$\\
$\frac{-\zeta^{2}-2-\sqrt{-7 \zeta^{4}+4 \zeta^{2}+36}}{4 \zeta^{2}+4}, \, \frac{\zeta^{2}}{\zeta^{2}+1}$
}
\\	\hline 	
$\mathrm{CP}_{B,5-2}$
&
$\frac{\sqrt{3}(2-\zeta^{2})}{6(\zeta^{2}+1)}$
&
$-\frac{\sqrt{\zeta^{2}+2}}{\sqrt{2}(\zeta^{2}+1)}$
&
$\frac{\zeta}{\zeta^{2}+1}$
&
$\zeta$
&
\makecell{
$\frac{-\zeta^{2}-2}{2 \zeta^{2}+2}, \, \frac{-\zeta^{2}-2+\sqrt{-7 \zeta^{4}+4 \zeta^{2}+36}}{4 \zeta^{2}+4},$\\
$\frac{-\zeta^{2}-2-\sqrt{-7 \zeta^{4}+4 \zeta^{2}+36}}{4 \zeta^{2}+4}, \, \frac{\zeta^{2}}{\zeta^{2}+1}$
}
\\
\hline		
$\mathrm{CP}_{B,6-1}$
&
$0$
&
$+\frac{\sqrt{-2 \zeta^{2}+12}}{6}$
&
$\frac{\zeta}{3}$
&
$\zeta$
&
$\frac{\zeta^{2}-2}{3}, \, \frac{\zeta^{2}}{3}, \, \frac{\zeta^{2}-6}{6}, \, \frac{\zeta^{2}-6}{6}$
\\	\hline 		
$\mathrm{CP}_{B,6-2}$
&
$0$
&
$-\frac{\sqrt{-2 \zeta^{2}+12}}{6}$
&
$\frac{\zeta}{3}$
&
$\zeta$
&
$\frac{\zeta^{2}-2}{3}, \, \frac{\zeta^{2}}{3}, \, \frac{\zeta^{2}-6}{6}, \, \frac{\zeta^{2}-6}{6}$
\\		
\hline
$\mathrm{CP}_{B,7-1}$	
&
$0$	
&
$+\frac{1}{\sqrt{3}}$	
&
$0$	
&
$0$
&
$-1, \, -\frac{2}{3}, \, -1, \, 0$
\\
\hline 	
$\mathrm{CP}_{B,7-2}$	
&
$0$	
&
$-\frac{1}{\sqrt{3}}$	
&
$0$	
&
$0$
&
$-1, \, -\frac{2}{3}, \, -1, \, 0$
\\
\hline 		
$\mathrm{CP}_{B,8}$	
&
$\frac{1}{\sqrt{3}}$	
&
$+1$	
&
$0$	
&
$0$
&
$1, \, -2, \, -1, \, 0$
\\
\hline
$\mathrm{CP}_{B,9}$	
&
$\frac{1}{\sqrt{3}}$	
&
$-1$	
&
$0$	
&
$0$
&
$1, \, -2, \, -1, \, 0$
\\
\hline \hline
\end{tabular}
\end{table}

\begin{table}[H]
\small
    \centering
    \caption{
        List of critical points and their corresponding eigenvalues for the potential $\mathrm{C}$.
    }
    \label{tableVC}
    \begin{tabular}{|c||c||c||c||c||c|}
        \hline \hline
        $\mathrm{CP}$
        &
        $S$
        &
        $U$
        &
        $P$
        &
        $\lambda$
        &
        $\varepsilon_{S}, \varepsilon_{U}, \varepsilon_{P}, \varepsilon_{\lambda}$
\\
\hline
$\mathrm{CP}_{C,1}$
&
$-\frac{\sqrt{3}}{6}$
&
$0$
&
$0$
&
$\lambda$
&
$-\frac{1}{2}, \, \frac{1}{2}, \, -\frac{1}{2}, \, 0$
\\
\hline
$\mathrm{CP}_{C,2-1}$
&
$+\frac{1}{\sqrt{3}}$
&
$0$
&
$0$
&
$\lambda$
&
$2, \, 1, \, 0, \, 0$
\\ \hline
$\mathrm{CP}_{C,2-2}$
&
$- \frac{1}{\sqrt{3}}$
&
$0$
&
$0$
&
$\lambda$
&
$\frac{2}{3}, \, 1, \, 0, \, 0$
\\
\hline
$\mathrm{CP}_{C,3-1}$
&
$+\sqrt{\frac{2-3P^2}{6}}$
&
$0$
&
$P$
&
$+\xi$
&
$\frac{4+\sqrt{4-6P^{2}}}{3},\, 1-\frac{P\xi}{2}, \, 0, \, +P\xi$
\\ \hline
$\mathrm{CP}_{C,3-2}$
&
$-\sqrt{\frac{2-3P^2}{6}}$
&
$0$
&
$P$
&
$+\xi$
&
$\frac{4-\sqrt{4-6P^{2}}}{3},\, 1-\frac{P\xi}{2}, \, 0, \, +P\xi$
\\ \hline
$\mathrm{CP}_{C,4-1}$
&
$+\sqrt{\frac{2-3P^2}{6}}$
&
$0$
&
$P$
&
$-\xi$
&
$\frac{4+\sqrt{4-6P^{2}}}{3},\, 1+\frac{P\xi}{2}, \, 0, \, -P\xi$
\\ \hline
$\mathrm{CP}_{C,4-2}$
&
$-\sqrt{\frac{2-3P^2}{6}}$
&
$0$
&
$P$
&
$-\xi$
&
$\frac{4-\sqrt{4-6P^{2}}}{3},\, 1+\frac{P\xi}{2}, \, 0, \, -P\xi$
\\
\hline
$\mathrm{CP}_{C,5}$
&
$\frac{1}{\sqrt{3}}$
&
$+1$
&
$0$
&
$0$
&
$1, \, -2, \, -1, \, 0$
\\
\hline
$\mathrm{CP}_{C,6}$
&
$\frac{1}{\sqrt{3}}$
&
$-1$
&
$0$
&
$0$
&
$1, \, -2, \, -1, \, 0$
\\ 
\hline
$\mathrm{CP}_{C,7-1}$
&
$0$
&
$+\frac{1}{\sqrt{3}}$
&
$0$
&
$0$
&
$-1, \, -\frac{2}{3}, \, -1, \, 0$
\\ \hline
$\mathrm{CP}_{C,7-2}$
&
$0$
&
$-\frac{1}{\sqrt{3}}$
&
$0$
&
$0$
&
$-1, \, -\frac{2}{3}, \, -1, \, 0$
\\
\hline 	
$\mathrm{CP}_{C,8-1}$
&
$\frac{\sqrt{3}(2-\xi^{2})}{6(\xi^{2}+1)}$
&
$+\frac{\sqrt{\xi^{2}+2}}{\sqrt{2}(\xi^{2}+1)}$
&
$+\frac{\xi}{\xi^{2}+1}$
&
$+\xi$
&
\makecell{
$\frac{-\xi^{2}-2}{2 \xi^{2}+2},
\,
\frac{-\xi^{2}-2+\sqrt{-7 \xi^{4}+4 \xi^{2}+36}}{4 \xi^{2}+4},$\\
$\frac{-\xi^{2}-2-\sqrt{-7 \xi^{4}+4 \xi^{2}+36}}{4 \xi^{2}+4}, 
\,
\frac{\xi^{2}}{\xi^{2}+1}$
}
\\	 \hline
$\mathrm{CP}_{C,8-2}$
&
$\frac{\sqrt{3}(2-\xi^{2})}{6(\xi^{2}+1)}$
&
$-\frac{\sqrt{\xi^{2}+2}}{\sqrt{2}(\xi^{2}+1)}$
&
$+\frac{\xi}{\xi^{2}+1}$
&
$+\xi$
&
\makecell{
$\frac{-\xi^{2}-2}{2 \xi^{2}+2},
\,
\frac{-\xi^{2}-2+\sqrt{-7 \xi^{4}+4 \xi^{2}+36}}{4 \xi^{2}+4},$
\\
$\frac{-\xi^{2}-2-\sqrt{-7 \xi^{4}+4 \xi^{2}+36}}{4 \xi^{2}+4}, 
\,
\frac{\xi^{2}}{\xi^{2}+1}$
}		
\\
\hline
$\mathrm{CP}_{C,9-1}$
&
$\frac{\sqrt{3}(2-\xi^{2})}{6(\xi^{2}+1)}$
&
$+\frac{\sqrt{\xi^{2}+2}}{\sqrt{2}(\xi^{2}+1)}$
&
$-\frac{\xi}{\xi^{2}+1}$
&
$-\xi$
&
\makecell{
$\frac{-\xi^{2}-2}{2 \xi^{2}+2},
\,
\frac{-\xi^{2}-2+\sqrt{-7 \xi^{4}+4 \xi^{2}+36}}{4 \xi^{2}+4},$\\
$\frac{-\xi^{2}-2-\sqrt{-7 \xi^{4}+4 \xi^{2}+36}}{4 \xi^{2}+4}, 
\,
\frac{\xi^{2}}{\xi^{2}+1}$
}
\\ \hline	
$\mathrm{CP}_{C,9-2}$
&
$\frac{\sqrt{3}(2-\xi^{2})}{6(\xi^{2}+1)}$
&
$-\frac{\sqrt{\xi^{2}+2}}{\sqrt{2}(\xi^{2}+1)}$
&
$-\frac{\xi}{\xi^{2}+1}$
&
$-\xi$
&
\makecell{
$\frac{-\xi^{2}-2}{2 \xi^{2}+2},
\,
\frac{-\xi^{2}-2+\sqrt{-7 \xi^{4}+4 \xi^{2}+36}}{4 \xi^{2}+4},$
\\
$\frac{-\xi^{2}-2-\sqrt{-7 \xi^{4}+4 \xi^{2}+36}}{4 \xi^{2}+4}, 
\,
\frac{\xi^{2}}{\xi^{2}+1}$
}		
\\
\hline
$\mathrm{CP}_{C,10-1}$
&
$0$
&
$+\frac{\sqrt{-2 \xi^{2}+12}}{6}$
&
$\frac{\xi}{3}$
&
$\xi$
&
$\frac{\xi^{2}-2}{3}, 
\,
\frac{\xi^{2}}{3}, 
\,
\frac{\xi^{2}-6}{6}, 
\,
\frac{\xi^{2}-6}{6}$
\\	\hline	
$\mathrm{CP}_{C,10-2}$
&
$0$
&
$-\frac{\sqrt{-2 \xi^{2}+12}}{6}$
&
$\frac{\xi}{3}$
&
$\xi$
&
$\frac{\xi^{2}-2}{3}, 
\,
\frac{\xi^{2}}{3}, 
\,
\frac{\xi^{2}-6}{6}, 
\,
\frac{\xi^{2}-6}{6}$
\\
\hline \hline
\end{tabular}
\end{table}

\begin{table}[H]
\small
    \centering
    \caption{
        List of critical points and their corresponding eigenvalues for the potential $\mathrm{D}$.
    }
    \label{tableVD}
    \begin{tabular}{|c||c||c||c||c||c|}
        \hline \hline
        $\mathrm{CP}$
        &
        $S$
        &
        $U$
        &
        $P$
        &
        $\lambda$
        &
        $\varepsilon_{S}, \varepsilon_{U}, \varepsilon_{P}, \varepsilon_{\lambda}$
\\
\hline

$\mathrm{CP}_{D,1-1}$
&
$+\sqrt{\frac{2-3P^2}{6}}$
&
$0$
&
$P$
&
$-\alpha \beta$
&
$\frac{4+\sqrt{4-6P^{2}}}{3},\, 1+\frac{P\alpha \beta}{2}, \, 0, \, +2P \beta$
\\ \hline
$\mathrm{CP}_{D,1-2}$
&
$-\sqrt{\frac{2-3P^2}{6}}$
&
$0$
&
$P$
&
$-\alpha \beta$
&
$\frac{4-\sqrt{4-6P^{2}}}{3},\, 1+\frac{P\alpha \beta}{2}, \, 0, \, +2P \beta$
\\ \hline
$\mathrm{CP}_{D,2-1}$
&
$+\sqrt{\frac{2-3P^2}{6}}$
&
$0$
&
$P$
&
$+\alpha \beta$
&
$\frac{4+\sqrt{4-6P^{2}}}{3},\, 1-\frac{P\alpha \beta}{2}, \, 0, \, -2P \beta$
\\ \hline
$\mathrm{CP}_{D,2-2}$
&
$-\sqrt{\frac{2-3P^2}{6}}$
&
$0$
&
$P$
&
$+\alpha \beta$
&
$\frac{4-\sqrt{4-6P^{2}}}{3},\, 1-\frac{P\alpha \beta}{2}, \, 0, \, -2P \beta$
\\
\hline
$\mathrm{CP}_{D,3}$
&
$-\frac{\sqrt{3}}{6}$
&
$0$
&
$0$
&
$\lambda$
&
$-\frac{1}{2}, \, \frac{1}{2}, \, -\frac{1}{2}, \, 0$
\\
\hline
$\mathrm{CP}_{D,4-1}$
&
$+\frac{1}{\sqrt{3}}$
&
$0$
&
$0$
&
$\lambda$
&
$2, \, 1, \, 0, \, 0$
\\ \hline
$\mathrm{CP}_{D,4-2}$
&
$- \frac{1}{\sqrt{3}}$
&
$0$
&
$0$
&
$\lambda$
&
$\frac{2}{3}, \, 1, \, 0, \, 0$
\\
\hline
$\mathrm{CP}_{D,5-1}$
&
$\frac{\sqrt{3}(2-\alpha^{2} \beta^{2})}{6(\alpha^{2} \beta^{2}+1)}$
&
$+\frac{\sqrt{\alpha^{2} \beta^{2}+2}}{\sqrt{2}(\alpha^{2} \beta^{2}+1)}$
&
$-\frac{\alpha \beta}{\alpha^{2} \beta^{2}+1}$
&
$-\alpha \beta$
&
\makecell{
$\frac{-\alpha^{2} \beta^{2}-2}{2 \alpha^{2} \beta^{2}+2},
\,
\frac{-\alpha^{2} \beta^{2}-2+\sqrt{-7 \alpha^{4} \beta^{4}+4 \alpha^{2} \beta^{2}+36}}{4 \alpha^{2} \beta^{2}+4},$
\\
$\frac{-\alpha^{2} \beta^{2}-2-\sqrt{-7 \alpha^{4} \beta^{4}+4 \alpha^{2} \beta^{2}+36}}{4 \alpha^{2} \beta^{2}+4}, 
\,
\frac{-2\alpha \beta^{2}}{\alpha^{2} \beta^{2}+1}$
}
\\	 \hline
$\mathrm{CP}_{D,5-2}$
&
$\frac{\sqrt{3}(2-\alpha^{2} \beta^{2})}{6(\alpha^{2} \beta^{2}+1)}$
&
$-\frac{\sqrt{\alpha^{2} \beta^{2}+2}}{\sqrt{2}(\alpha^{2} \beta^{2}+1)}$
&
$-\frac{\alpha \beta}{\alpha^{2} \beta^{2}+1}$
&
$-\alpha \beta$
&
\makecell{
$\frac{-\alpha^{2} \beta^{2}-2}{2 \alpha^{2} \beta^{2}+2},
\,
\frac{-\alpha^{2} \beta^{2}-2+\sqrt{-7 \alpha^{4} \beta^{4}+4 \alpha^{2} \beta^{2}+36}}{4 \alpha^{2} \beta^{2}+4},$
\\
$\frac{-\alpha^{2} \beta^{2}-2-\sqrt{-7 \alpha^{4} \beta^{4}+4 \alpha^{2} \beta^{2}+36}}{4 \alpha^{2} \beta^{2}+4}, 
\,
\frac{-2\alpha \beta^{2}}{\alpha^{2} \beta^{2}+1}$
}
\\
\hline
$\mathrm{CP}_{D,6-1}$
&
$\frac{\sqrt{3}(2-\alpha^{2} \beta^{2})}{6(\alpha^{2} \beta^{2}+1)}$
&
$+\frac{\sqrt{\alpha^{2} \beta^{2}+2}}{\sqrt{2}(\alpha^{2} \beta^{2}+1)}$
&
$+\frac{\alpha \beta}{\alpha^{2} \beta^{2}+1}$
&
$+\alpha \beta$
&
\makecell{
$\frac{-\alpha^{2} \beta^{2}-2}{2 \alpha^{2} \beta^{2}+2},
\,
\frac{-\alpha^{2} \beta^{2}-2+\sqrt{-7 \alpha^{4} \beta^{4}+4 \alpha^{2} \beta^{2}+36}}{4 \alpha^{2} \beta^{2}+4},$
\\
$\frac{-\alpha^{2} \beta^{2}-2-\sqrt{-7 \alpha^{4} \beta^{4}+4 \alpha^{2} \beta^{2}+36}}{4 \alpha^{2} \beta^{2}+4}, 
\,
\frac{-2\alpha \beta^{2}}{\alpha^{2} \beta^{2}+1}$
}
\\ \hline	
$\mathrm{CP}_{D,6-2}$
&
$\frac{\sqrt{3}(2-\alpha^{2} \beta^{2})}{6(\alpha^{2} \beta^{2}+1)}$
&
$-\frac{\sqrt{\alpha^{2} \beta^{2}+2}}{\sqrt{2}(\alpha^{2} \beta^{2}+1)}$
&
$+\frac{\alpha \beta}{\alpha^{2} \beta^{2}+1}$
&
$+\alpha \beta$
&
\makecell{
$\frac{-\alpha^{2} \beta^{2}-2}{2 \alpha^{2} \beta^{2}+2},
\,
\frac{-\alpha^{2} \beta^{2}-2+\sqrt{-7 \alpha^{4} \beta^{4}+4 \alpha^{2} \beta^{2}+36}}{4 \alpha^{2} \beta^{2}+4},$
\\
$\frac{-\alpha^{2} \beta^{2}-2-\sqrt{-7 \alpha^{4} \beta^{4}+4 \alpha^{2} \beta^{2}+36}}{4 \alpha^{2} \beta^{2}+4}, 
\,
\frac{-2\alpha \beta^{2}}{\alpha^{2} \beta^{2}+1}$
}
\\
\hline
$\mathrm{CP}_{D,7-1}$
&
$0$
&
$+\frac{\sqrt{-2 \alpha^{2}\beta^{2}+12}}{6}$
&
$+\frac{\alpha \beta}{3}$
&
$+\alpha \beta$
&
$\frac{\alpha^{2}\beta^{2}-2}{3}, 
\,
-\frac{2\alpha \beta^{2}}{3}, 
\,
\frac{\alpha^{2} \beta^{2}-6}{6}, 
\,
\frac{\alpha^{2}\beta^{2}-6}{6}$
\\	\hline	
$\mathrm{CP}_{D,7-2}$
&
$0$
&
$-\frac{\sqrt{-2 \alpha^{2}\beta^{2}+12}}{6}$
&
$+\frac{\alpha \beta}{3}$
&
$+\alpha \beta$
&
$\frac{\alpha^{2}\beta^{2}-2}{3}, 
\,
-\frac{2\alpha \beta^{2}}{3}, 
\,
\frac{\alpha^{2}\beta^{2}-6}{6}, 
\,
\frac{\alpha^{2}\beta^{2}-6}{6}$
\\
\hline
$\mathrm{CP}_{D,8-1}$
&
$0$
&
$+\frac{\sqrt{-2 \alpha^{2}\beta^{2}+12}}{6}$
&
$-\frac{\alpha \beta}{3}$
&
$-\alpha \beta$
&
$\frac{\alpha^{2}\beta^{2}-2}{3}, 
\,
-\frac{2\alpha \beta^{2}}{3}, 
\,
\frac{\alpha^{2}\beta^{2}-6}{6}, 
\,
\frac{\alpha^{2}\beta^{2}-6}{6}$
\\	\hline	
$\mathrm{CP}_{D,8-2}$
&
$0$
&
$-\frac{\sqrt{-2 \alpha^{2}\beta^{2}+12}}{6}$
&
$-\frac{\alpha \beta}{3}$
&
$-\alpha \beta$
&
$\frac{\alpha^{2}\beta^{2}-2}{3}, 
\,
-\frac{2\alpha \beta^{2}}{3}, 
\,
\frac{\alpha^{2}\beta^{2}-6}{6}, 
\,
\frac{\alpha^{2}\beta^{2}-6}{6}$
\\
\hline
$\mathrm{CP}_{D,9-1}$
&
$0$
&
$+\frac{1}{\sqrt{3}}$
&
$0$
&
$0$
&
$-1, \, -\frac{2}{3}, \, \frac{-1}{2}+\frac{\sqrt{12 \alpha \beta^{2}+9}}{6}, \, \frac{-1}{2}-\frac{\sqrt{12 \alpha \beta^{2}+9}}{6}$
\\ \hline
$\mathrm{CP}_{D,9-2}$
&
$0$
&
$-\frac{1}{\sqrt{3}}$
&
$0$
&
$0$
&
$-1, \, -\frac{2}{3}, \, \frac{-1}{2}+\frac{\sqrt{12 \alpha \beta^{2}+9}}{6}, \, \frac{-1}{2}-\frac{\sqrt{12 \alpha \beta^{2}+9}}{6}$
\\
\hline
$\mathrm{CP}_{D,10-1}$
&
$0$
&
$+1$
&
$0$
&
$0$
&
$-\frac{5}{3}, \, -\frac{8}{3},  \, \frac{-5}{6}+\frac{\sqrt{36 \alpha \beta^{2}+25}}{6}, \, \frac{-5}{6}-\frac{\sqrt{36 \alpha \beta^{2}+25}}{6}$
\\ \hline
$\mathrm{CP}_{D,10-2}$
&
$0$
&
$-1$
&
$0$
&
$0$
&
$-\frac{5}{3}, \, -\frac{8}{3},  \, \frac{-5}{6}+\frac{\sqrt{36 \alpha \beta^{2}+25}}{6}, \, \frac{-5}{6}-\frac{\sqrt{36 \alpha \beta^{2}+25}}{6}$
\\
\hline \hline
\end{tabular}
\end{table}

\begin{table}[H]
\small
    \centering
    \caption{
        List of critical points and their corresponding eigenvalues for the potential $\mathrm{E}$.
    }
    \label{tableVE}
    \begin{tabular}{|c||c||c||c||c||c|}
        \hline \hline
        $\mathrm{CP}$
        &
        $S$
        &
        $U$
        &
        $P$
        &
        $\lambda$
        &
        $\varepsilon_{S}, \varepsilon_{U}, \varepsilon_{P}, \varepsilon_{\lambda}$
\\
\hline
$\mathrm{CP}_{E,1}$
&
$-\frac{\sqrt{3}}{6}$
&
$0$
&
$0$
&
$\lambda$
&
$-\frac{1}{2}, \, \frac{1}{2}, \, -\frac{1}{2}, \, 0$
\\
\hline
$\mathrm{CP}_{E,2-1}$
&
$+\frac{1}{\sqrt{3}}$
&
$0$
&
$0$
&
$\lambda$
&
$2, \, 1, \, 0, \, 0$
\\ \hline
$\mathrm{CP}_{E,2-2}$
&
$- \frac{1}{\sqrt{3}}$
&
$0$
&
$0$
&
$\lambda$
&
$\frac{2}{3}, \, 1, \, 0, \, 0$
\\
\hline
$\mathrm{CP}_{E,3-1}$
&
$+\sqrt{\frac{2-3P^{2}}{6}}$
&
$0$
&
$P$
&
$-\frac{i}{l}$
&
$\frac{4+\sqrt{4-6P^{2}}}{3},\, 0, \, +\frac{iP}{l}, \,  1+\frac{iP}{2l}$
\\ \hline
$\mathrm{CP}_{E,3-2}$
&
$-\sqrt{\frac{2-3P^2}{6}}$
&
$0$
&
$P$
&
$-\frac{i}{l}$
&
$\frac{4-\sqrt{4-6P^{2}}}{3},\, 0, \, +\frac{iP}{l}, \,  1+\frac{iP}{2l}$
\\ \hline
$\mathrm{CP}_{E,3-3}$
&
$+\sqrt{\frac{2-3P^2}{6}}$
&
$0$
&
$P$
&
$+\frac{i}{l}$
&
$\frac{4+\sqrt{4-6P^{2}}}{3},\, 0, \, -\frac{iP}{l}, \,  1-\frac{iP}{2l}$
\\ \hline
$\mathrm{CP}_{E,3-4}$
&
$-\sqrt{\frac{2-3P^2}{6}}$
&
$0$
&
$P$
&
$+\frac{i}{l}$
&
$\frac{4-\sqrt{4-6P^{2}}}{3},\, 0, \, -\frac{iP}{l}, \,  1-\frac{iP}{2l}$
\\	
\hline
$\mathrm{CP}_{E,4-1}$
&
$0$
&
$+\frac{1}{\sqrt{3}}$
&
$0$
&
$0$
&
$-1, \, -\frac{2}{3}, \, \frac{-1}{2}+\frac{\sqrt{9l^{2}+6}}{6l}, \, \frac{-1}{2}-\frac{\sqrt{9l^{2}+6}}{6l}$
\\ \hline
$\mathrm{CP}_{E,4-2}$
&
$0$
&
$-\frac{1}{\sqrt{3}}$
&
$0$
&
$0$
&
$-1, \, -\frac{2}{3}, \, \frac{-1}{2}+\frac{\sqrt{9l^{2}+6}}{6l}, \, \frac{-1}{2}-\frac{\sqrt{9l^{2}+6}}{6l}$
\\
\hline
$\mathrm{CP}_{E,5}$
&
$\frac{1}{\sqrt{3}}$
&
$+1$
&
$0$
&
$0$
&
$1, \, -2, \, \frac{-1}{2}+\frac{\sqrt{l^{2}+2}}{2l}, \, \frac{-1}{2}-\frac{\sqrt{l^{2}+2}}{2l}$
\\
\hline
$\mathrm{CP}_{E,6}$
&
$\frac{1}{\sqrt{3}}$
&
$-1$
&
$0$
&
$0$
&
$1, \, -2, \, \frac{-1}{2}+\frac{\sqrt{l^{2}+2}}{2l}, \, \frac{-1}{2}-\frac{\sqrt{l^{2}+2}}{2l}$
\\ 
\hline
$\mathrm{CP}_{E,7-1}$
&
$+\frac{\sqrt{3}\, \left(2 l^{2}+1\right)}{6 l^{2}-6}$
&
$+\frac{\sqrt{4 l^{2}-2}\, l}{2 l^{2}-2}$
&
$+\frac{l i}{l^{2}-1}$
&
$+\frac{i}{l}$
&
\makecell{
$\frac{-2 l^{2}+1}{2 l^{2}-2}, \, \frac{-2 l^{2}+\sqrt{36 l^{4}-4 l^{2}-7}+1}{4 l^{2}-4},$ \\
$\frac{-2 l^{2}-\sqrt{36 l^{4}-4 l^{2}-7}+1}{4 l^{2}-4}, \, \frac{1}{\left(l^{2}-1\right)}$
}
\\ \hline
$\mathrm{CP}_{E,7-2}$
&
$+\frac{\sqrt{3}\, \left(2 l^{2}+1\right)}{6 l^{2}-6}$
&
$-\frac{\sqrt{4 l^{2}-2}\, l}{2 l^{2}-2}$
&
$+\frac{l i}{l^{2}-1}$
&
$+\frac{i}{l}$
&
\makecell{
$\frac{-2 l^{2}+1}{2 l^{2}-2}, \, \frac{-2 l^{2}+\sqrt{36 l^{4}-4 l^{2}-7}+1}{4 l^{2}-4},$ \\
$\frac{-2 l^{2}-\sqrt{36 l^{4}-4 l^{2}-7}+1}{4 l^{2}-4}, \, \frac{1}{\left(l^{2}-1\right)}$
}
\\ \hline
$\mathrm{CP}_{E,7-3}$
&
$+\frac{\sqrt{3}\, \left(2 l^{2}+1\right)}{6 l^{2}-6}$
&
$-\frac{\sqrt{4 l^{2}-2}\, l}{2 l^{2}-2}$
&
$-\frac{l i}{l^{2}-1}$
&
$-\frac{i}{l}$
&
\makecell{
$\frac{-2 l^{2}+1}{2 l^{2}-2}, \, \frac{-2 l^{2}+\sqrt{36 l^{4}-4 l^{2}-7}+1}{4 l^{2}-4},$ \\
$\frac{-2 l^{2}-\sqrt{36 l^{4}-4 l^{2}-7}+1}{4 l^{2}-4}, \, \frac{1}{\left(l^{2}-1\right)}$
}
\\ \hline
$\mathrm{CP}_{E,7-4}$
&
$+\frac{\sqrt{3}\, \left(2 l^{2}+1\right)}{6 l^{2}-6}$
&
$+\frac{\sqrt{4 l^{2}-2}\, l}{2 l^{2}-2}$
&
$-\frac{l i}{l^{2}-1}$
&
$-\frac{i}{l}$
&
\makecell{
$\frac{-2 l^{2}+1}{2 l^{2}-2}, \, \frac{-2 l^{2}+\sqrt{36 l^{4}-4 l^{2}-7}+1}{4 l^{2}-4},$ \\
$\frac{-2 l^{2}-\sqrt{36 l^{4}-4 l^{2}-7}+1}{4 l^{2}-4}, \, \frac{1}{\left(l^{2}-1\right)}$
}
\\ \hline
$\mathrm{CP}_{E,7-5}$
&
$+\frac{\sqrt{3}\, \left(2 l^{2}+1\right)}{6 l^{2}-6}$
&
$+\frac{\sqrt{4 l^{2}-2}\, l}{2 l^{2}-2}$
&
$+\frac{l i}{l^{2}-1}$
&
$-\frac{i}{l}$
&
``Very Long Terms''
\\ \hline
$\mathrm{CP}_{E,7-6}$
&
$+\frac{\sqrt{3}\, \left(2 l^{2}+1\right)}{6 l^{2}-6}$
&
$-\frac{\sqrt{4 l^{2}-2}\, l}{2 l^{2}-2}$
&
$+\frac{l i}{l^{2}-1}$
&
$-\frac{i}{l}$
&
``Very Long Terms''
\\ \hline
$\mathrm{CP}_{E,7-7}$
&
$+\frac{\sqrt{3}\, \left(2 l^{2}+1\right)}{6 l^{2}-6}$
&
$+\frac{\sqrt{4 l^{2}-2}\, l}{2 l^{2}-2}$
&
$-\frac{l i}{l^{2}-1}$
&
$+\frac{i}{l}$
&
``Very Long Terms''
\\ \hline
$\mathrm{CP}_{E,7-8}$
&
$+\frac{\sqrt{3}\, \left(2 l^{2}+1\right)}{6 l^{2}-6}$
&
$-\frac{\sqrt{4 l^{2}-2}\, l}{2 l^{2}-2}$
&
$-\frac{l i}{l^{2}-1}$
&
$+\frac{i}{l}$
&
``Very Long Terms''
\\
\hline
$\mathrm{CP}_{E,8-1}$
&
$0$
&
$+\frac{\sqrt{12 l^{2}+2}\, l}{6l}$
&
$+\frac{i}{3l}$
&
$-\frac{i}{l}$
&
\makecell{
$\frac{10 l^{2}+5+\sqrt{4 l^{4}+148 l^{2}+25}}{-12 l^{2}}, \, \frac{6 l^{2}+1}{-6 l^{2}},$ \\
$\frac{10 l^{2}+5-\sqrt{4 l^{4}+148 l^{2}+25}}{-12 l^{2}}, \, \frac{-1}{3 l^{2}}$
}
\\ \hline
$\mathrm{CP}_{E,8-2}$
&
$0$
&
$-\frac{\sqrt{12 l^{2}+2}\, l}{6l}$
&
$+\frac{i}{3l}$
&
$-\frac{i}{l}$
&
\makecell{
$\frac{10 l^{2}+5+\sqrt{4 l^{4}+148 l^{2}+25}}{-12 l^{2}}, \, \frac{6 l^{2}+1}{-6 l^{2}},$ \\
$\frac{10 l^{2}+5-\sqrt{4 l^{4}+148 l^{2}+25}}{-12 l^{2}}, \, \frac{-1}{3 l^{2}}$
}
\\ \hline
$\mathrm{CP}_{E,8-3}$
&
$0$
&
$+\frac{\sqrt{12 l^{2}+2}\, l}{6l}$
&
$-\frac{i}{3l}$
&
$+\frac{i}{l}$
&
\makecell{
$\frac{10 l^{2}+5+\sqrt{4 l^{4}+148 l^{2}+25}}{-12 l^{2}}, \, \frac{6 l^{2}+1}{-6 l^{2}},$ \\
$\frac{10 l^{2}+5-\sqrt{4 l^{4}+148 l^{2}+25}}{-12 l^{2}}, \, \frac{-1}{3 l^{2}}$
}
\\ \hline
$\mathrm{CP}_{E,8-4}$
&
$0$
&
$-\frac{\sqrt{12 l^{2}+2}\, l}{6l}$
&
$-\frac{i}{3l}$
&
$+\frac{i}{l}$
&
\makecell{
$\frac{10 l^{2}+5+\sqrt{4 l^{4}+148 l^{2}+25}}{-12 l^{2}}, \, \frac{6 l^{2}+1}{-6 l^{2}},$ \\
$\frac{10 l^{2}+5-\sqrt{4 l^{4}+148 l^{2}+25}}{-12 l^{2}}, \, \frac{-1}{3 l^{2}}$
}
\\ \hline
$\mathrm{CP}_{E,8-5}$
&
$0$
&
$+\frac{\sqrt{12 l^{2}+2}}{6l}$
&
$+\frac{i}{3l}$
&
$+\frac{i}{l}$
&
$\frac{6 l^{2}+1}{-6l^{2}}, \, \frac{6 l^{2}+1}{-6l^{2}}, \, \frac{2 l^{2}+1}{-3l^{2}}, \, \frac{1}{3l^{2}}$
\\ \hline
$\mathrm{CP}_{E,8-6}$
&
$0$
&
$-\frac{\sqrt{12 l^{2}+2}}{6l}$
&
$+\frac{i}{3l}$
&
$+\frac{i}{l}$
&
$\frac{6 l^{2}+1}{-6l^{2}}, \, \frac{6 l^{2}+1}{-6l^{2}}, \, \frac{2 l^{2}+1}{-3l^{2}}, \, \frac{1}{3l^{2}}$
\\ \hline
$\mathrm{CP}_{E,8-7}$
&
$0$
&
$+\frac{\sqrt{12 l^{2}+2}}{6l}$
&
$-\frac{i}{3l}$
&
$-\frac{i}{l}$
&
$\frac{6 l^{2}+1}{-6l^{2}}, \, \frac{6 l^{2}+1}{-6l^{2}}, \, \frac{2 l^{2}+1}{-3l^{2}}, \, \frac{1}{3l^{2}}$
\\ \hline
$\mathrm{CP}_{E,8-8}$
&
$0$
&
$-\frac{\sqrt{12 l^{2}+2}}{6l}$
&
$-\frac{i}{3l}$
&
$-\frac{i}{l}$
&
$\frac{6 l^{2}+1}{-6l^{2}}, \, \frac{6 l^{2}+1}{-6l^{2}}, \, \frac{2 l^{2}+1}{-3l^{2}}, \, \frac{1}{3l^{2}}$
\\
\hline \hline
\end{tabular}
\end{table}

\subsection{Tables of Critical Points and Their Corresponding Important Information\label{appendix_tables2}}
In the tables below, the analytical solutions for the critical points, together with their other relevant and necessary properties, are listed.

For clarity in reading the tables, the following points should be noted:

\begin{enumerate}
    \item In these tables, the term ``Univ.'' is an abbreviation for ``Universe'', indicating the universe in which the corresponding critical point may occur. If a critical point may occur. If a critical point can occur in two of the three metrics, depending both metrics are indicated using a slash. However, if a critical point can occur in all three universes under consideration, the term ``All'' is used.

    \item The term ``Inf. E.'' is an abbreviation for ``Inflationary Expansion'', indicating whether the corresponding critical point belongs to an expanding, inflationary universe. Clearly, even if inflation occurs in a contracting universe, the answer in this column is negative. In this column, the symbol ``$\checkmark$'' denotes ``yes under all conditions'', whereas the symbol ``$\times$'' denotes ``no under all conditions''. If either possibility may occur, depending on the chosen values of the constants, the term ``Both'' is used.

    \item The abbreviations ``S. Iso.'' and ``W. Iso.'' stand for ``Strong Isotropization'' and ``Weak Isotropization'', respectively. They indicate whether strong and weak isotropization (without paying any attention to its phase, i.e. expanding/contracting) can occur, or whether both are possible under specific choices of The term ``Sta.?'' is an abbreviation for ``Stationary Situation'' and specifies the type of stability associated with the critical point. In this column, ``$\mathcal{A.}$'', ``$\mathcal{R.}$'', and ``$\mathcal{S.}$'' denote ``$\mathcal{A}$ttractor'', ``$\mathcal{R}$epeller'', and ``$\mathcal{S}$addle'', respectively. If more than one possibility exists, the alternatives are separated by a slash, indicating that both listed situations can occur under suitable conditions. If all three possibilities are allowed, the term ``All'' is used.

    \item For the sake of brevity, the following symbols are used throughout the tables:
\end{enumerate}

\begin{equation*}
	\begin{split}
&\zeta^{2}+(n) \equiv \chi_{(n)},\\
&\xi^{2}+(n) \equiv \mu_{(n)},\\
&\alpha^{2} \beta^{2} +(n) \equiv \nu_{(n)},\\
&l^{2}+(n) \equiv \kappa_{(n)},\\
&\kappa^{\star}_{2}\equiv \frac{
4\kappa_{(0)}
\kappa_{(-1)}^{2}\kappa_{(-1/2)}}{9\kappa_{(1/3)}^{2}\,t^{2}},\\
&\kappa^{\star}_{1} \equiv \frac{+2l\kappa_{(-1)}}{3\kappa_{(1/3)}}.
	\end{split}
\end{equation*}

\begin{landscape}
\small
\setlength{\tabcolsep}{2pt}
\renewcommand{\arraystretch}{0.85}
\begin{longtable}{|c||c||c||c||c||c||c||c||c||c||c||c||c|}
    \caption{
       Properties of the critical points associated with potential $\mathrm{A}$.
    }
    \label{tableSVA}
    \\
    \hline \hline
    $\mathrm{CP}$
    &
    $\Theta$
    &
    $\sigma$
    &
    $\psi$
    &
    $a_{\mathrm{ave.}}(t)$
    &
    $\varphi(t)$
    &
    $V$
    &
    $\mathrm{Univ.}$
    &
    $\mathrm{Inf. E.}$
    &
    $\sigma / \Theta$
    &
    $\mathrm{S. Iso.}$
    &
    $\mathrm{W. Iso.}$
    &
    $\mathrm{Sta.?}$
    \\
    \hline \hline
    \endfirsthead

    \multicolumn{13}{c}{
        \tablename\ \thetable\ \; Continued from the table on the previous page.
    }
    \\
    \hline
    $\mathrm{CP}$
    &
    $\Theta$
    &
    $\sigma$
    &
    $\psi$
    &
    $a_{\mathrm{ave.}}(t)$
    &
    $\varphi(t)$
    &
    $V$
    &
    $\mathrm{Univ.}$
    &
    $\mathrm{Inf. E.}$
    &
    $\sigma / \Theta$
    &
    $\mathrm{S. Iso.}$
    &
    $\mathrm{W. Iso.}$
    &
    $\mathrm{Sta.?}$
    \\
    \hline \hline
    \endhead

    \hline
    \multicolumn{13}{c}{Continued on the next page} \\
    \endfoot

    \hline
    \endlastfoot

    $\mathrm{CP}_{A,1}$
    &
    $2/t$
    &
    $-1/(\sqrt{3}\,t)$
    &
    $0$
    &
    $c_{0}t^{2/3}$
    &
    $c_{1}$
    &
    $0$
    &
    $\mathrm{KS}$
    &
    $\times$
    &
    $-\sqrt{3}/6$
    &
    $\times$
    &
    $\checkmark$
    &
    $\mathcal{S.}$
    \\
    \hline

    $\mathrm{CP}_{A,2-1}$
    &
    $1/t$
    &
    $+1/(\sqrt{3}\,t)$
    &
    $0$
    &
    $c_{0}t^{1/3}$
    &
    $c_{1}$
    &
    $0$
    &
    $\mathrm{BI}$
    &
    $\times$
    &
    $+1/\sqrt{3}$
    &
    $\times$
    &
    $\checkmark$
    &
    $\mathcal{S.}$
    \\
    \hline

    $\mathrm{CP}_{A,2-2}$
    &
    $1/t$
    &
    $-1/(\sqrt{3}\,t)$
    &
    $0$
    &
    $c_{0}t^{1/3}$
    &
    $c_{1}$
    &
    $0$
    &
    $\mathrm{BI}$
    &
    $\times$
    &
    $-1/\sqrt{3}$
    &
    $\times$
    &
    $\checkmark$
    &
    $\mathcal{S.}$
    \\
    \hline

    $\mathrm{CP}_{A,3-1}$
    &
    $\frac{1}{(1-2P^{2})t}$
    &
    $-\frac{\sqrt{12-18P^{2}}}
    {6(2P^{2}-1)t}$
    &
    $\frac{P}{(1-2P^{2})t}$
    &
    $c_{0}t^{1/(3-6P^{2})}$
    &
    $\frac{P\ln(t)}{1-2P^{2}}+c_{1}$
    &
    $0$
    &
    $\mathrm{BI}$
    &
    Both
    &
    $\frac{\sqrt{12-18P^{2}}}{+6}$
    &
    Both
    &
    $\checkmark$
    &
    $\mathcal{R.}/\mathcal{S.}$
    \\
    \hline

    $\mathrm{CP}_{A,3-2}$
    &
    $\frac{1}{(1-2P^{2})t}$
    &
    $+\frac{\sqrt{12-18P^{2}}}
    {6(2P^{2}-1)t}$
    &
    $\frac{P}{(1-2P^{2})t}$
    &
    $c_{0}t^{1/(3-6P^{2})}$
    &
    $\frac{P\ln(t)}{1-2P^{2}}+c_{1}$
    &
    $0$
    &
    $\mathrm{BI}$
    &
    Both
    &
    $\frac{\sqrt{12-18P^{2}}}{-6}$
    &
    Both
    &
    $\checkmark$
    &
    $\mathcal{R.}/\mathcal{S.}$
    \\
    \hline

    $\mathrm{CP}_{A,4}$
    &
    $\theta_{0}$
    &
    $\theta_{0}/\sqrt{3}$
    &
    $0$
    &
    $e^{\theta_{0}t/3}$
    &
    $c_{1}$
    &
    $\theta_{0}^{2}$
    &
    $\mathrm{BIII}$
    &
    $\checkmark$
    &
    $1/\sqrt{3}$
    &
    $\times$
    &
    $\times$
    &
    $\mathcal{S.}$
    \\
    \hline

    $\mathrm{CP}_{A,5}$
    &
    $-|\theta_{0}|$
    &
    $-|\theta_{0}|/\sqrt{3}$
    &
    $0$
    &
    $e^{-|\theta_{0}|t/3}$
    &
    $c_{1}$
    &
    $\theta_{0}^{2}$
    &
    $\mathrm{BIII}$
    &
    $\times$
    &
    $1/\sqrt{3}$
    &
    $\times$
    &
    $\times$
    &
    $\mathcal{S.}$
    \\
    \hline

    $\mathrm{CP}_{A,6-1}$
    &
    $\theta_{0}$
    &
    $0$
    &
    $0$
    &
    $e^{\theta_{0}t/3}$
    &
    $c_{1}$
    &
    $\theta_{0}^{2}/3$
    &
    $\mathrm{BI}$
    &
    $\checkmark$
    &
    $0$
    &
    $\checkmark$
    &
    $\checkmark$
    &
    $\mathcal{A.}$
    \\
    \hline

    $\mathrm{CP}_{A,6-2}$
    &
    $-|\theta_{0}|$
    &
    $0$
    &
    $0$
    &
    $e^{-|\theta_{0}|t/3}$
    &
    $c_{1}$
    &
    $\theta_{0}^{2}/3$
    &
    $\mathrm{BI}$
    &
    $\times$
    &
    $0$
    &
    $\checkmark$
    &
    $\checkmark$
    &
    $\mathcal{A.}$
    \\
    \hline \hline
\end{longtable}
\end{landscape}
\begin{landscape}
\small
\setlength{\tabcolsep}{2pt}
\renewcommand{\arraystretch}{0.85}
\begin{longtable}{|c||c||c||c||c||c||c||c||c||c||c||c||c|}
    \caption{
        Properties of the critical points associated with potential $\mathrm{B}$.
    }
    \label{tableSVB}
    \\
    \hline \hline
    $\mathrm{CP}$
    &
    $\Theta$
    &
    $\sigma$
    &
    $\psi$
    &
    $a_{\mathrm{ave.}}(t)$
    &
    $\varphi(t)$
    &
    $V$
    &
    $\mathrm{Univ.}$
    &
    $\mathrm{Inf. E.}$
    &
    $\sigma / \Theta$
    &
    $\mathrm{S. Iso.}$
    &
    $\mathrm{W. Iso.}$
    &
    $\mathrm{Sta.?}$
    \\
    \hline \hline
    \endfirsthead

    \multicolumn{13}{c}{
        \tablename\ \thetable\ \; Continued from the table on the previous page.
    }
    \\
    \hline
    $\mathrm{CP}$
    &
    $\Theta$
    &
    $\sigma$
    &
    $\psi$
    &
    $a_{\mathrm{ave.}}(t)$
    &
    $\varphi(t)$
    &
    $V$
    &
    $\mathrm{Univ.}$
    &
    $\mathrm{Inf. E.}$
    &
    $\sigma / \Theta$
    &
    $\mathrm{S. Iso.}$
    &
    $\mathrm{W. Iso.}$
    &
    $\mathrm{Sta.?}$
    \\
    \hline \hline
    \endhead

    \hline
    \multicolumn{13}{c}{Continued on the next page} \\
    \endfoot

    \hline
    \endlastfoot
$\mathrm{CP_{B,1-1}}$
&
$\frac{1}{\left(1-2 P^{2}\right) t}$
&
$- \, \frac{\sqrt{-18 P^{2}+12}}{6\left(2 P^{2} -1\right) t}$
&
$\frac{P}{\left(1-2 P^{2}\right) t}$
&
$c_{0} \, t^{\frac{1}{3-6 P^{2}}}$
&
$\frac{P \ln (t)}{1-2 P^{2}}+c_{1}$
&
$0$
&
$\mathrm{BI}$
&
Both
&
$\frac{\sqrt{-18 P^{2}+12}}{+6}$
&
Both
&
$\checkmark$
&
$\mathcal{R.}/\mathcal{S.}$
\\
\hline
$\mathrm{CP_{B,1-2}}$
&
$\frac{1}{\left(1-2 P^{2}\right) t}$
&
$+ \, \frac{\sqrt{-18 P^{2}+12}}{6\left(2 P^{2} -1\right) t}$
&
$\frac{P}{\left(1-2 P^{2}\right) t}$
&
$c_{0} \, t^{\frac{1}{3-6 P^{2}}}$
&
$\frac{P \ln (t)}{1-2 P^{2}}+c_{1}$
&
$0$
&
$\mathrm{BI}$
&
Both
&
$\frac{\sqrt{-18 P^{2}+12}}{-6}$
&
Both
&
$\checkmark$
&
$\mathcal{R.}/\mathcal{S.}$
\\
\hline
$\mathrm{CP_{B,2-1}}$
&
$\frac{1}{\left(1-2 P^{2}\right) t}$
&
$- \, \frac{\sqrt{-18 P^{2}+12}}{6\left(2 P^{2} -1\right) t}$
&
$\frac{P}{\left(1-2 P^{2}\right) t}$
&
$c_{0} \, t^{\frac{1}{3-6 P^{2}}}$
&
$\frac{P \ln (t)}{1-2 P^{2}}+c_{1}$
&
$0$
&
$\mathrm{BI}$
&
Both
&
$\frac{\sqrt{-18 P^{2}+12}}{+6}$
&
Both
&
$\checkmark$
&
$\mathcal{R.}/\mathcal{S.}$
\\
\hline
$\mathrm{CP_{B,2-2}}$
&
$\frac{1}{\left(1-2 P^{2}\right) t}$
&
$+ \, \frac{\sqrt{-18 P^{2}+12}}{6\left(2 P^{2} -1\right) t}$
&
$\frac{P}{\left(1-2 P^{2}\right) t}$
&
$c_{0} \, t^{\frac{1}{3-6 P^{2}}}$
&
$\frac{P \ln (t)}{1-2 P^{2}}+c_{1}$
&
$0$
&
$\mathrm{BI}$
&
Both
&
$\frac{\sqrt{-18 P^{2}+12}}{-6}$
&
Both
&
$\checkmark$
&
$\mathcal{R.}/\mathcal{S.}$
\\
\hline
$\mathrm{CP_{B,3}}$
&
$2/t$
&
$-1/(\sqrt{3}\, t)$
&
$0$
&
$c_{0} \, t^{2/3}$
&
$c_{1}$
&
$0$
&
$\mathrm{KS}$
&
$\times$
&
$-\sqrt{3}/6$
&
$\times$
&
$\checkmark$
&
$\mathcal{S.}$
\\
\hline
$\mathrm{CP_{B,4-1}}$
&
$1/t$
&
$+1/(\sqrt{3}\, t)$
&
$0$
&
$c_{0}\, t^{1/3}$
&
$c_{1}$
&
$0$
&
$\mathrm{BI}$
&
$\times$
&
$+1/\sqrt{3}$
&
$\times$
&
$\checkmark$
&
$\mathcal{S.}$
\\
\hline
$\mathrm{CP_{B,4-2}}$
&
$1/t$
&
$-1/(\sqrt{3}\, t)$
&
$0$
&
$c_{0}\, t^{1/3}$
&
$c_{1}$
&
$0$
&
$\mathrm{BI}$
&
$\times$
&
$-1/\sqrt{3}$
&
$\times$
&
$\checkmark$
&
$\mathcal{S.}$
\\
\hline
$\mathrm{CP_{B,5-1}}$
&
$\frac{2 \chi_{(+1)}^{2}}{\chi_{(0)} \chi_{(-3)} t}$
&
$\frac{ \sqrt{3} \chi_{(+1)} \chi_{(-2)}}{-3 \chi_{(0)} \chi_{(-3)} t}$
&
$\frac{2 \chi_{(+2)}}{\zeta \chi_{(-3)} t}$
&
$c_{0}t^{\frac{2 \chi_{(+1)}^{2}}{3 \chi_{(0)} \chi_{(-3)}}}$
&
$\frac{2 \chi_{(+1)} \ln (t)}{\zeta \chi_{(-3)}}+c_{1}$
&
$\frac{2 \chi_{(+2)} \chi_{(+1)}^{2}}{\chi_{(0)}^{2} \chi_{(-3)}^{2} t^{2}}$
&
$\mathrm{KS/BI}$
&
Both
&
$\frac{\sqrt{3}\, \chi_{(-2)}}{-6 \chi_{(+1)}}$
&
Both
&
$\checkmark$
&
$\mathcal{S.}$
\\
\hline
$\mathrm{CP_{B,5-2}}$
&
$\frac{2 \chi_{(+1)}^{2}}{\chi_{(0)} \chi_{(-3)} t}$
&
$\frac{ \sqrt{3} \chi_{(+1)} \chi_{(-2)}}{-3 \chi_{(0)} \chi_{(-3)} t}$
&
$\frac{2 \chi_{(+2)}}{\zeta \chi_{(-3)} t}$
&
$c_{0}t^{\frac{2 \chi_{(+1)}^{2}}{3 \chi_{(0)} \chi_{(-3)}}}$
&
$\frac{2 \chi_{(+1)} \ln (t)}{\zeta \chi_{(-3)}}+c_{1}$
&
$\frac{2 \chi_{(+2)} \chi_{(+1)}^{2}}{\chi_{(0)}^{2} \chi_{(-3)}^{2} t^{2}}$
&
$\mathrm{KS/BI}$
&
$\times$
&
$\frac{\sqrt{3}\, \chi_{(-2)}}{-6 \chi_{(+1)}}$
&
Both
&
$\checkmark$
&
$\mathcal{S.}$
\\
\hline
$\mathrm{CP_{B,6-1}}$
&
$-18/(\zeta^{2}t)$
&
$0$
&
$-6/(\zeta t)$
&
$c_{0} \,t^{-6/\zeta^{2}}$
&
$\frac{6 \ln (t)}{-\zeta}+c_{1}$
&
$\frac{-18 \zeta^{2}+108}{\zeta^{4} t^{2}}$
&
$\mathrm{BI}$
&
$\times$
&
$0$
&
$\checkmark$
&
$\checkmark$
&
All
\\
\hline
$\mathrm{CP_{B,6-2}}$
&
$-18/(\zeta^{2}t)$
&
$0$
&
$-6/(\zeta t)$
&
$c_{0} \,t^{-6/\zeta^{2}}$
&
$\frac{6 \ln (t)}{-\zeta}+c_{1}$
&
$\frac{-18 \zeta^{2}+108}{\zeta^{4} t^{2}}$
&
$\mathrm{BI}$
&
$\times$
&
$0$
&
$\checkmark$
&
$\checkmark$
&
All
\\
\hline
$\mathrm{CP_{B,7-1}}$
&
$\theta_{0}$
&
$0$
&
$0$
&
$e^{\theta_{0} t/3}$
&
$c_{1}$
&
$\theta_{0}^{2}/3$
&
$\mathrm{BI}$
&
$\checkmark$
&
$0$
&
$\checkmark$
&
$\checkmark$
&
$\mathcal{A.}$
\\
\hline
$\mathrm{CP_{B,7-2}}$
&
$- \left| \theta_{0} \right|$
&
$0$
&
$0$
&
$e^{- \left| \theta_{0} \right| t/3}$
&
$c_{1}$
&
$\theta_{0}^{2}/3$
&
$\mathrm{BI}$
&
$\times$
&
$0$
&
$\checkmark$
&
$\checkmark$
&
$\mathcal{A.}$
\\
\hline
$\mathrm{CP_{B,8}}$
&
$\theta_{0}$
&
$\theta_{0}/\sqrt{3}$
&
$0$
&
$e^{\theta_{0} t/3}$
&
$c_{1}$
&
$\theta_{0}^{2}$
&
$\mathrm{BIII}$
&
$\checkmark$
&
$1/\sqrt{3}$
&
$\times$
&
$\times$
&
$\mathcal{S.}$
\\
\hline
$\mathrm{CP_{B,9}}$
&
$-\left| \theta_{0} \right|$
&
$-\left| \theta_{0} \right| /\sqrt{3}$
&
$0$
&
$e^{-\left|\theta_{0} \right| t/3}$
&
$c_{1}$
&
$\theta_{0}^{2}$
&
$\mathrm{BIII}$
&
$\times$
&
$1/\sqrt{3}$
&
$\times$
&
$\times$
&
$\mathcal{S.}$
    \\
    \hline \hline
\end{longtable}
\end{landscape}

\begin{landscape}
\small
\setlength{\tabcolsep}{2pt}
\renewcommand{\arraystretch}{0.85}
\begin{longtable}{|c||c||c||c||c||c||c||c||c||c||c||c||c|}
    \caption{
     Properties of the critical points associated with potential $\mathrm{C}$.
    }
    \label{tableSVC}
    \\
    \hline \hline
    $\mathrm{CP}$
    &
    $\Theta$
    &
    $\sigma$
    &
    $\psi$
    &
    $a_{\mathrm{ave.}}(t)$
    &
    $\varphi(t)$
    &
    $V$
    &
    $\mathrm{Univ.}$
    &
    $\mathrm{Inf. E.}$
    &
    $\sigma / \Theta$
    &
    $\mathrm{S. Iso.}$
    &
    $\mathrm{W. Iso.}$
    &
    $\mathrm{Sta.?}$
    \\
    \hline \hline
    \endfirsthead

    \multicolumn{13}{c}{
        \tablename\ \thetable\ \; Continued from the table on the previous page.
    }
    \\
    \hline
    $\mathrm{CP}$
    &
    $\Theta$
    &
    $\sigma$
    &
    $\psi$
    &
    $a_{\mathrm{ave.}}(t)$
    &
    $\varphi(t)$
    &
    $V$
    &
    $\mathrm{Univ.}$
    &
    $\mathrm{Inf. E.}$
    &
    $\sigma / \Theta$
    &
    $\mathrm{S. Iso.}$
    &
    $\mathrm{W. Iso.}$
    &
    $\mathrm{Sta.?}$
    \\
    \hline \hline
    \endhead

    \hline
    \multicolumn{13}{c}{Continued on the next page} \\
    \endfoot

    \hline
    \endlastfoot

$\mathrm{CP_{C,1}}$
&
$2/t$
&
$-1/(\sqrt{3}\, t)$
&
$0$
&
$c_{0}t^{2/3}$
&
$c_{1}$
&
$0$
&
$\mathrm{KS}$
&
$\times$
&
$-\sqrt{3}/6$
&
$\times$
&
$\checkmark$
&
$\mathcal{S.}$
\\
\hline
$\mathrm{CP_{C,2-1}}$
&
$1/t$
&
$+1/(\sqrt{3}\, t)$
&
$0$
&
$c_{0}t^{1/3}$
&
$c_{1}$
&
$0$
&
$\mathrm{BI}$
&
$\times$
&
$+1/\sqrt{3}$
&
$\times$
&
$\checkmark$
&
$\mathcal{S.}$
\\
\hline
$\mathrm{CP_{C,2-2}}$
&
$1/t$
&
$-1/(\sqrt{3}\, t)$
&
$0$
&
$c_{0}t^{1/3}$
&
$c_{1}$
&
$0$
&
$\mathrm{BI}$
&
$\times$
&
$-1/\sqrt{3}$
&
$\times$
&
$\checkmark$
&
$\mathcal{S.}$
\\
\hline
$\mathrm{CP_{C,3-1}}$
&
$\frac{1}{\left(1-2 P^{2}\right) t}$
&
$- \, \frac{\sqrt{-18 P^{2}+12}}{6\left(2 P^{2} -1\right) t}$
&
$\frac{P}{\left(1-2 P^{2}\right) t}$
&
$c_{0} \, t^{\frac{1}{3-6 P^{2}}}$
&
$\frac{P \ln (t)}{1-2 P^{2}}+c_{1}$
&
$0$
&
$\mathrm{BI}$
&
Both
&
$\frac{\sqrt{-18 P^{2}+12}}{+6}$
&
Both
&
$\checkmark$
&
$\mathcal{R.}/\mathcal{S.}$
\\
\hline
$\mathrm{CP_{C,3-2}}$
&
$\frac{1}{\left(1-2 P^{2}\right) t}$
&
$+ \, \frac{\sqrt{-18 P^{2}+12}}{6\left(2 P^{2} -1\right) t}$
&
$\frac{P}{\left(1-2 P^{2}\right) t}$
&
$c_{0} \, t^{\frac{1}{3-6 P^{2}}}$
&
$\frac{P \ln (t)}{1-2 P^{2}}+c_{1}$
&
$0$
&
$\mathrm{BI}$
&
Both
&
$\frac{\sqrt{-18 P^{2}+12}}{-6}$
&
Both
&
$\checkmark$
&
$\mathcal{R.}/\mathcal{S.}$
\\
\hline
$\mathrm{CP_{C,4-1}}$
&
$\frac{1}{\left(1-2 P^{2}\right) t}$
&
$- \, \frac{\sqrt{-18 P^{2}+12}}{6\left(2 P^{2} -1\right) t}$
&
$\frac{P}{\left(1-2 P^{2}\right) t}$
&
$c_{0} \, t^{\frac{1}{3-6 P^{2}}}$
&
$\frac{P \ln (t)}{1-2 P^{2}}+c_{1}$
&
$0$
&
$\mathrm{BI}$
&
Both
&
$\frac{\sqrt{-18 P^{2}+12}}{+6}$
&
Both
&
$\checkmark$
&
$\mathcal{R.}/\mathcal{S.}$
\\
\hline
$\mathrm{CP_{C,4-2}}$
&
$\frac{1}{\left(1-2 P^{2}\right) t}$
&
$+ \, \frac{\sqrt{-18 P^{2}+12}}{6\left(2 P^{2} -1\right) t}$
&
$\frac{P}{\left(1-2 P^{2}\right) t}$
&
$c_{0} \, t^{\frac{1}{3-6 P^{2}}}$
&
$\frac{P \ln (t)}{1-2 P^{2}}+c_{1}$
&
$0$
&
$\mathrm{BI}$
&
Both
&
$\frac{\sqrt{-18 P^{2}+12}}{-6}$
&
Both
&
$\checkmark$
&
$\mathcal{R.}/\mathcal{S.}$
\\
\hline
$\mathrm{CP_{C,5}}$
&
$\theta_{0}$
&
$\theta_{0}/\sqrt{3}$
&
$0$
&
$e^{\theta_{0}t/3}$
&
$c_{1}$
&
$\theta_{0}^{2}$
&
$\mathrm{BIII}$
&
$\checkmark$
&
$1/\sqrt{3}$
&
$\times$
&
$\times$
&
$\mathcal{S.}$
\\
\hline
$\mathrm{CP_{C,6}}$
&
$-|\theta_{0}|$
&
$- |\theta_{0}|/\sqrt{3}$
&
$0$
&
$e^{-|\theta_{0}| \, t/3}$
&
$c_{1}$
&
$\theta_{0}^{2}$
&
$\mathrm{BIII}$
&
$\times$
&
$1/\sqrt{3}$
&
$\times$
&
$\times$
&
$\mathcal{S.}$
\\
\hline
$\mathrm{CP_{C,7-1}}$
&
$\theta_{0}$
&
$0$
&
$0$
&
$e^{\theta_{0} t/3}$
&
$c_{1}$
&
$\theta_{0}^{2}/3$
&
$\mathrm{BI}$
&
$\checkmark$
&
$0$
&
$\checkmark$
&
$\checkmark$
&
$\mathcal{A.}$
\\
\hline
$\mathrm{CP_{C,7-2}}$
&
$-|\theta_{0}|$
&
$0$
&
$0$
&
$e^{-|\theta_{0}| t/3}$
&
$c_{1}$
&
$\theta_{0}^{2}/3$
&
$\mathrm{BI}$
&
$\times$
&
$0$
&
$\checkmark$
&
$\checkmark$
&
$\mathcal{A.}$
\\
\hline
$\mathrm{CP_{C,8-1}}$
&
$\frac{2 \mu_{(+1)}^{2}}{\mu_{(0)} \mu_{(-3)} t}$
&
$\frac{ \sqrt{3} \mu_{(+1)} \mu_{(-2)}}{-3 \mu_{(0)} \mu_{(-3)} t}$
&
$\frac{2 \mu_{(+2)}}{\xi \mu_{(-3)} t}$
&
$c_{0}t^{\frac{2 \mu_{(+1)}^{2}}{3 \mu_{(0)} \mu_{(-3)}}}$
&
$\frac{2 \mu_{(+1)} \ln (t)}{\xi \mu_{(-3)}}+c_{1}$
&
$\frac{2 \mu_{(+2)} \mu_{(+1)}^{2}}{\mu_{(0)}^{2} \mu_{(-3)}^{2} t^{2}}$
&
All
&
Both
&
$\frac{\sqrt{3}\, \mu_{(-2)}}{-6 \mu_{(+1)}}$
&
Both
&
$\checkmark$
&
$\mathcal{S.}$
\\
\hline
$\mathrm{CP_{C,8-2}}$
&
$\frac{2 \mu_{(+1)}^{2}}{\mu_{(0)} \mu_{(-3)} t}$
&
$\frac{ \sqrt{3} \mu_{(+1)} \mu_{(-2)}}{-3 \mu_{(0)} \mu_{(-3)} t}$
&
$\frac{2 \mu_{(+2)}}{\xi \mu_{(-3)} t}$
&
$c_{0}t^{\frac{2 \mu_{(+1)}^{2}}{3 \mu_{(0)} \mu_{(-3)}}}$
&
$\frac{2 \mu_{(+1)} \ln (t)}{\xi \mu_{(-3)}}+c_{1}$
&
$\frac{2 \mu_{(+2)} \mu_{(+1)}^{2}}{\mu_{(0)}^{2} \mu_{(-3)}^{2} t^{2}}$
&
All
&
$\times$
&
$\frac{\sqrt{3}\, \mu_{(-2)}}{-6 \mu_{(+1)}}$
&
Both
&
$\checkmark$
&
$\mathcal{S.}$
\\
\hline
$\mathrm{CP_{C,9-1}}$
&
$\frac{2 \mu_{(+1)}^{2}}{\mu_{(0)} \mu_{(-3)} t}$
&
$\frac{ \sqrt{3} \mu_{(+1)} \mu_{(-2)}}{-3 \mu_{(0)} \mu_{(-3)} t}$
&
$\frac{-2 \mu_{(+2)}}{\xi \mu_{(-3)} t}$
&
$c_{0}t^{\frac{2 \mu_{(+1)}^{2}}{3 \mu_{(0)} \mu_{(-3)}}}$
&
$\frac{2 \mu_{(+1)} \ln (t)}{- \xi \mu_{(-3)}}+c_{1}$
&
$\frac{2 \mu_{(+2)} \mu_{(+1)}^{2}}{\mu_{(0)}^{2} \mu_{(-3)}^{2} t^{2}}$
&
All
&
Both
&
$\frac{\sqrt{3}\, \mu_{(-2)}}{-6 \mu_{(+1)}}$
&
Both
&
$\checkmark$
&
$\mathcal{S.}$
\\
\hline
$\mathrm{CP_{C,9-2}}$
&
$\frac{2 \mu_{(+1)}^{2}}{\mu_{(0)} \mu_{(-3)} t}$
&
$\frac{ \sqrt{3} \mu_{(+1)} \mu_{(-2)}}{-3 \mu_{(0)} \mu_{(-3)} t}$
&
$\frac{-2 \mu_{(+2)}}{\xi \mu_{(-3)} t}$
&
$c_{0}t^{\frac{2 \mu_{(+1)}^{2}}{3 \mu_{(0)} \mu_{(-3)}}}$
&
$\frac{2 \mu_{(+1)} \ln (t)}{- \xi \mu_{(-3)}}+c_{1}$
&
$\frac{2 \mu_{(+2)} \mu_{(+1)}^{2}}{\mu_{(0)}^{2} \mu_{(-3)}^{2} t^{2}}$
&
All
&
$\times$
&
$\frac{\sqrt{3}\, \mu_{(-2)}}{-6 \mu_{(+1)}}$
&
Both
&
$\checkmark$
&
$\mathcal{S.}$
\\
\hline
$\mathrm{CP_{C,10-1}}$
&
$-18/(\xi^{2}t)$
&
$0$
&
$-6/(\xi t)$
&
$c_{0} \,t^{-6/\xi^{2}}$
&
$\frac{6 \ln (t)}{-\xi}+c_{1}$
&
$\frac{-18 \xi^{2}+108}{\xi^{4} t^{2}}$
&
$\mathrm{BI}$
&
$\times$
&
$0$
&
$\checkmark$
&
$\checkmark$
&
$\mathcal{R.}/\mathcal{S.}$
\\
\hline
$\mathrm{CP_{C,10-2}}$
&
$-18/(\xi^{2}t)$
&
$0$
&
$-6/(\xi t)$
&
$c_{0} \,t^{-6/\xi^{2}}$
&
$\frac{6 \ln (t)}{-\xi}+c_{1}$
&
$\frac{-18 \xi^{2}+108}{\xi^{4} t^{2}}$
&
$\mathrm{BI}$
&
$\times$
&
$0$
&
$\checkmark$
&
$\checkmark$
&
$\mathcal{R.}/\mathcal{S.}$
    \\
    \hline \hline
\end{longtable}
\end{landscape}

\begin{landscape}
\small
\setlength{\tabcolsep}{2pt}
\renewcommand{\arraystretch}{0.85}
\begin{longtable}{|c||c||c||c||c||c||c||c||c||c||c||c||c|}
    \caption{
        Properties of the critical points associated with potential $\mathrm{D}$.
    }
    \label{tableSVD}
    \\
    \hline \hline
    $\mathrm{CP}$
    &
    $\Theta$
    &
    $\sigma$
    &
    $\psi$
    &
    $a_{\mathrm{ave.}}(t)$
    &
    $\varphi(t)$
    &
    $V$
    &
    $\mathrm{Univ.}$
    &
    $\mathrm{Inf. E.}$
    &
    $\sigma / \Theta$
    &
    $\mathrm{S. Iso.}$
    &
    $\mathrm{W. Iso.}$
    &
    $\mathrm{Sta.?}$
    \\
    \hline \hline
    \endfirsthead

    \multicolumn{13}{c}{
        \tablename\ \thetable\ \; Continued from the table on the previous page.
    }
    \\
    \hline
    $\mathrm{CP}$
    &
    $\Theta$
    &
    $\sigma$
    &
    $\psi$
    &
    $a_{\mathrm{ave.}}(t)$
    &
    $\varphi(t)$
    &
    $V$
    &
    $\mathrm{Univ.}$
    &
    $\mathrm{Inf. E.}$
    &
    $\sigma / \Theta$
    &
    $\mathrm{S. Iso.}$
    &
    $\mathrm{W. Iso.}$
    &
    $\mathrm{Sta.?}$
    \\
    \hline \hline
    \endhead

    \hline
    \multicolumn{13}{c}{Continued on the next page} \\
    \endfoot

    \hline
    \endlastfoot

$\mathrm{CP_{D,1-1}}$
&
$\frac{1}{\left(1-2 P^{2}\right) t}$
&
$- \, \frac{\sqrt{-18 P^{2}+12}}{6\left(2 P^{2} -1\right) t}$
&
$\frac{P}{\left(1-2 P^{2}\right) t}$
&
$c_{0} \, t^{\frac{1}{3-6 P^{2}}}$
&
$\frac{P \ln (t)}{1-2 P^{2}}+c_{1}$
&
$0$
&
$\mathrm{BI}$
&
Both
&
$\frac{\sqrt{-18 P^{2}+12}}{+6}$
&
Both
&
$\checkmark$
&
$\mathcal{R.}/\mathcal{S.}$
\\
\hline
$\mathrm{CP_{D,1-2}}$
&
$\frac{1}{\left(1-2 P^{2}\right) t}$
&
$+ \, \frac{\sqrt{-18 P^{2}+12}}{6\left(2 P^{2} -1\right) t}$
&
$\frac{P}{\left(1-2 P^{2}\right) t}$
&
$c_{0} \, t^{\frac{1}{3-6 P^{2}}}$
&
$\frac{P \ln (t)}{1-2 P^{2}}+c_{1}$
&
$0$
&
$\mathrm{BI}$
&
Both
&
$\frac{\sqrt{-18 P^{2}+12}}{-6}$
&
Both
&
$\checkmark$
&
$\mathcal{R.}/\mathcal{S.}$
\\
\hline
$\mathrm{CP_{D,2-1}}$
&
$\frac{1}{\left(1-2 P^{2}\right) t}$
&
$- \, \frac{\sqrt{-18 P^{2}+12}}{6\left(2 P^{2} -1\right) t}$
&
$\frac{P}{\left(1-2 P^{2}\right) t}$
&
$c_{0} \, t^{\frac{1}{3-6 P^{2}}}$
&
$\frac{P \ln (t)}{1-2 P^{2}}+c_{1}$
&
$0$
&
$\mathrm{BI}$
&
Both
&
$\frac{\sqrt{-18 P^{2}+12}}{+6}$
&
Both
&
$\checkmark$
&
$\mathcal{R.}/\mathcal{S.}$
\\
\hline
$\mathrm{CP_{D,2-2}}$
&
$\frac{1}{\left(1-2 P^{2}\right) t}$
&
$+ \, \frac{\sqrt{-18 P^{2}+12}}{6\left(2 P^{2} -1\right) t}$
&
$\frac{P}{\left(1-2 P^{2}\right) t}$
&
$c_{0} \, t^{\frac{1}{3-6 P^{2}}}$
&
$\frac{P \ln (t)}{1-2 P^{2}}+c_{1}$
&
$0$
&
$\mathrm{BI}$
&
Both
&
$\frac{\sqrt{-18 P^{2}+12}}{-6}$
&
Both
&
$\checkmark$
&
$\mathcal{R.}/\mathcal{S.}$
\\
\hline
$\mathrm{CP_{D,3}}$
&
$2/t$
&
$-1/(\sqrt{3}\, t)$
&
$0$
&
$c_{0}t^{2/3}$
&
$c_{1}$
&
$0$
&
$\mathrm{KS}$
&
$\times$
&
$-\sqrt{3}/6$
&
$\times$
&
$\checkmark$
&
$\mathcal{S.}$
\\
\hline
$\mathrm{CP_{D,4-1}}$
&
$1/t$
&
$+1/(\sqrt{3}\, t)$
&
$0$
&
$c_{0}t^{1/3}$
&
$c_{1}$
&
$0$
&
$\mathrm{BI}$
&
$\times$
&
$+1/\sqrt{3}$
&
$\times$
&
$\checkmark$
&
$\mathcal{S.}$
\\
\hline
$\mathrm{CP_{D,4-2}}$
&
$1/t$
&
$-1/(\sqrt{3}\, t)$
&
$0$
&
$c_{0}t^{1/3}$
&
$c_{1}$
&
$0$
&
$\mathrm{BI}$
&
$\times$
&
$-1/\sqrt{3}$
&
$\times$
&
$\checkmark$
&
$\mathcal{S.}$
\\
\hline
$\mathrm{CP_{D,5-1}}$
&
$\frac{2 \nu_{(+1)}^{2}}{\nu_{(0)} \nu_{(-3)} t}$
&
$\frac{ \sqrt{3} \nu_{(+1)} \nu_{(-2)}}{-3 \nu_{(0)} \nu_{(-3)} t}$
&
$\frac{-2 \nu_{(+2)}}{\alpha \beta \nu_{(-3)} t}$
&
$c_{0}t^{\frac{2 \nu_{(+1)}^{2}}{3 \nu_{(0)} \nu_{(-3)}}}$
&
$\frac{2 \nu_{(+1)} \ln (t)}{- \alpha \beta \nu_{(-3)}}+c_{1}$
&
$\frac{2 \nu_{(+2)} \nu_{(+1)}^{2}}{\nu_{(0)}^{2} \nu_{(-3)}^{2} t^{2}}$
&
All
&
Both
&
$\frac{\sqrt{3}\, \nu_{(-2)}}{-6 \nu_{(+1)}}$
&
Both
&
$\checkmark$
&
$\mathcal{S.}/\mathcal{A.}$
\\
\hline
$\mathrm{CP_{D,5-2}}$
&
$\frac{2 \nu_{(+1)}^{2}}{\nu_{(0)} \nu_{(-3)} t}$
&
$\frac{ \sqrt{3} \nu_{(+1)} \nu_{(-2)}}{-3 \nu_{(0)} \nu_{(-3)} t}$
&
$\frac{-2 \nu_{(+2)}}{\alpha \beta \nu_{(-3)} t}$
&
$c_{0}t^{\frac{2 \nu_{(+1)}^{2}}{3 \nu_{(0)} \nu_{(-3)}}}$
&
$\frac{2 \nu_{(+1)} \ln (t)}{- \alpha \beta \nu_{(-3)}}+c_{1}$
&
$\frac{2 \nu_{(+2)} \nu_{(+1)}^{2}}{\nu_{(0)}^{2} \nu_{(-3)}^{2} t^{2}}$
&
All
&
$\times$
&
$\frac{\sqrt{3}\, \nu_{(-2)}}{-6 \nu_{(+1)}}$
&
Both
&
$\checkmark$
&
$\mathcal{S.}/\mathcal{A.}$
\\
\hline
$\mathrm{CP_{D,6-1}}$
&
$\frac{2 \nu_{(+1)}^{2}}{\nu_{(0)} \nu_{(-3)} t}$
&
$\frac{ \sqrt{3} \nu_{(+1)} \nu_{(-2)}}{-3 \nu_{(0)} \nu_{(-3)} t}$
&
$\frac{2 \nu_{(+2)}}{\alpha \beta \nu_{(-3)} t}$
&
$c_{0}t^{\frac{2 \nu_{(+1)}^{2}}{3 \nu_{(0)} \nu_{(-3)}}}$
&
$\frac{2 \nu_{(+1)} \ln (t)}{\alpha \beta \nu_{(-3)}}+c_{1}$
&
$\frac{2 \nu_{(+2)} \nu_{(+1)}^{2}}{\nu_{(0)}^{2} \nu_{(-3)}^{2} t^{2}}$
&
All
&
Both
&
$\frac{\sqrt{3}\, \nu_{(-2)}}{-6 \nu_{(+1)}}$
&
Both
&
$\checkmark$
&
$\mathcal{S.}$
\\
\hline
$\mathrm{CP_{D,6-2}}$
&
$\frac{2 \nu_{(+1)}^{2}}{\nu_{(0)} \nu_{(-3)} t}$
&
$\frac{ \sqrt{3} \nu_{(+1)} \nu_{(-2)}}{-3 \nu_{(0)} \nu_{(-3)} t}$
&
$\frac{2 \nu_{(+2)}}{\alpha \beta \nu_{(-3)} t}$
&
$c_{0}t^{\frac{2 \nu_{(+1)}^{2}}{3 \nu_{(0)} \nu_{(-3)}}}$
&
$\frac{2 \nu_{(+1)} \ln (t)}{\alpha \beta \nu_{(-3)}}+c_{1}$
&
$\frac{2 \nu_{(+2)} \nu_{(+1)}^{2}}{\nu_{(0)}^{2} \nu_{(-3)}^{2} t^{2}}$
&
All
&
$\times$
&
$\frac{\sqrt{3}\, \nu_{(-2)}}{-6 \nu_{(+1)}}$
&
Both
&
$\checkmark$
&
$\mathcal{S.}$
\\
\hline
$\mathrm{CP_{D,7-1}}$
&
$\frac{-18}{\alpha^{2} \beta^{2} t}$
&
$0$
&
$\frac{-6}{\alpha \beta t}$
&
$c_{0}t^{-6/(\alpha^{2}\beta^{2})}$
&
$\frac{-6\ln (t)}{\alpha \beta}+c_{1}$
&
$\frac{18(6-\alpha^{2}\beta^{2})}{\alpha^{4}\beta^{4}t^{2}}$
&
$\mathrm{BI}$
&
Both
&
$0$
&
Both
&
$\checkmark$
&
$\mathcal{S.}/\mathcal{A.}$
\\
\hline
$\mathrm{CP_{D,7-2}}$
&
$\frac{-18}{\alpha^{2} \beta^{2} t}$
&
$0$
&
$\frac{-6}{\alpha \beta t}$
&
$c_{0}t^{-6/(\alpha^{2}\beta^{2})}$
&
$\frac{-6\ln (t)}{\alpha \beta}+c_{1}$
&
$\frac{18(6-\alpha^{2}\beta^{2})}{\alpha^{4}\beta^{4}t^{2}}$
&
$\mathrm{BI}$
&
$\times$
&
$0$
&
Both
&
$\checkmark$
&
$\mathcal{S.}/\mathcal{A.}$
\\
\hline
$\mathrm{CP_{D,8-1}}$
&
$\frac{-18}{\alpha^{2} \beta^{2} t}$
&
$0$
&
$\frac{-6}{\alpha \beta t}$
&
$c_{0}t^{-6/(\alpha^{2}\beta^{2})}$
&
$\frac{-6\ln (t)}{\alpha \beta}+c_{1}$
&
$\frac{18(6-\alpha^{2}\beta^{2})}{\alpha^{4}\beta^{4}t^{2}}$
&
$\mathrm{BI}$
&
Both
&
$0$
&
Both
&
$\checkmark$
&
$\mathcal{S.}/\mathcal{A.}$
\\
\hline
$\mathrm{CP_{D,8-2}}$
&
$\frac{-18}{\alpha^{2} \beta^{2} t}$
&
$0$
&
$\frac{-6}{\alpha \beta t}$
&
$c_{0}t^{-6/(\alpha^{2}\beta^{2})}$
&
$\frac{-6\ln (t)}{\alpha \beta}+c_{1}$
&
$\frac{18(6-\alpha^{2}\beta^{2})}{\alpha^{4}\beta^{4}t^{2}}$
&
$\mathrm{BI}$
&
$\times$
&
$0$
&
Both
&
$\checkmark$
&
$\mathcal{S.}/\mathcal{A.}$
\\
\hline
$\mathrm{CP_{D,9-1}}$
&
$\theta_{0}$
&
$0$
&
$0$
&
$e^{\theta_{0} t/3}$
&
$c_{1}$
&
$\theta_{0}^{2}/3$
&
$\mathrm{BI}$
&
Both
&
$0$
&
$\checkmark$
&
$\checkmark$
&
$\mathcal{A.}$
\\
\hline
$\mathrm{CP_{D,9-2}}$
&
$- \left| \theta_{0} \right|$
&
$0$
&
$0$
&
$e^{- \left| \theta_{0} \right| t/3}$
&
$c_{1}$
&
$\theta_{0}^{2}/3$
&
$\mathrm{BI}$
&
$\checkmark$
&
$0$
&
$\times$
&
$\checkmark$
&
$\mathcal{A.}$
\\
\hline
$\mathrm{CP_{D,10-1}}$
&
$-3/(2t)$
&
$0$
&
$0$
&
$c_{0}/\sqrt{t}$
&
$c_{1}$
&
$9/(4t^{2})$
&
$\mathrm{BIII}$
&
$\checkmark$
&
$0$
&
$\times$
&
$\checkmark$
&
$\mathcal{A.}$
\\
\hline
$\mathrm{CP_{D,10-2}}$
&
$-3/(2t)$
&
$0$
&
$0$
&
$c_{0}/\sqrt{t}$
&
$c_{1}$
&
$9/(4t^{2})$
&
$\mathrm{BIII}$
&
$\checkmark$
&
$0$
&
$\times$
&
$\checkmark$
&
$\mathcal{A.}$
    \\
    \hline \hline
\end{longtable}
\end{landscape}

\begin{landscape}
\small
\setlength{\tabcolsep}{2pt}
\renewcommand{\arraystretch}{0.85}
\begin{longtable}{|c||c||c||c||c||c||c||c||c||c||c||c||c|}
    \caption{
        Properties of the critical points associated with potential $\mathrm{E}$.
    }
    \label{tableSVE}
    \\
    \hline \hline
    $\mathrm{CP}$
    &
    $\Theta$
    &
    $\sigma$
    &
    $\psi$
    &
    $a_{\mathrm{ave.}}(t)$
    &
    $\varphi(t)$
    &
    $V$
    &
    $\mathrm{Univ.}$
    &
    $\mathrm{Inf. E.}$
    &
    $\sigma / \Theta$
    &
    $\mathrm{S. Iso.}$
    &
    $\mathrm{W. Iso.}$
    &
    $\mathrm{Sta.?}$
    \\
    \hline \hline
    \endfirsthead

    \multicolumn{13}{c}{
        \tablename\ \thetable\ \; Continued from the table on the previous page.
    }
    \\
    \hline
    $\mathrm{CP}$
    &
    $\Theta$
    &
    $\sigma$
    &
    $\psi$
    &
    $a_{\mathrm{ave.}}(t)$
    &
    $\varphi(t)$
    &
    $V$
    &
    $\mathrm{Univ.}$
    &
    $\mathrm{Inf. E.}$
    &
    $\sigma / \Theta$
    &
    $\mathrm{S. Iso.}$
    &
    $\mathrm{W. Iso.}$
    &
    $\mathrm{Sta.?}$
    \\
    \hline \hline
    \endhead

    \hline
    \multicolumn{13}{c}{Continued on the next page} \\
    \endfoot

    \hline
    \endlastfoot
    
   $\mathrm{CP_{E,1}}$
&
$2/t$
&
$-1/(\sqrt{3}\, t)$
&
$0$
&
$c_{0}t^{2/3}$
&
$c_{1}$
&
$0$
&
$\mathrm{KS}$
&
$\times$
&
$-\sqrt{3}/6$
&
$\times$
&
$\checkmark$
&
$\mathcal{S.}$
\\
\hline
$\mathrm{CP_{E,2-1}}$
&
$1/t$
&
$+1/(\sqrt{3}\, t)$
&
$0$
&
$c_{0}t^{1/3}$
&
$c_{1}$
&
$0$
&
$\mathrm{BI}$
&
$\times$
&
$+1/\sqrt{3}$
&
$\times$
&
$\checkmark$
&
$\mathcal{S.}$
\\
\hline
$\mathrm{CP_{E,2-2}}$
&
$1/t$
&
$-1/(\sqrt{3}\, t)$
&
$0$
&
$c_{0}t^{1/3}$
&
$c_{1}$
&
$0$
&
$\mathrm{BI}$
&
$\times$
&
$-1/\sqrt{3}$
&
$\times$
&
$\checkmark$
&
$\mathcal{S.}$
\\
\hline
$\mathrm{CP_{E,3-1}}$
&
$\frac{1}{\left(1-2 P^{2}\right) t}$
&
$- \, \frac{\sqrt{-18 P^{2}+12}}{6\left(2 P^{2} -1\right) t}$
&
$\frac{P}{\left(1-2 P^{2}\right) t}$
&
$c_{0} \, t^{\frac{1}{3-6 P^{2}}}$
&
$\frac{P \ln (t)}{1-2 P^{2}}+c_{1}$
&
$0$
&
$\mathrm{BI}$
&
Both
&
$\frac{\sqrt{-18 P^{2}+12}}{+6}$
&
Both
&
$\checkmark$
&
$\mathcal{R.}/\mathcal{S.}$
\\
\hline
$\mathrm{CP_{E,3-2}}$
&
$\frac{1}{\left(1-2 P^{2}\right) t}$
&
$+ \, \frac{\sqrt{-18 P^{2}+12}}{6\left(2 P^{2} -1\right) t}$
&
$\frac{P}{\left(1-2 P^{2}\right) t}$
&
$c_{0} \, t^{\frac{1}{3-6 P^{2}}}$
&
$\frac{P \ln (t)}{1-2 P^{2}}+c_{1}$
&
$0$
&
$\mathrm{BI}$
&
Both
&
$\frac{\sqrt{-18 P^{2}+12}}{-6}$
&
Both
&
$\checkmark$
&
$\mathcal{R.}/\mathcal{S.}$
\\
\hline
$\mathrm{CP_{E,3-3}}$
&
$\frac{1}{\left(1-2 P^{2}\right) t}$
&
$- \, \frac{\sqrt{-18 P^{2}+12}}{6\left(2 P^{2} -1\right) t}$
&
$\frac{P}{\left(1-2 P^{2}\right) t}$
&
$c_{0} \, t^{\frac{1}{3-6 P^{2}}}$
&
$\frac{P \ln (t)}{1-2 P^{2}}+c_{1}$
&
$0$
&
$\mathrm{BI}$
&
Both
&
$\frac{\sqrt{-18 P^{2}+12}}{+6}$
&
Both
&
$\checkmark$
&
$\mathcal{R.}/\mathcal{S.}$
\\
\hline
$\mathrm{CP_{E,3-4}}$
&
$\frac{1}{\left(1-2 P^{2}\right) t}$
&
$+ \, \frac{\sqrt{-18 P^{2}+12}}{6\left(2 P^{2} -1\right) t}$
&
$\frac{P}{\left(1-2 P^{2}\right) t}$
&
$c_{0} \, t^{\frac{1}{3-6 P^{2}}}$
&
$\frac{P \ln (t)}{1-2 P^{2}}+c_{1}$
&
$0$
&
$\mathrm{BI}$
&
Both
&
$\frac{\sqrt{-18 P^{2}+12}}{-6}$
&
Both
&
$\checkmark$
&
$\mathcal{R.}/\mathcal{S.}$
\\
\hline
$\mathrm{CP_{E,4-1}}$
&
$\theta_{0}$
&
$0$
&
$0$
&
$e^{\theta_{0} t/3}$
&
$c_{1}$
&
$\theta_{0}^{2}/3$
&
$\mathrm{BI}$
&
$\checkmark$
&
$0$
&
$\checkmark$
&
$\checkmark$
&
$\mathcal{A.}$
\\
\hline
$\mathrm{CP_{E,4-2}}$
&
$-|\theta_{0}|$
&
$0$
&
$0$
&
$e^{-|\theta_{0}| t/3}$
&
$c_{1}$
&
$\theta_{0}^{2}/3$
&
$\mathrm{BI}$
&
$\times$
&
$0$
&
$\checkmark$
&
$\checkmark$
&
$\mathcal{A.}$
\\
\hline
$\mathrm{CP_{E,5}}$
&
$\theta_{0}$
&
$\theta_{0}/\sqrt{3}$
&
$0$
&
$e^{\theta_{0}t/3}$
&
$c_{1}$
&
$\theta_{0}^{2}$
&
$\mathrm{BIII}$
&
$\checkmark$
&
$1/\sqrt{3}$
&
$\times$
&
$\times$
&
$\mathcal{S.}$
\\
\hline
$\mathrm{CP_{E,6}}$
&
$-|\theta_{0}|$
&
$- |\theta_{0}|/\sqrt{3}$
&
$0$
&
$e^{-|\theta_{0}| \, t/3}$
&
$c_{1}$
&
$\theta_{0}^{2}$
&
$\mathrm{BIII}$
&
$\times$
&
$1/\sqrt{3}$
&
$\times$
&
$\times$
&
$\mathcal{S.}$
\\
\hline
$\mathrm{CP_{E,7-1}}$
&
$\frac{2\kappa_{(-1)}^{2}}{3\kappa_{(1/3)}\,t}$
&
$\frac{2\kappa_{(1/2)}\kappa_{(-1)}}{3\sqrt{3}\kappa_{(1/3)}\,t}$
&
$\frac{i\, \kappa^{\star}_{1}}{t}$
&
$c_{0} t^{\frac{2\kappa_{(-1)}^{2}}{9\kappa_{(1/3)}}}$
&
$i\, \kappa^{\star}_{1}\ln (t)+c_{1}$
&
$\kappa^{\star}_{2}$
&
All
&
All
&
$\frac{\kappa_{(1/2)}}{\sqrt{3}\, \kappa_{(-1)}}$
&
$\times$
&
$\checkmark$
&
$\mathcal{R.}/\mathcal{A.}$
\\
\hline
$\mathrm{CP_{E,7-2}}$
&
$\frac{2\kappa_{(-1)}^{2}}{3\kappa_{(1/3)}\,t}$
&
$\frac{2\kappa_{(1/2)}\kappa_{(-1)}}{3\sqrt{3}\kappa_{(1/3)}\,t}$
&
$\frac{i\, \kappa^{\star}_{1}}{t}$
&
$c_{0} t^{\frac{2\kappa_{(-1)}^{2}}{9\kappa_{(1/3)}}}$
&
$i\, \kappa^{\star}_{1}\ln (t)+c_{1}$
&
$\kappa^{\star}_{2}$
&
All
&
All
&
$\frac{\kappa_{(1/2)}}{\sqrt{3}\, \kappa_{(-1)}}$
&
$\times$
&
$\checkmark$
&
$\mathcal{R.}/\mathcal{A.}$
\\
\hline
$\mathrm{CP_{E,7-3}}$
&
$\frac{2\kappa_{(-1)}^{2}}{3\kappa_{(1/3)}\,t}$
&
$\frac{2\kappa_{(1/2)}\kappa_{(-1)}}{3\sqrt{3}\kappa_{(1/3)}\,t}$
&
$\frac{i\, \kappa^{\star}_{1}}{t}$
&
$c_{0} t^{\frac{2\kappa_{(-1)}^{2}}{9\kappa_{(1/3)}}}$
&
$i\, \kappa^{\star}_{1}\ln (t)+c_{1}$
&
$\kappa^{\star}_{2}$
&
All
&
All
&
$\frac{\kappa_{(1/2)}}{\sqrt{3}\, \kappa_{(-1)}}$
&
$\times$
&
$\checkmark$
&
$\mathcal{R.}/\mathcal{A.}$
\\
\hline
$\mathrm{CP_{E,7-4}}$
&
$\frac{2\kappa_{(-1)}^{2}}{3\kappa_{(1/3)}\,t}$
&
$\frac{2\kappa_{(1/2)}\kappa_{(-1)}}{3\sqrt{3}\kappa_{(1/3)}\,t}$
&
$\frac{i\, \kappa^{\star}_{1}}{t}$
&
$c_{0} t^{\frac{2\kappa_{(-1)}^{2}}{9\kappa_{(1/3)}}}$
&
$i\, \kappa^{\star}_{1}\ln (t)+c_{1}$
&
$\kappa^{\star}_{2}$
&
All
&
All
&
$\frac{\kappa_{(1/2)}}{\sqrt{3}\, \kappa_{(-1)}}$
&
$\times$
&
$\checkmark$
&
$\mathcal{R.}/\mathcal{A.}$
\\
\hline
$\mathrm{CP_{E,7-5}}$
&
$\frac{2\kappa_{(-1)}^{2}}{3\kappa_{(1/3)}\,t}$
&
$\frac{2\kappa_{(1/2)}\kappa_{(-1)}}{3\sqrt{3}\kappa_{(1/3)}\,t}$
&
$-\frac{i\, \kappa^{\star}_{1}}{t}$
&
$c_{0} t^{\frac{2\kappa_{(-1)}^{2}}{9\kappa_{(1/3)}}}$
&
$-i\, \kappa^{\star}_{1}\ln (t)+c_{1}$
&
$\kappa^{\star}_{2}$
&
All
&
All
&
$\frac{\kappa_{(1/2)}}{\sqrt{3}\, \kappa_{(-1)}}$
&
$\times$
&
$\checkmark$
&
$\mathcal{R.}/\mathcal{A.}$
\\
\hline
$\mathrm{CP_{E,7-6}}$
&
$\frac{2\kappa_{(-1)}^{2}}{3\kappa_{(1/3)}\,t}$
&
$\frac{2\kappa_{(1/2)}\kappa_{(-1)}}{3\sqrt{3}\kappa_{(1/3)}\,t}$
&
$-\frac{i\, \kappa^{\star}_{1}}{t}$
&
$c_{0} t^{\frac{2\kappa_{(-1)}^{2}}{9\kappa_{(1/3)}}}$
&
$-i\, \kappa^{\star}_{1}\ln (t)+c_{1}$
&
$\kappa^{\star}_{2}$
&
All
&
All
&
$\frac{\kappa_{(1/2)}}{\sqrt{3}\, \kappa_{(-1)}}$
&
$\times$
&
$\checkmark$
&
$\mathcal{R.}/\mathcal{A.}$
\\
\hline
$\mathrm{CP_{E,7-7}}$
&
$\frac{2\kappa_{(-1)}^{2}}{3\kappa_{(1/3)}\,t}$
&
$\frac{2\kappa_{(1/2)}\kappa_{(-1)}}{3\sqrt{3}\kappa_{(1/3)}\,t}$
&
$-\frac{i\, \kappa^{\star}_{1}}{t}$
&
$c_{0} t^{\frac{2\kappa_{(-1)}^{2}}{9\kappa_{(1/3)}}}$
&
$-i\, \kappa^{\star}_{1}\ln (t)+c_{1}$
&
$\kappa^{\star}_{2}$
&
All
&
All
&
$\frac{\kappa_{(1/2)}}{\sqrt{3}\, \kappa_{(-1)}}$
&
$\times$
&
$\checkmark$
&
$\mathcal{R.}/\mathcal{A.}$
\\
\hline
$\mathrm{CP_{E,7-8}}$
&
$\frac{2\kappa_{(-1)}^{2}}{3\kappa_{(1/3)}\,t}$
&
$\frac{2\kappa_{(1/2)}\kappa_{(-1)}}{3\sqrt{3}\kappa_{(1/3)}\,t}$
&
$-\frac{i\, \kappa^{\star}_{1}}{t}$
&
$c_{0} t^{\frac{2\kappa_{(-1)}^{2}}{9\kappa_{(1/3)}}}$
&
$-i\, \kappa^{\star}_{1}\ln (t)+c_{1}$
&
$\kappa^{\star}_{2}$
&
All
&
All
&
$\frac{\kappa_{(1/2)}}{\sqrt{3}\, \kappa_{(-1)}}$
&
$\times$
&
$\checkmark$
&
$\mathcal{R.}/\mathcal{A.}$
\\
\hline
$\mathrm{CP_{E,8-1}}$
&
$18 l^{2}/t$
&
$0$
&
$+6il/t$
&
$c_{0}t^{6l^{2}}$
&
$+6il\, \ln(t)+c_{1}$
&
$\frac{108 \, l^{2}\kappa_{(18/108)}}{t^{2}}$
&
$\mathrm{BI}$
&
All
&
$0$
&
$\checkmark$
&
$\checkmark$
&
$\mathcal{A.}$
\\
\hline
$\mathrm{CP_{E,8-2}}$
&
$18 l^{2}/t$
&
$0$
&
$+6il/t$
&
$c_{0}t^{6l^{2}}$
&
$+6il\, \ln(t)+c_{1}$
&
$\frac{108 \, l^{2}\kappa_{(18/108)}}{t^{2}}$
&
$\mathrm{BI}$
&
All
&
$0$
&
$\checkmark$
&
$\checkmark$
&
$\mathcal{A.}$
\\
\hline
$\mathrm{CP_{E,8-3}}$
&
$18 l^{2}/t$
&
$0$
&
$+6il/t$
&
$c_{0}t^{6l^{2}}$
&
$+6il\, \ln(t)+c_{1}$
&
$\frac{108 \, l^{2}\kappa_{(18/108)}}{t^{2}}$
&
$\mathrm{BI}$
&
All
&
$0$
&
$\checkmark$
&
$\checkmark$
&
$\mathcal{A.}$
\\
\hline
$\mathrm{CP_{E,8-4}}$
&
$18 l^{2}/t$
&
$0$
&
$+6il/t$
&
$c_{0}t^{6l^{2}}$
&
$+6il\, \ln(t)+c_{1}$
&
$\frac{108 \, l^{2}\kappa_{(18/108)}}{t^{2}}$
&
$\mathrm{BI}$
&
All
&
$0$
&
$\checkmark$
&
$\checkmark$
&
$\mathcal{A.}$
\\
\hline
$\mathrm{CP_{E,8-5}}$
&
$18 l^{2}/t$
&
$0$
&
$-6il/t$
&
$c_{0}t^{6l^{2}}$
&
$-6il\, \ln(t)+c_{1}$
&
$\frac{108 \, l^{2}\kappa_{(18/108)}}{t^{2}}$
&
$\mathrm{BI}$
&
All
&
$0$
&
$\checkmark$
&
$\checkmark$
&
$\mathcal{S.}$
\\
\hline
$\mathrm{CP_{E,8-6}}$
&
$18 l^{2}/t$
&
$0$
&
$-6il/t$
&
$c_{0}t^{6l^{2}}$
&
$-6il\, \ln(t)+c_{1}$
&
$\frac{108 \, l^{2}\kappa_{(18/108)}}{t^{2}}$
&
$\mathrm{BI}$
&
All
&
$0$
&
$\checkmark$
&
$\checkmark$
&
$\mathcal{S.}$
\\
\hline
$\mathrm{CP_{E,8-7}}$
&
$18 l^{2}/t$
&
$0$
&
$-6il/t$
&
$c_{0}t^{6l^{2}}$
&
$-6il\, \ln(t)+c_{1}$
&
$\frac{108 \, l^{2}\kappa_{(18/108)}}{t^{2}}$
&
$\mathrm{BI}$
&
All
&
$0$
&
$\checkmark$
&
$\checkmark$
&
$\mathcal{S.}$
\\
\hline
$\mathrm{CP_{E,8-8}}$
&
$18 l^{2}/t$
&
$0$
&
$-6il/t$
&
$c_{0}t^{6l^{2}}$
&
$-6il\, \ln(t)+c_{1}$
&
$\frac{108 \, l^{2}\kappa_{(18/108)}}{t^{2}}$
&
$\mathrm{BI}$
&
All
&
$0$
&
$\checkmark$
&
$\checkmark$
&
$\mathcal{S.}$
    \\
    \hline \hline
\end{longtable}
\end{landscape}

\section{Figures Produced by the Perturbation Method for the Potential $\mathrm{A}$\label{appendix_figures}}
The following figures present the results obtained by applying the perturbative method to all critical points of potential $\mathrm{A}$. All possible cases have been examined. For $\mathrm{CP}_{3-1}$ and $\mathrm{CP}_{3-2}$, two figures are provided because, under the conditions associated with these critical points, both ``inflation without strong isotropization'' and ``strong isotropization without inflation'' are possible. These two cases must therefore be analyzed separately.

	\begin{landscape}
	\begin{figure}
		\centering
		\subfloat[]{
			\includegraphics[width=1.9 in, height= 1.9 in]{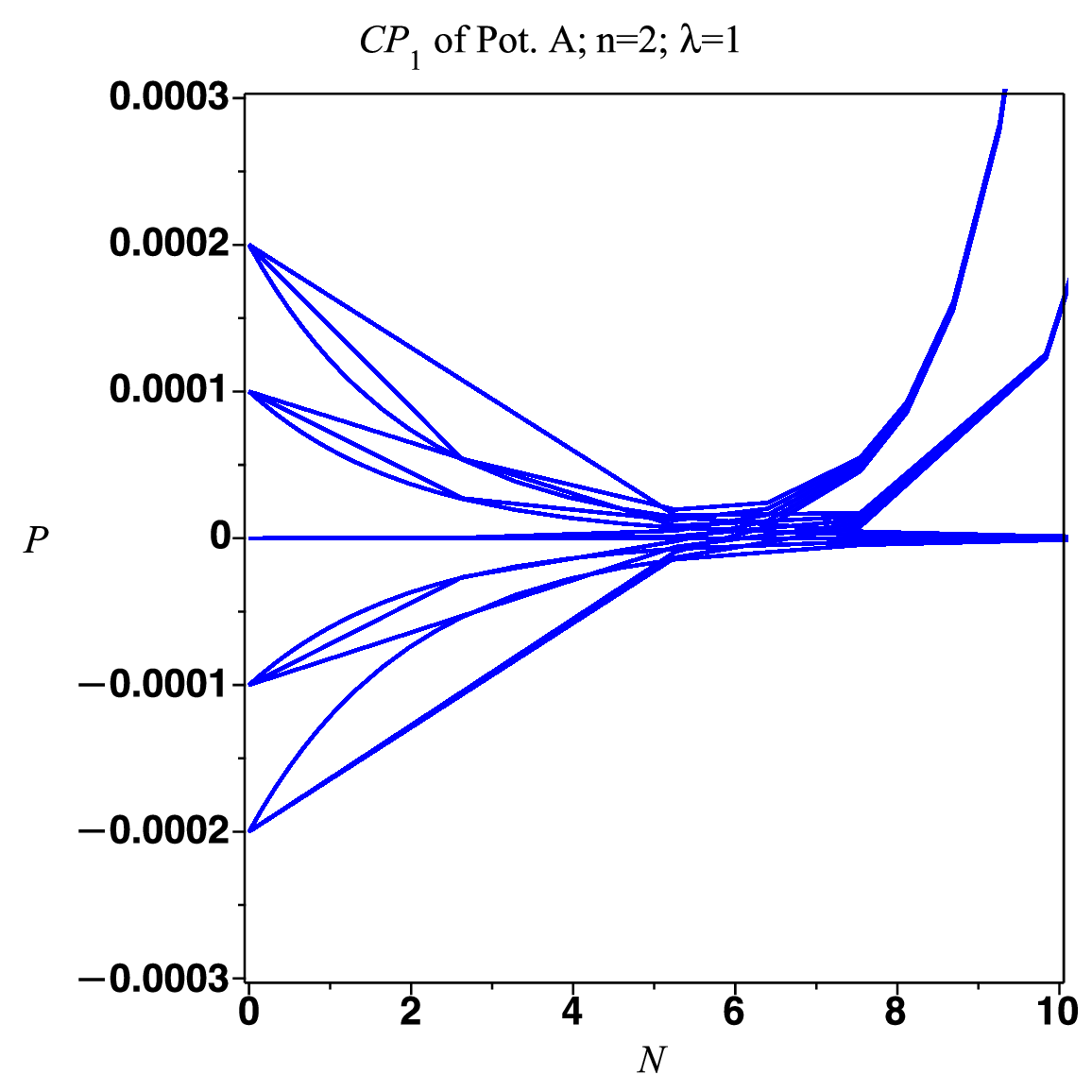}
		}
		\quad
		\subfloat[]{
			\includegraphics[width=1.9 in, height= 1.9 in]{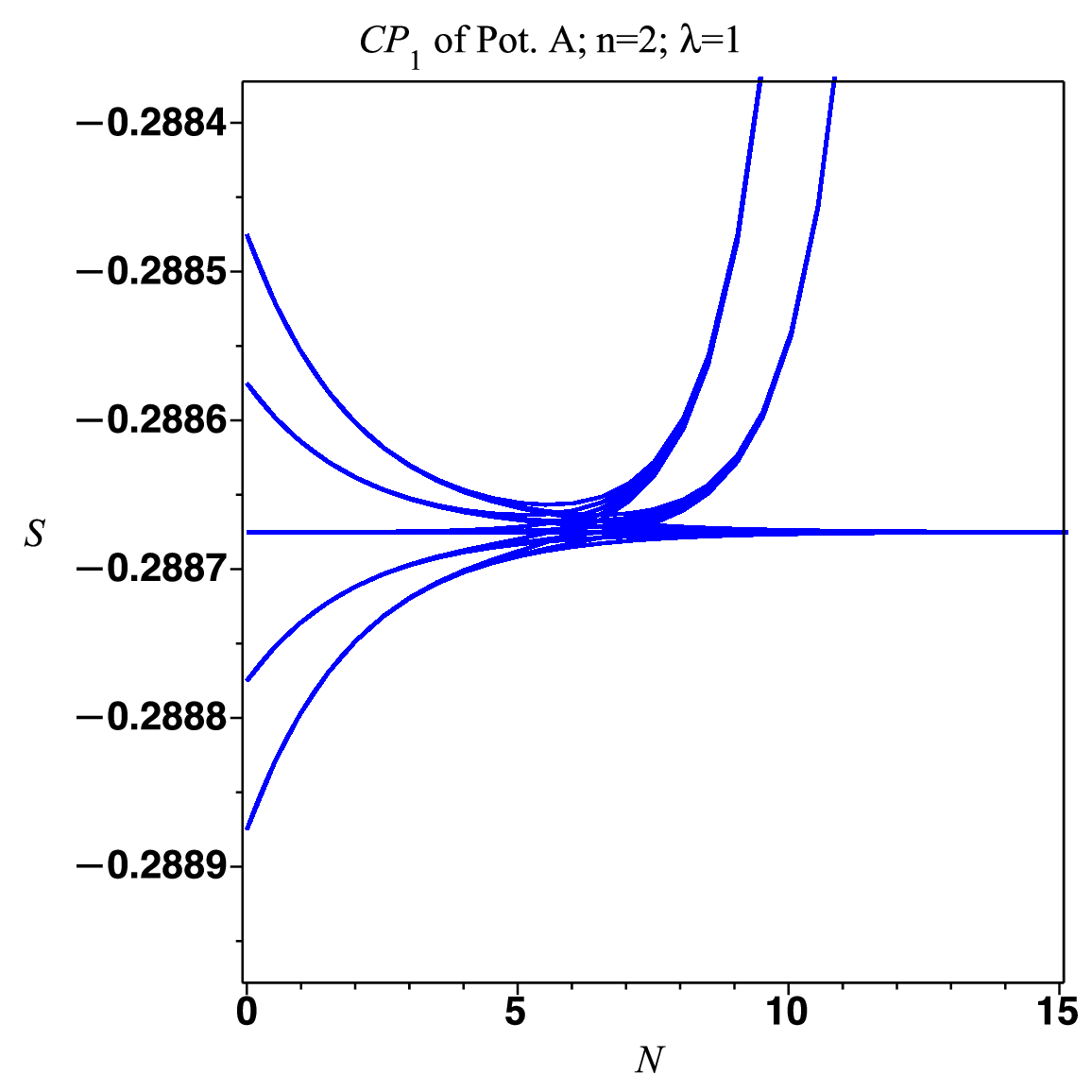}
		}
		\quad
		\subfloat[]{
			\includegraphics[width=1.9 in, height= 1.9 in]{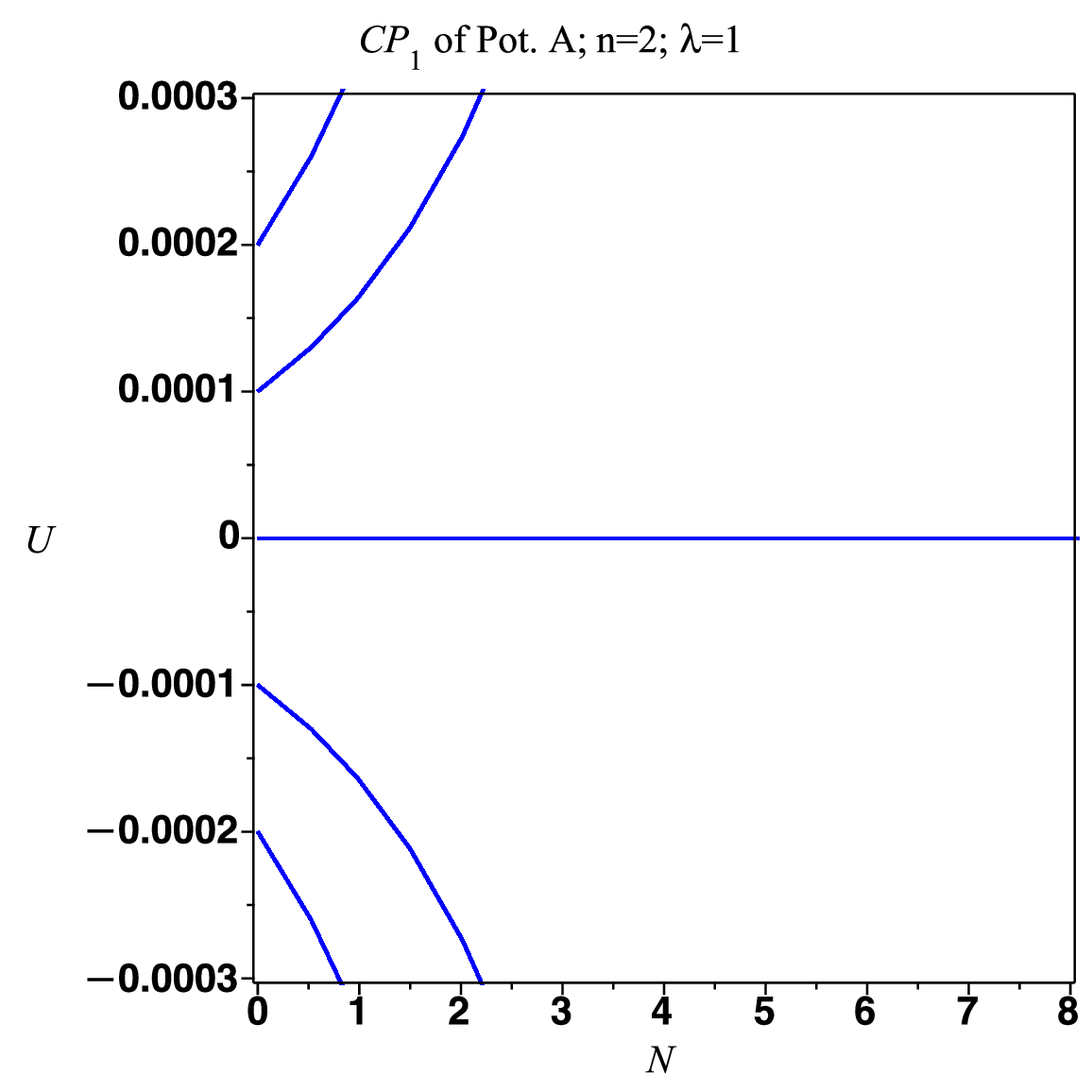}
		}
		\quad
		\subfloat[]{
			\includegraphics[width=1.9 in, height= 1.9 in]{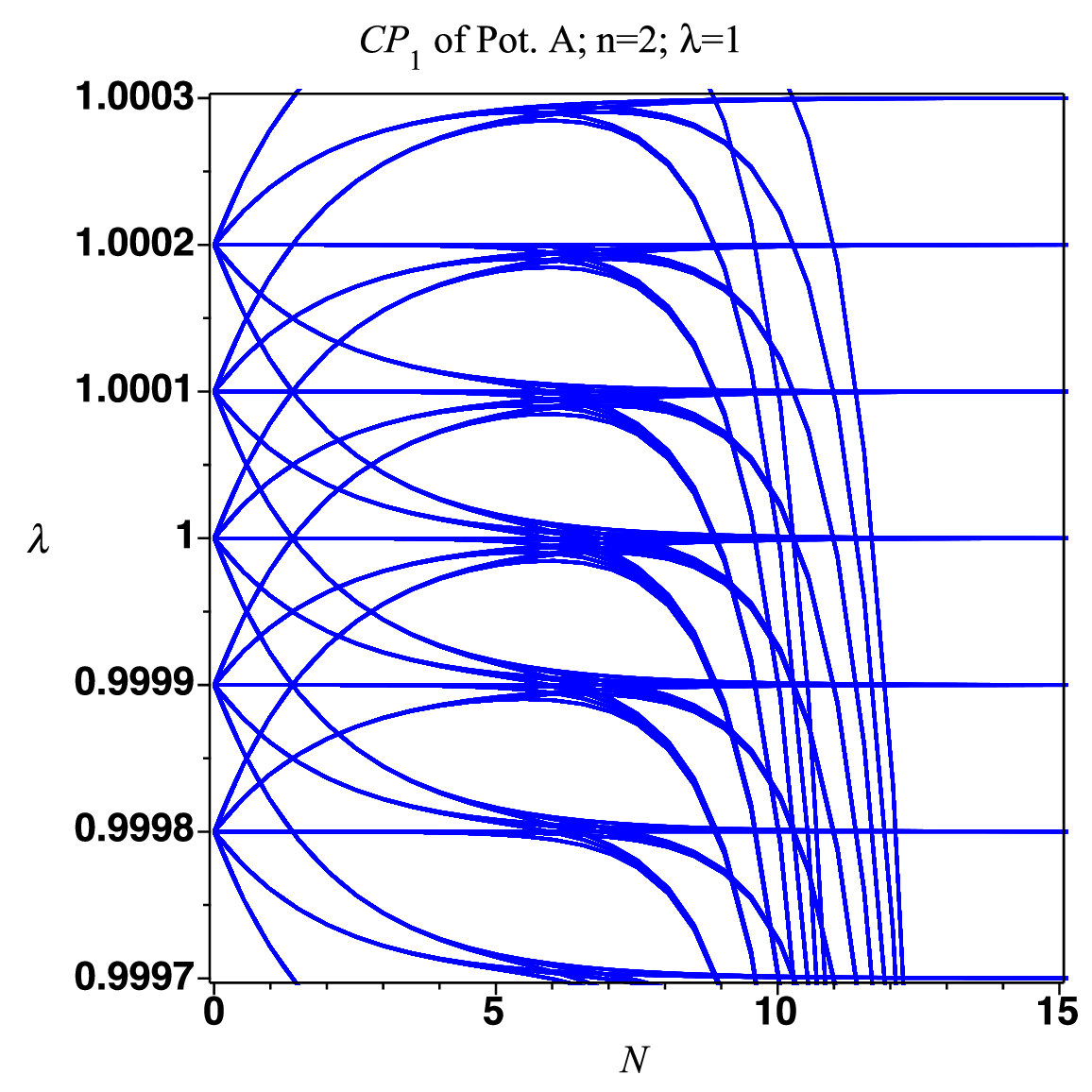}
		}
		\caption{
			Perturbative behavior of potential $\mathrm{A}$ around the equilibrium point $\mathrm{CP}_{A,1}$ for $\lambda=1$ and $n=2$.
		}
		\label{FigPotA1}
	\end{figure}
\end{landscape}
	\begin{landscape}
	\begin{figure}
		\centering
		\subfloat[]{
			\includegraphics[width=1.9 in, height= 1.9 in]{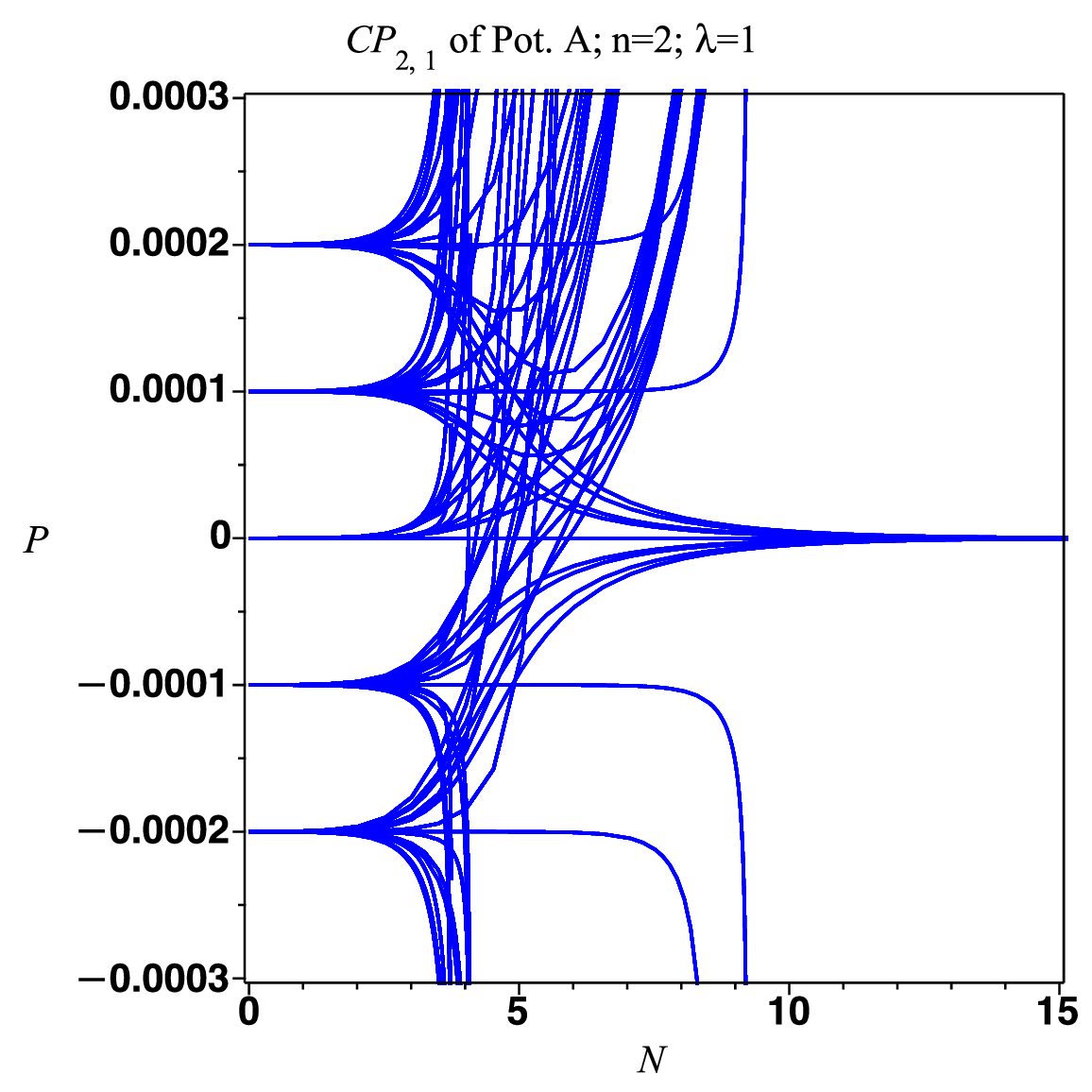}
		}
		\quad
		\subfloat[]{
			\includegraphics[width=1.9 in, height= 1.9 in]{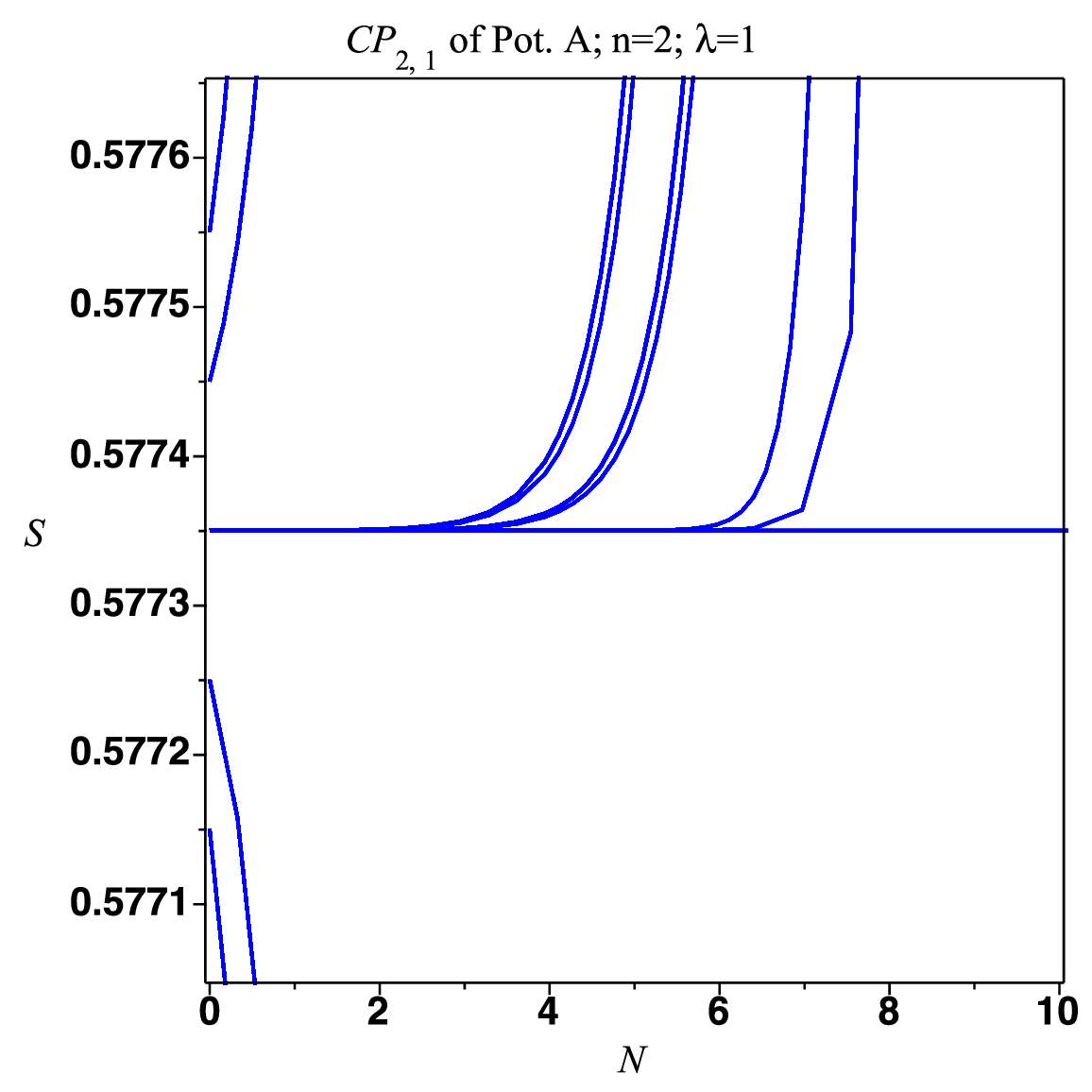}
		}
		\quad
		\subfloat[]{
			\includegraphics[width=1.9 in, height= 1.9 in]{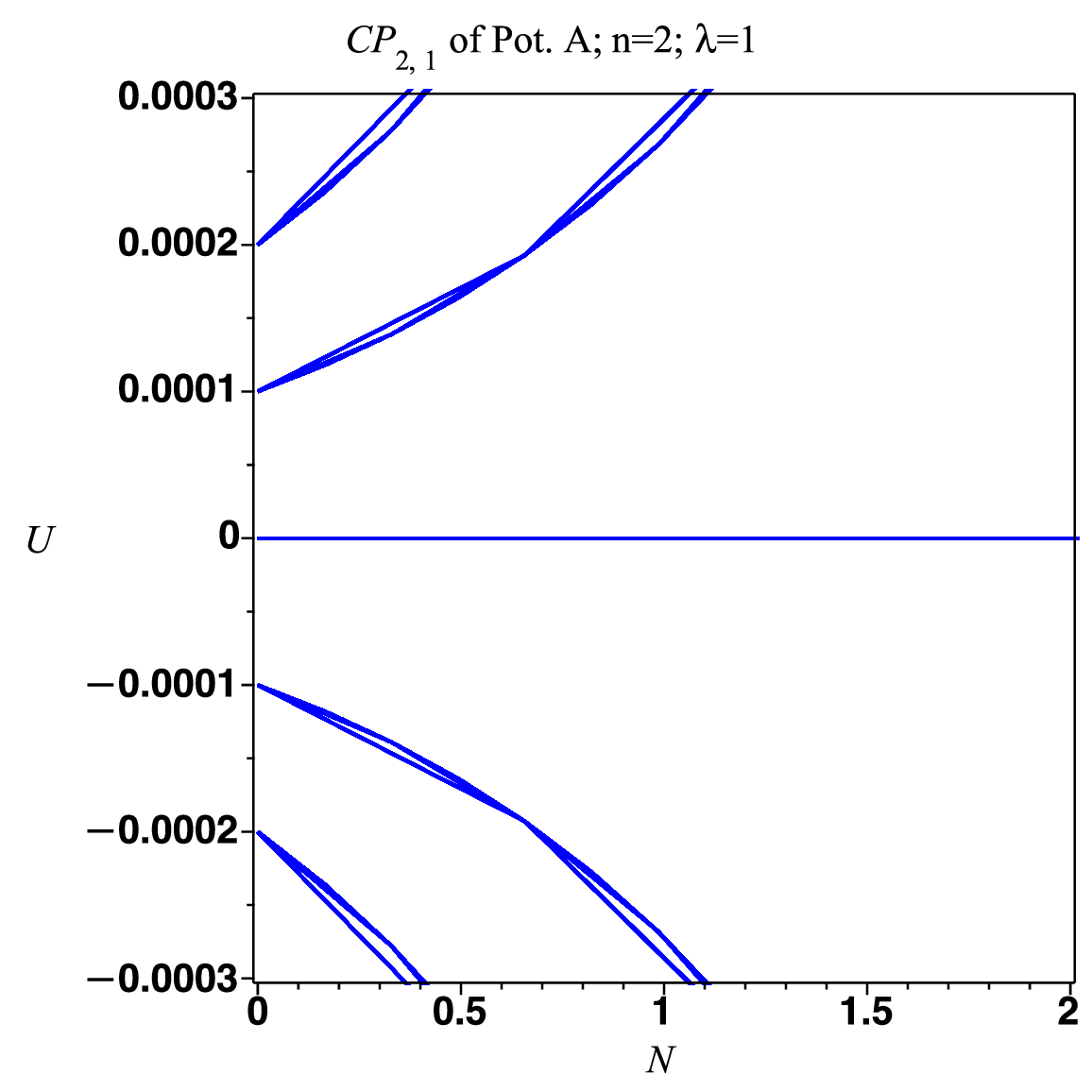}
		}
		\quad
		\subfloat[]{
			\includegraphics[width=1.9 in, height= 1.9 in]{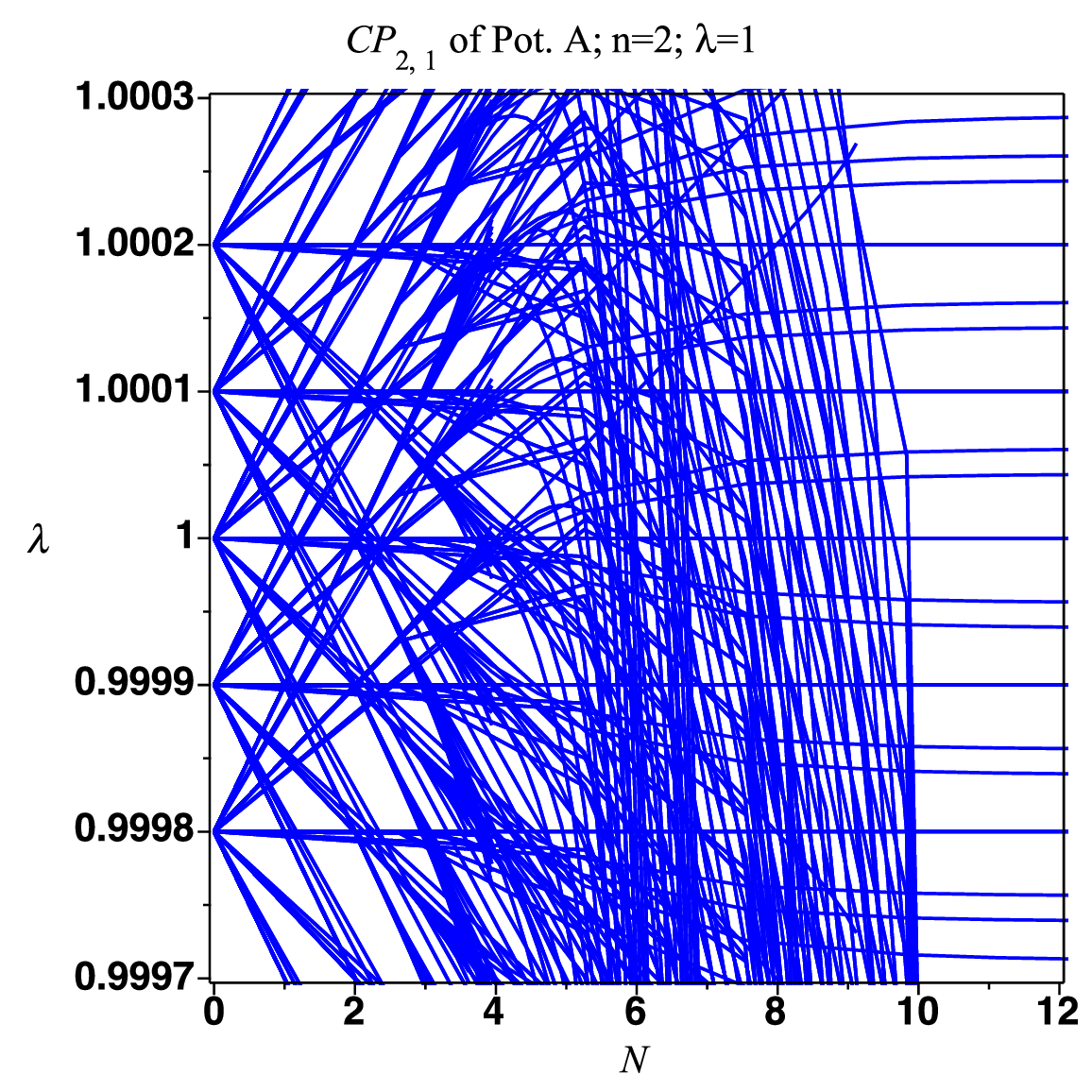}
		}
		\caption{
			Perturbative behavior of potential $\mathrm{A}$ around the equilibrium point $\mathrm{CP}_{A,2-1}$ for $\lambda=1$ and $n=2$.
		}
		\label{FigPotA21}
	\end{figure}
	\begin{figure}
		\centering
		\subfloat[]{
			\includegraphics[width=1.9 in, height= 1.9 in]{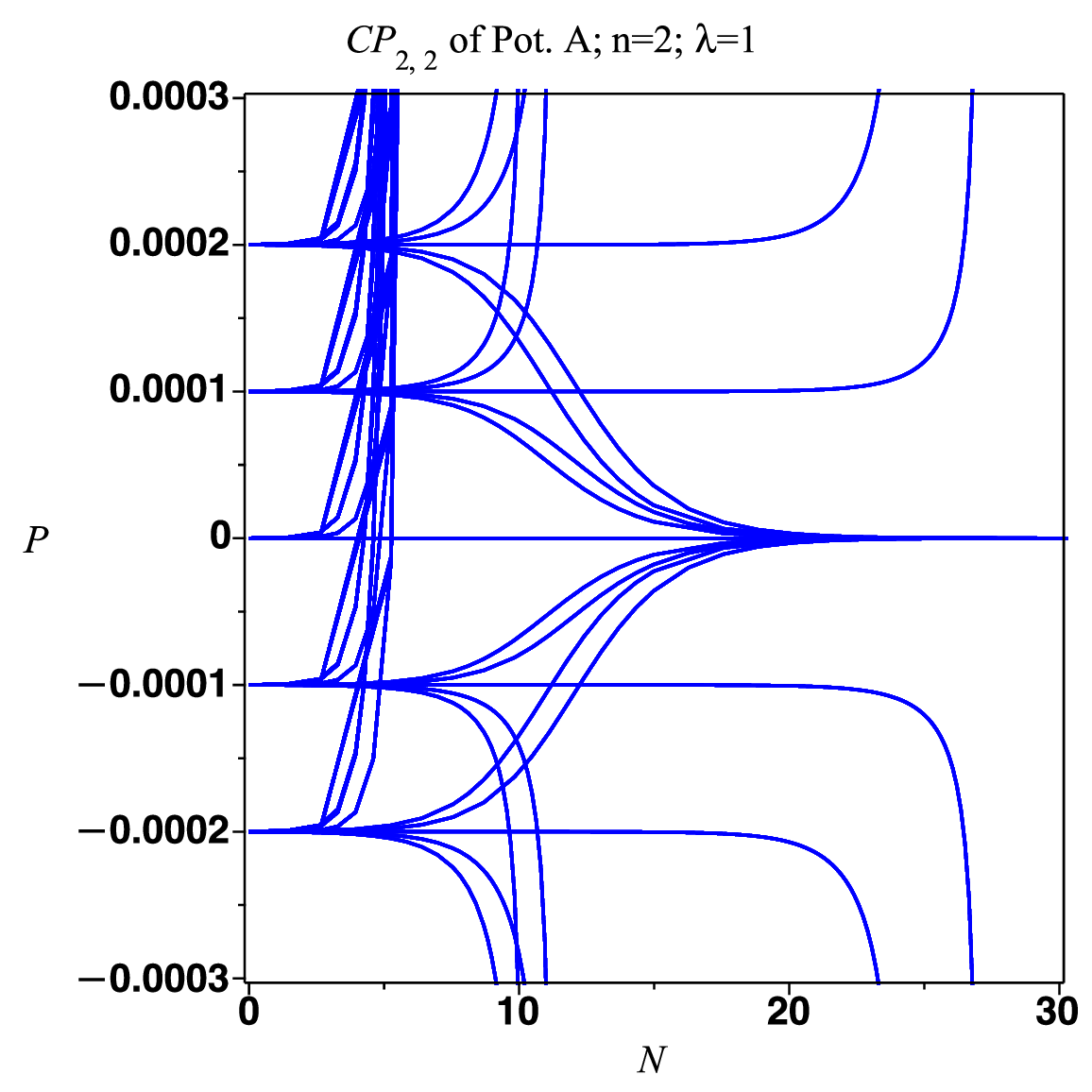}
		}
		\quad
		\subfloat[]{
			\includegraphics[width=1.9 in, height= 1.9 in]{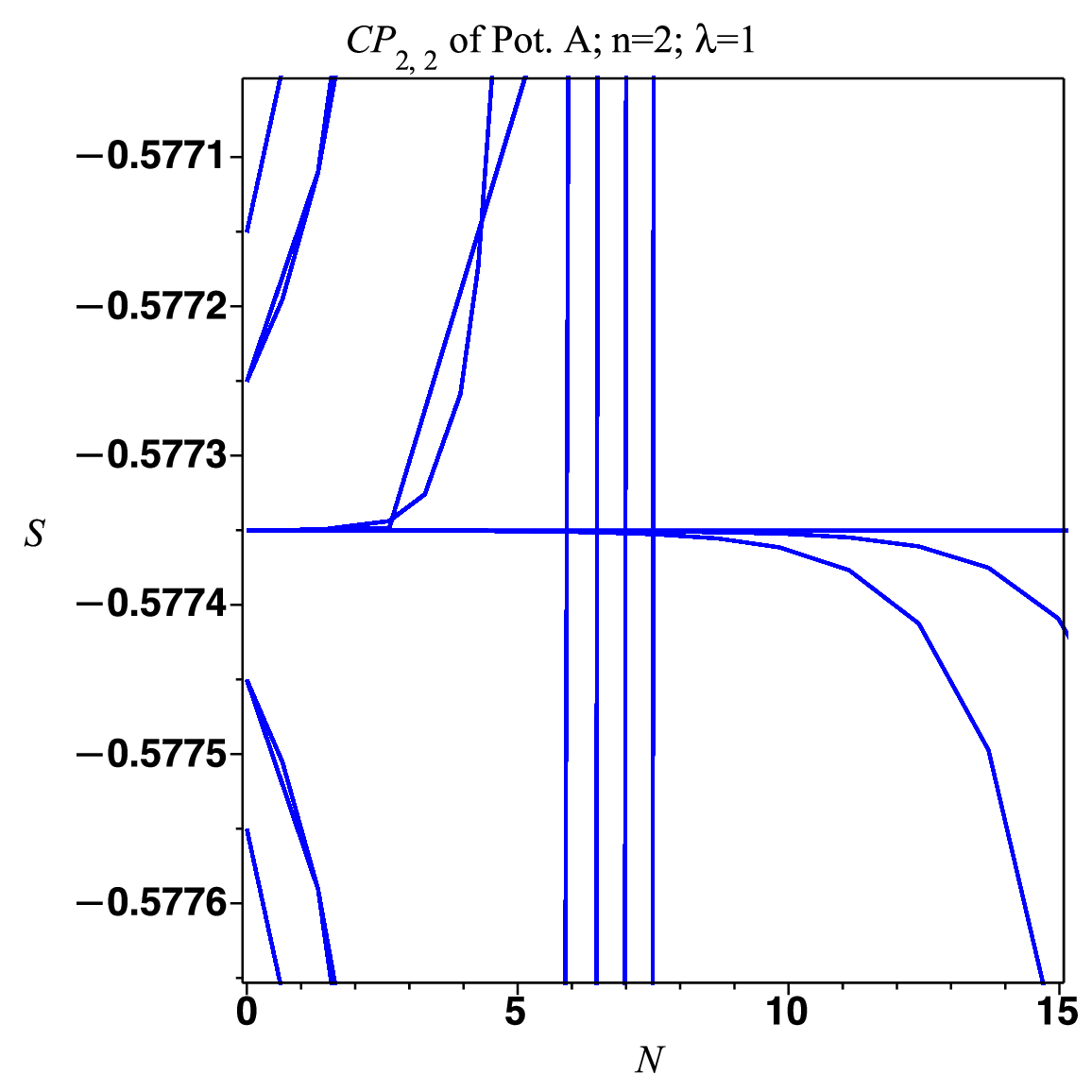}
		}
		\quad
		\subfloat[]{
			\includegraphics[width=1.9 in, height= 1.9 in]{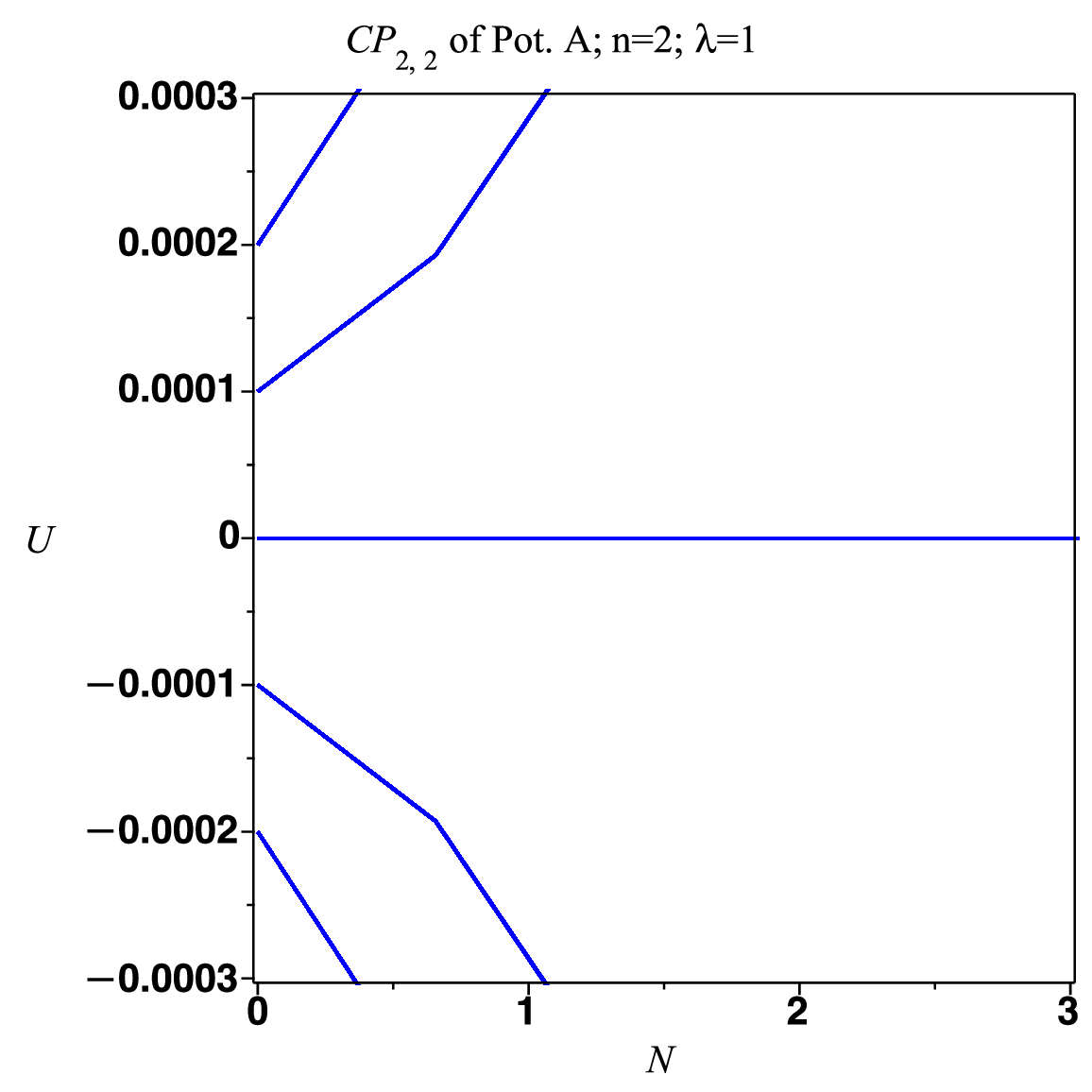}
		}
		\quad
		\subfloat[]{
			\includegraphics[width=1.9 in, height= 1.9 in]{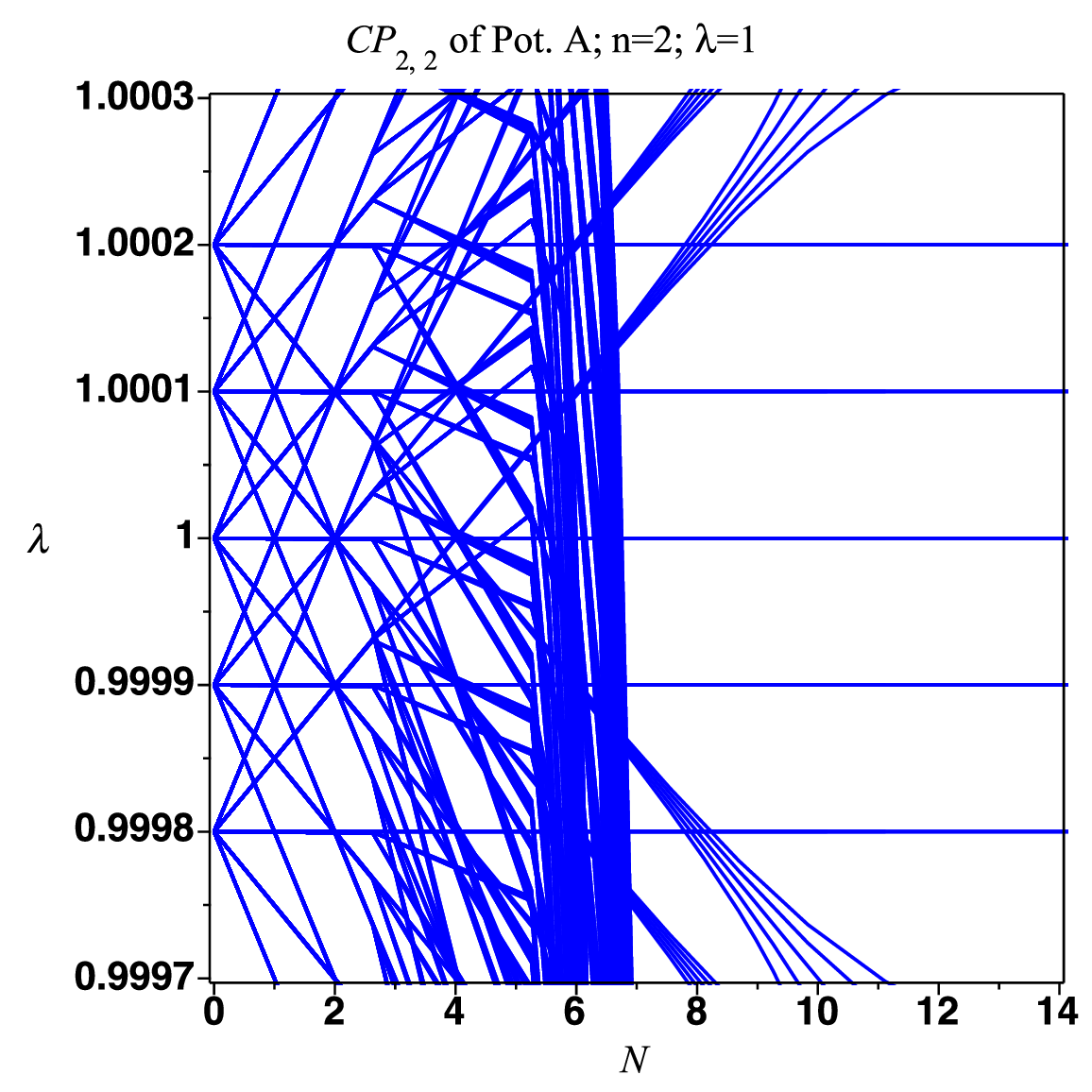}
		}
		\caption{
				Perturbative behavior of potential $\mathrm{A}$ around the equilibrium point $\mathrm{CP}_{A,2-2}$ for $\lambda=1$ and $n=2$.
		}
		\label{FigPotA22}
	\end{figure}
\end{landscape}
	\begin{landscape}
	\begin{figure}
		\centering
		\subfloat[]{
			\includegraphics[width=1.9 in, height= 1.9 in]{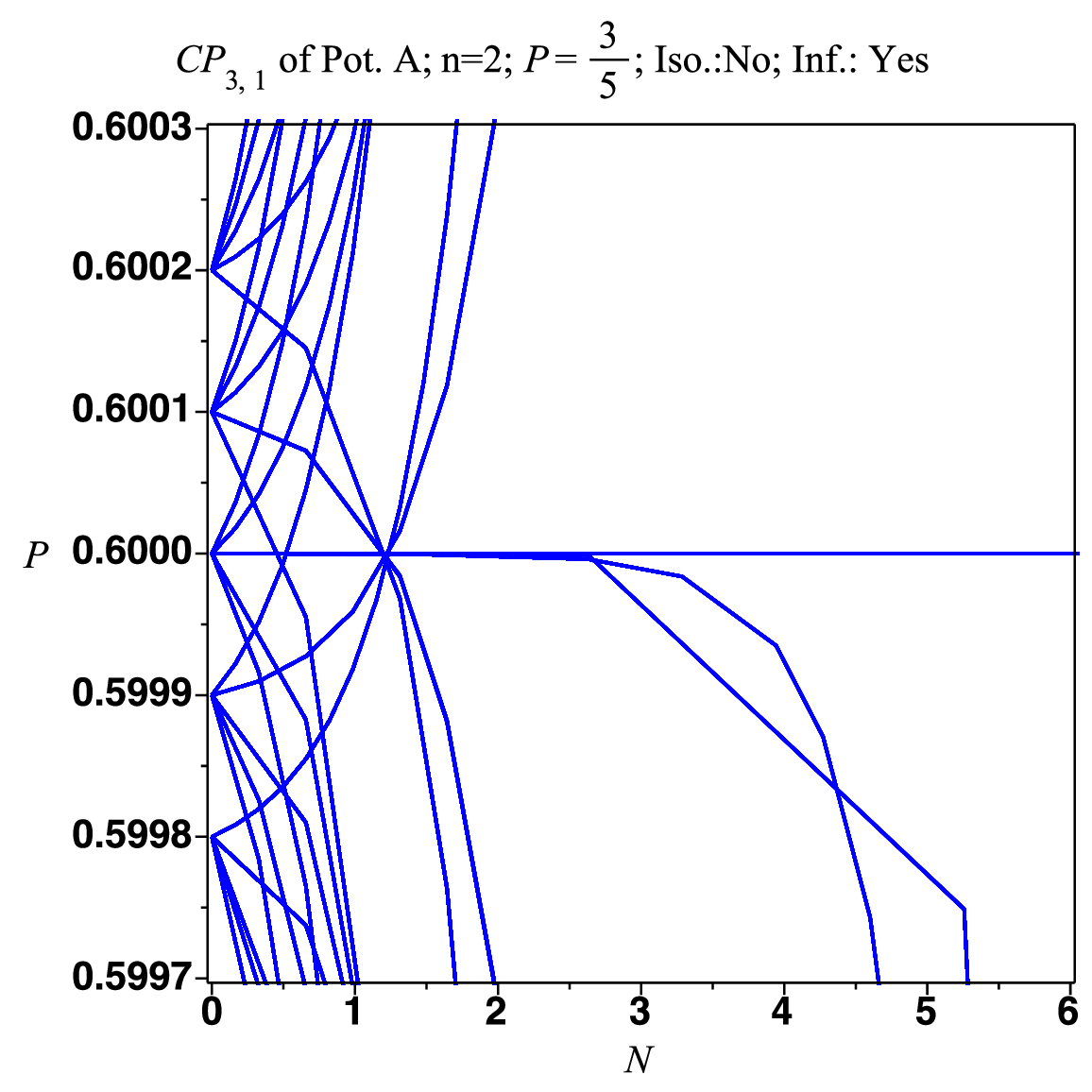}
		}
		\quad
		\subfloat[]{
			\includegraphics[width=1.9 in, height= 1.9 in]{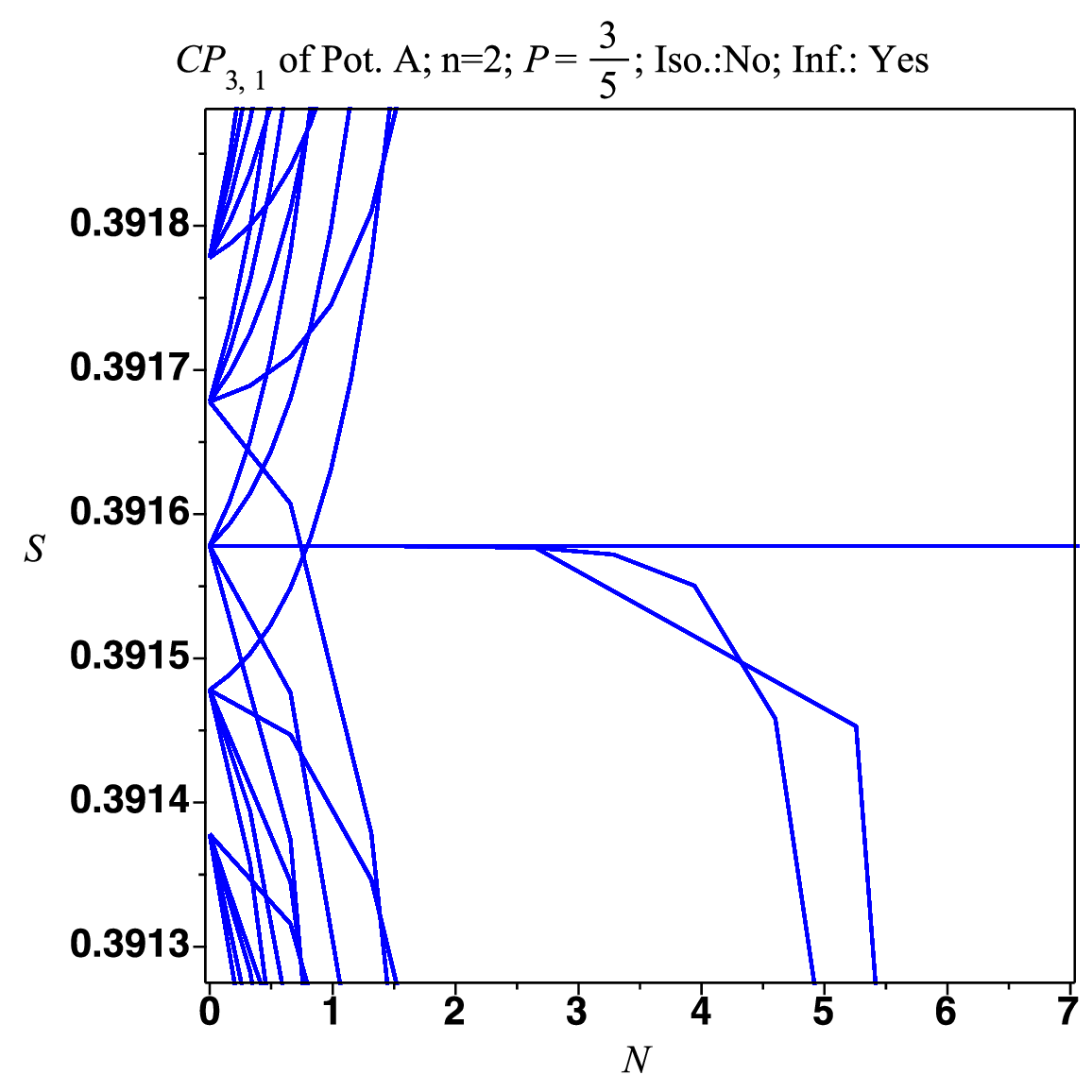}
		}
		\quad
		\subfloat[]{
			\includegraphics[width=1.9 in, height= 1.9 in]{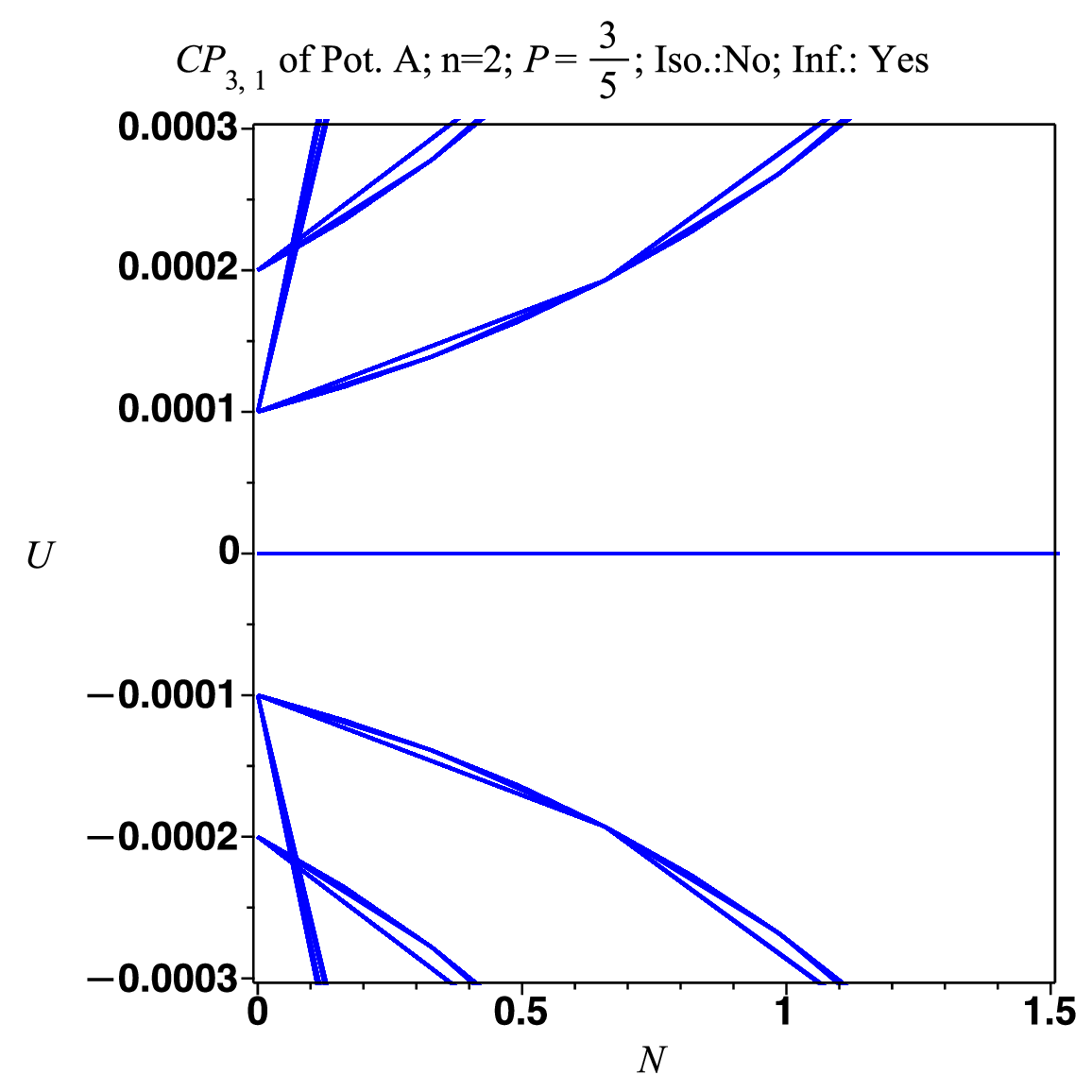}
		}
		\quad
		\subfloat[]{
			\includegraphics[width=1.9 in, height= 1.9 in]{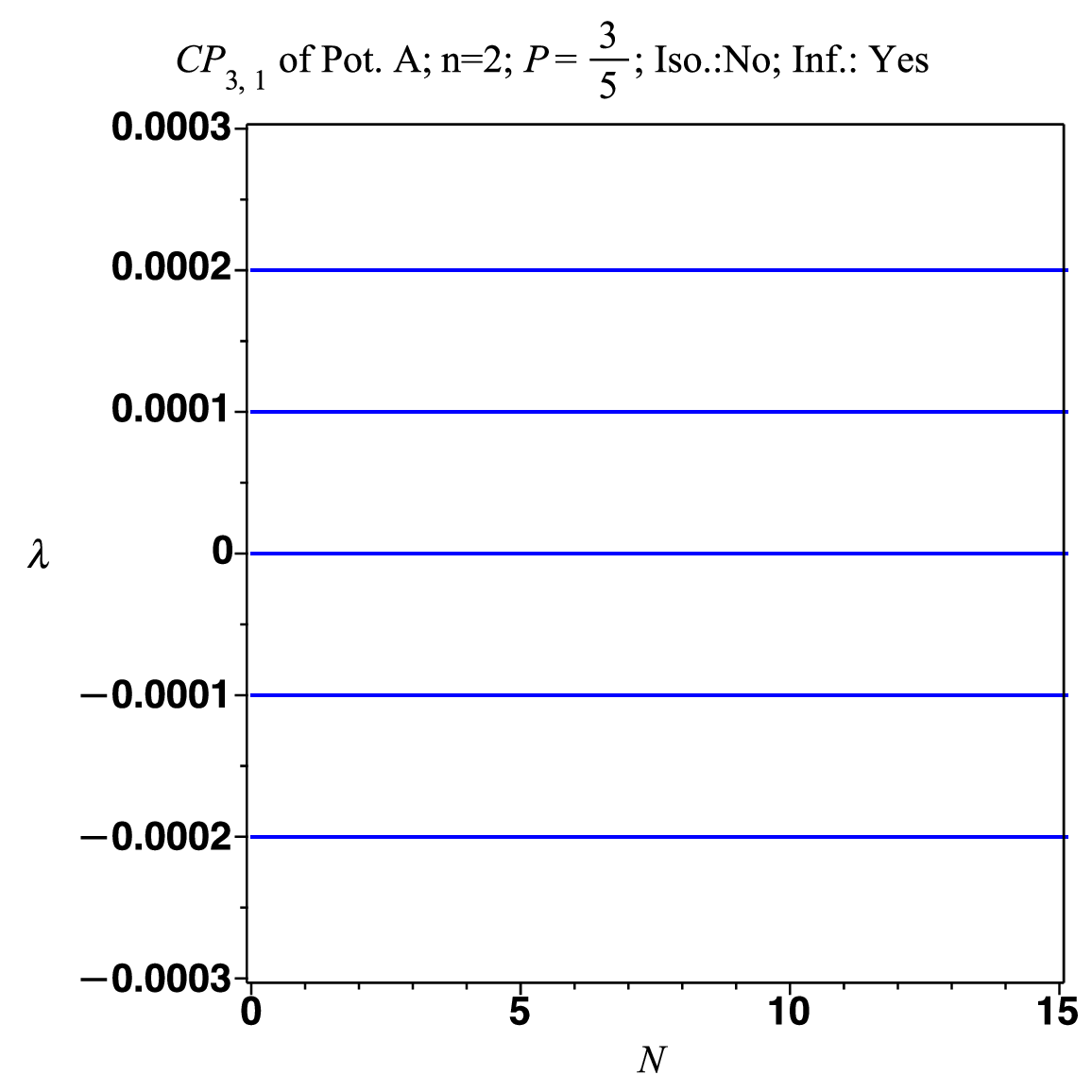}
		}
		\caption{
				Perturbative behavior of potential $\mathrm{A}$ around the equilibrium point $\mathrm{CP}_{A,3-1}$ for $\lambda=1$, $n=2$, and $P=3/5$ (Inflation without Strong Isotropization).
		}
		\label{FigPotA31-inf}
	\end{figure}
	\begin{figure}
		\centering
		\subfloat[]{
			\includegraphics[width=1.9 in, height= 1.9 in]{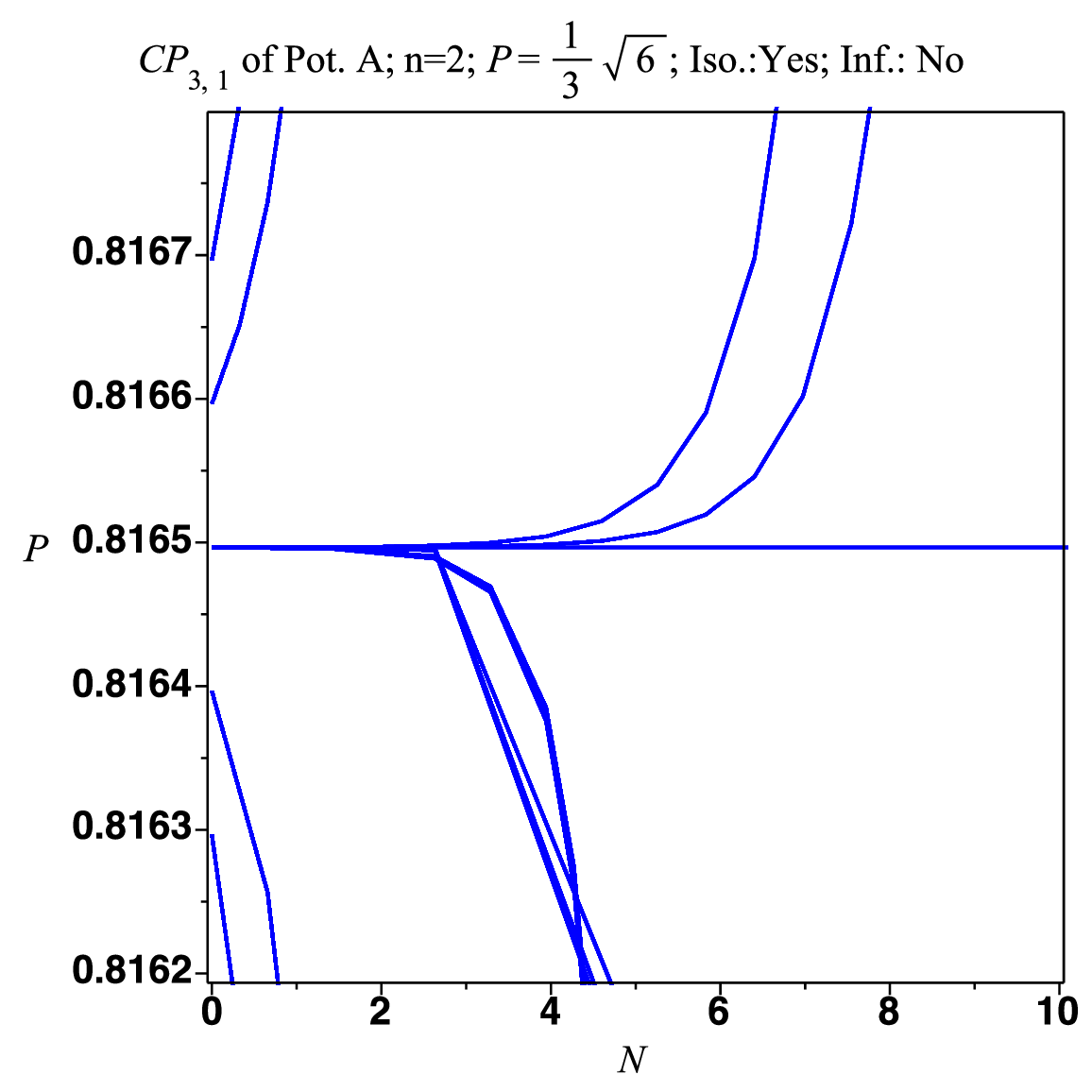}
		}
		\quad
		\subfloat[]{
			\includegraphics[width=1.9 in, height= 1.9 in]{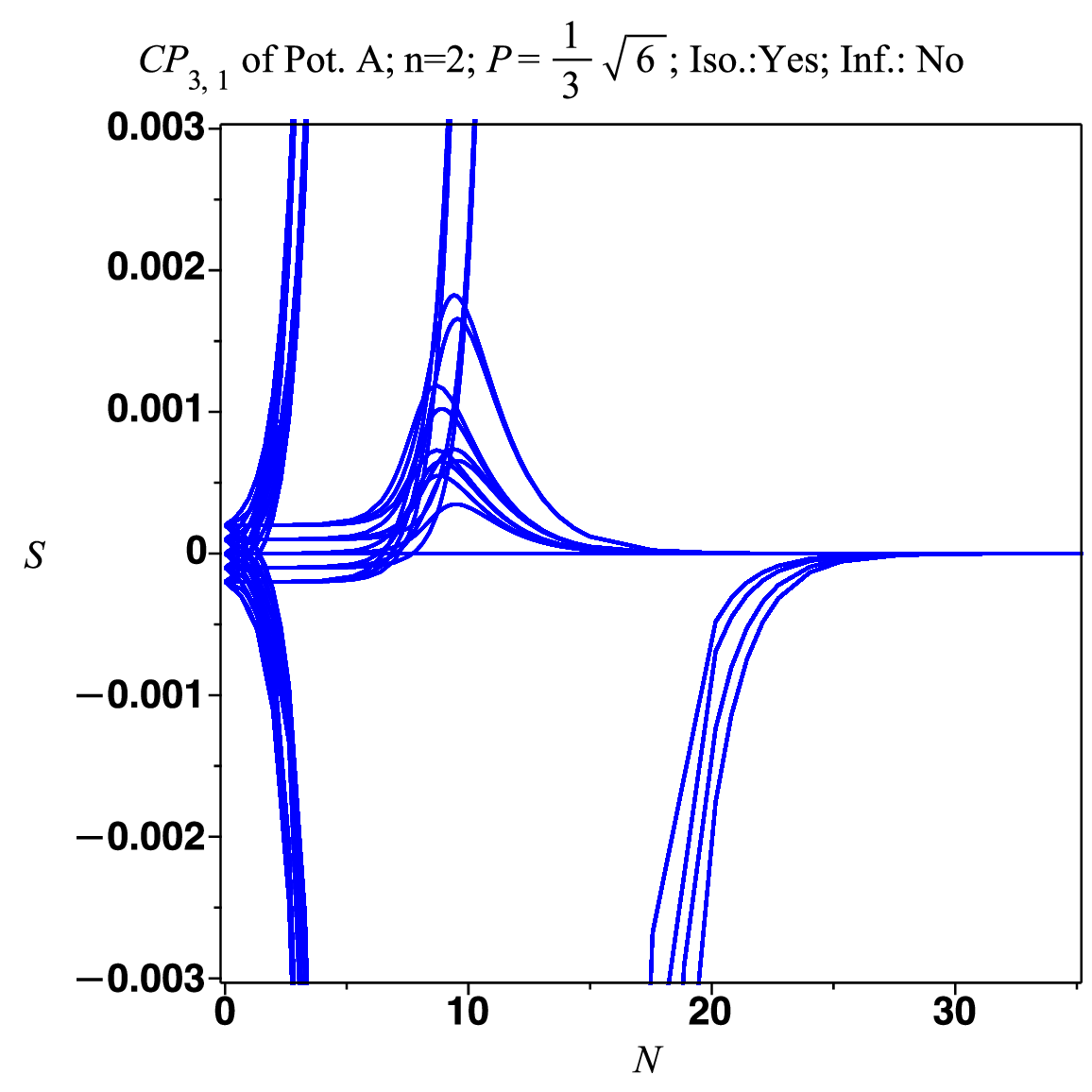}
		}
		\quad
		\subfloat[]{
			\includegraphics[width=1.9 in, height= 1.9 in]{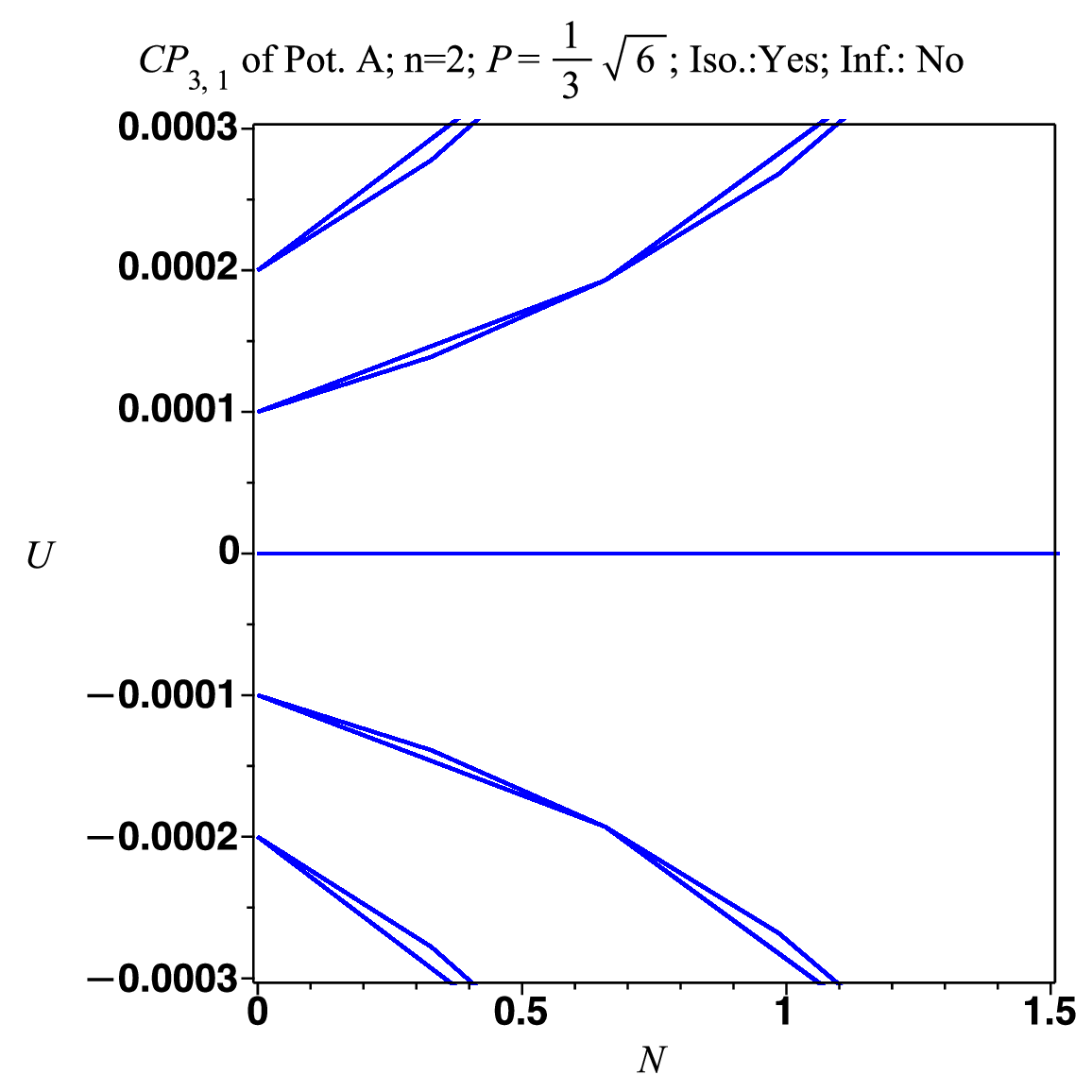}
		}
		\quad
		\subfloat[]{
			\includegraphics[width=1.9 in, height= 1.9 in]{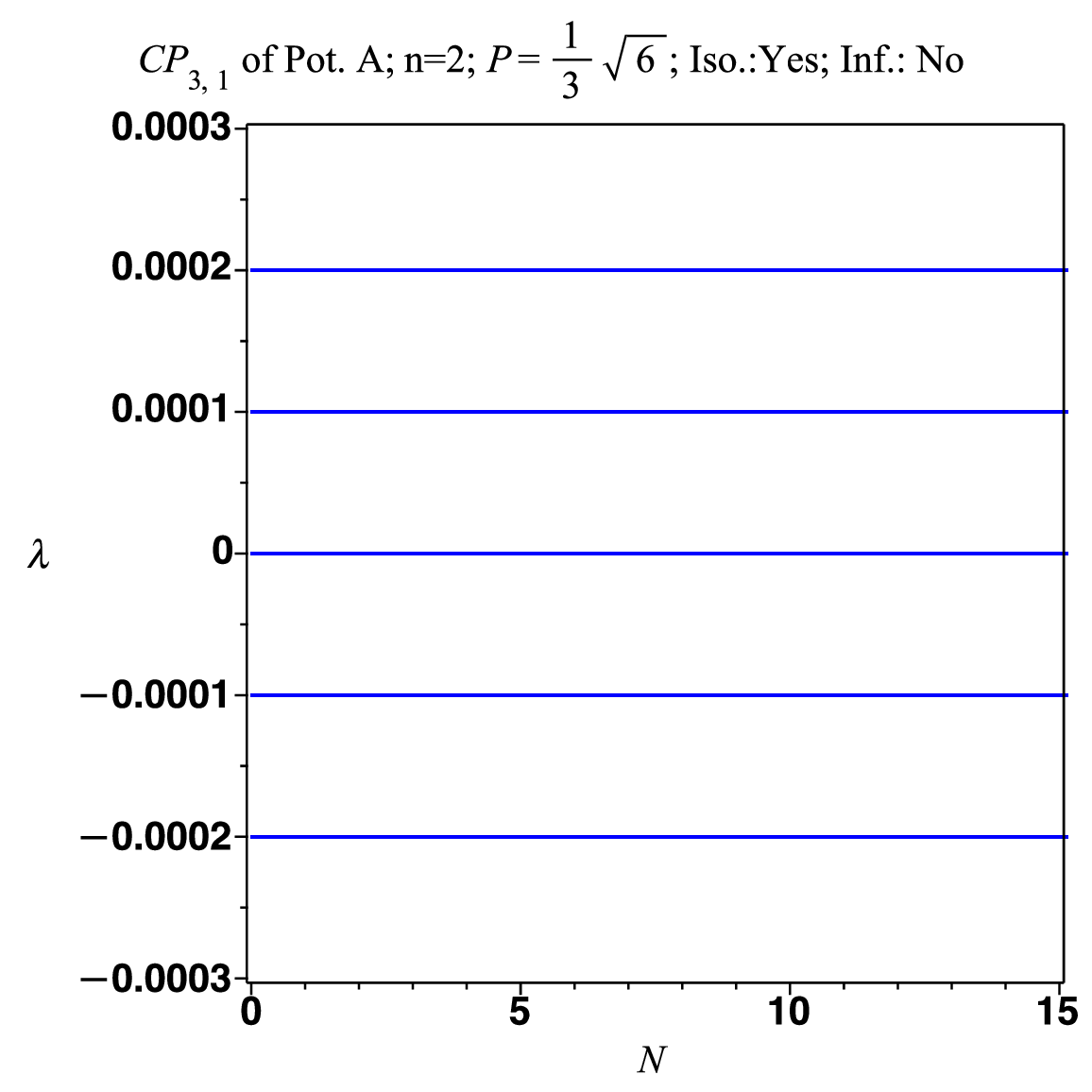}
		}
		\caption{
				Perturbative behavior of potential $\mathrm{A}$ around the equilibrium point $\mathrm{CP}_{A,3-1}$ for $\lambda=1$, $n=2$, and $\sqrt{6}/3$ (Strong Isotropization without Inflation).
		}
		\label{FigPotA31-iso}
	\end{figure}
\end{landscape}
	\begin{landscape}
	\begin{figure}
		\centering
		\subfloat[]{
			\includegraphics[width=1.9 in, height= 1.9 in]{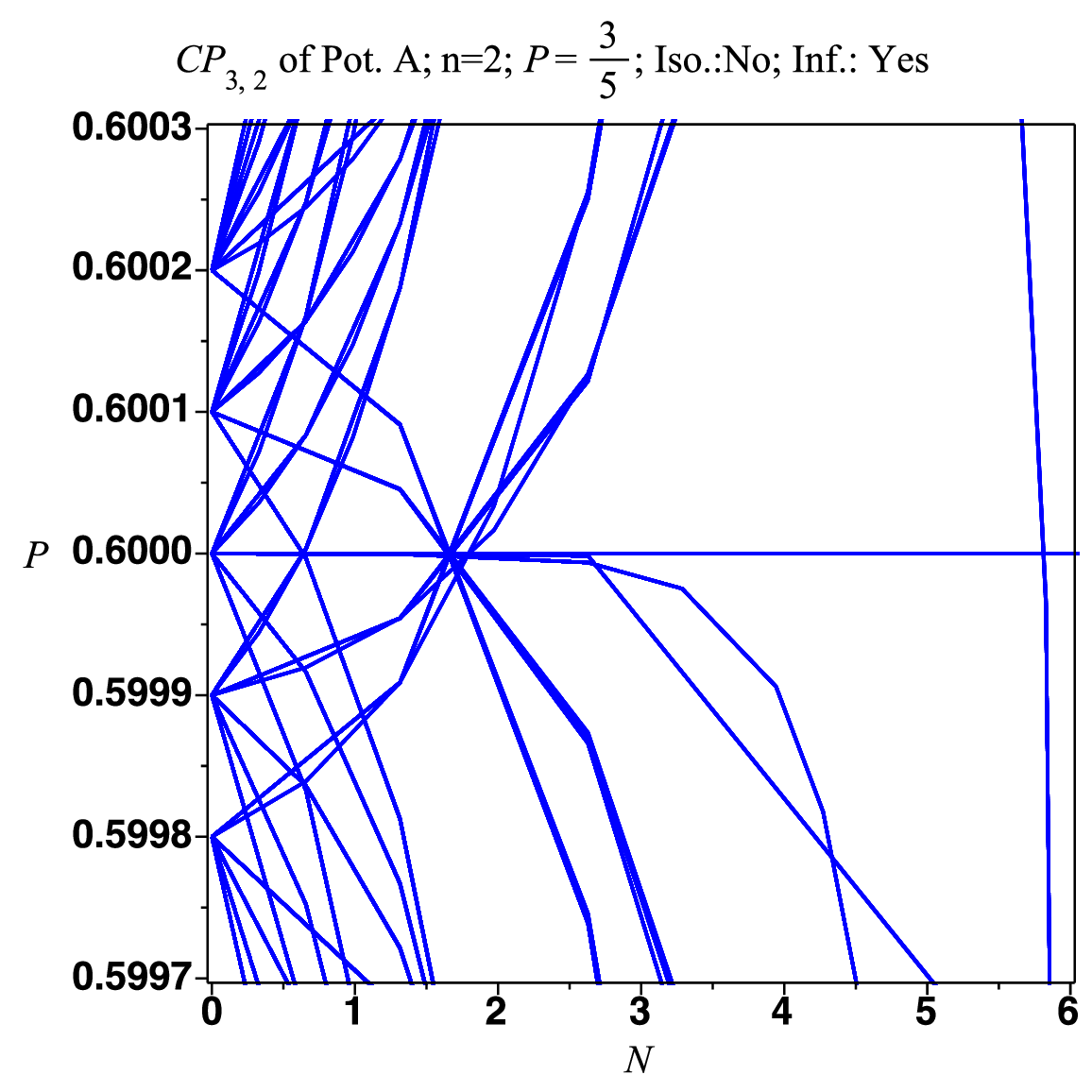}
		}
		\quad
		\subfloat[]{
			\includegraphics[width=1.9 in, height= 1.9 in]{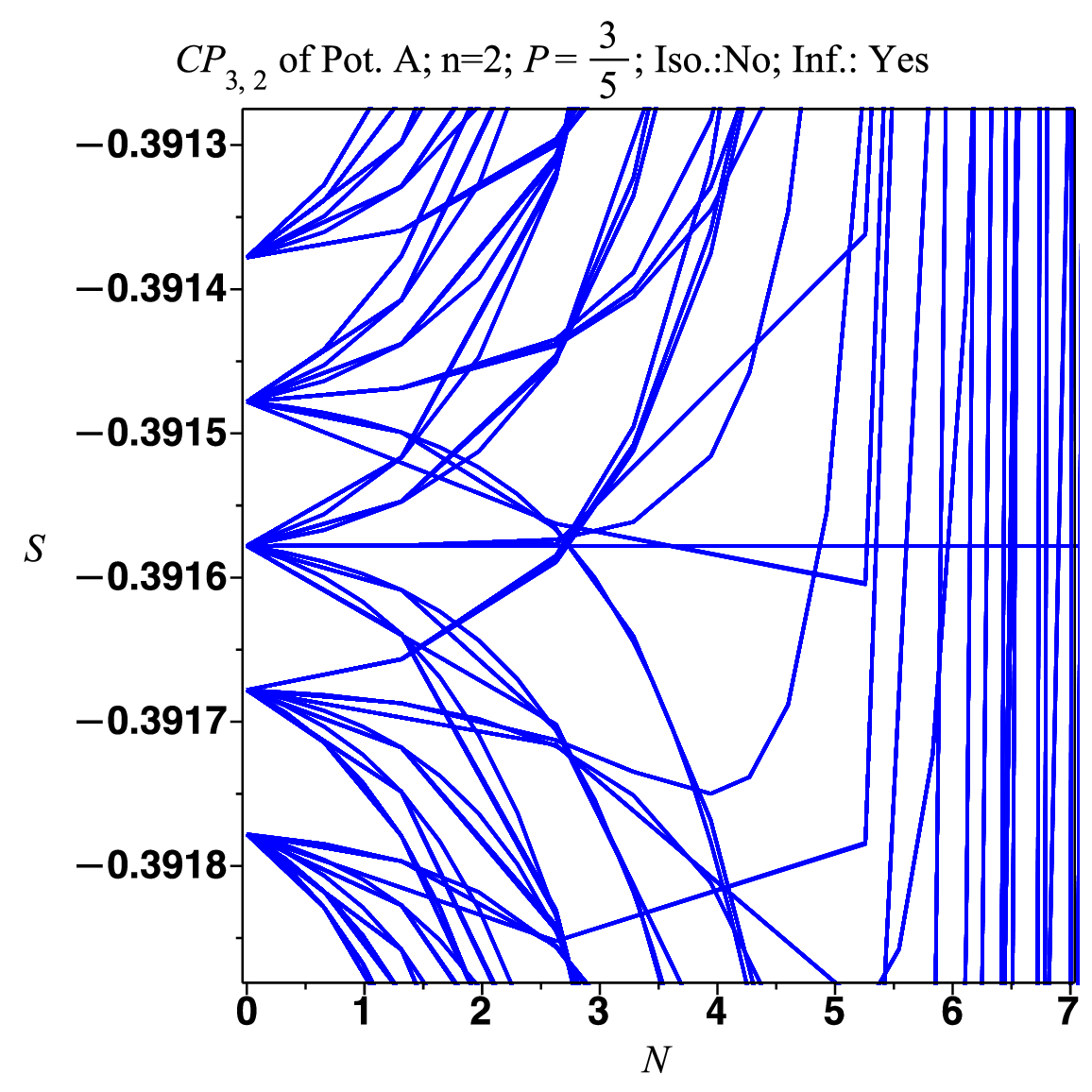}
		}
		\quad
		\subfloat[]{
			\includegraphics[width=1.9 in, height= 1.9 in]{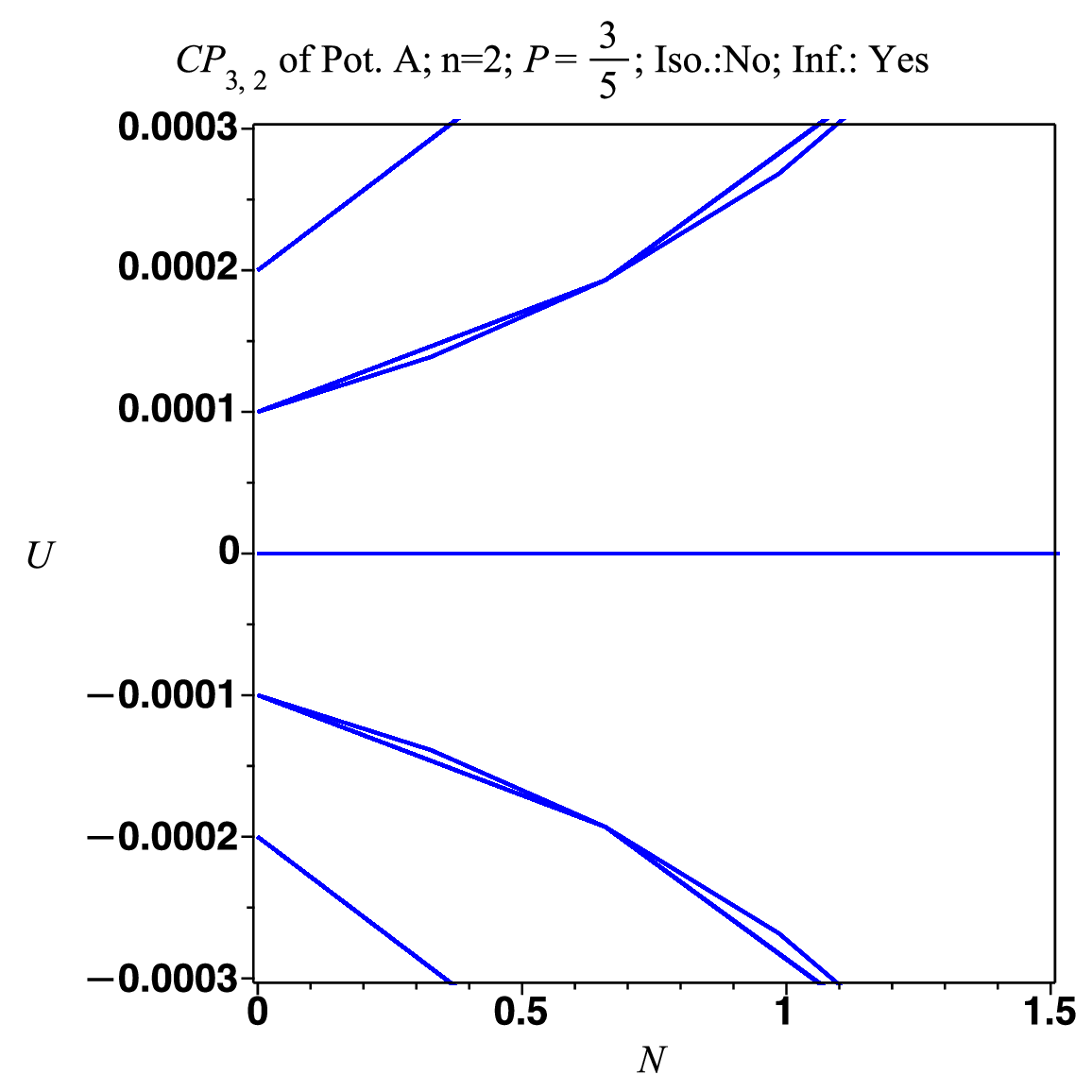}
		}
		\quad
		\subfloat[]{
			\includegraphics[width=1.9 in, height= 1.9 in]{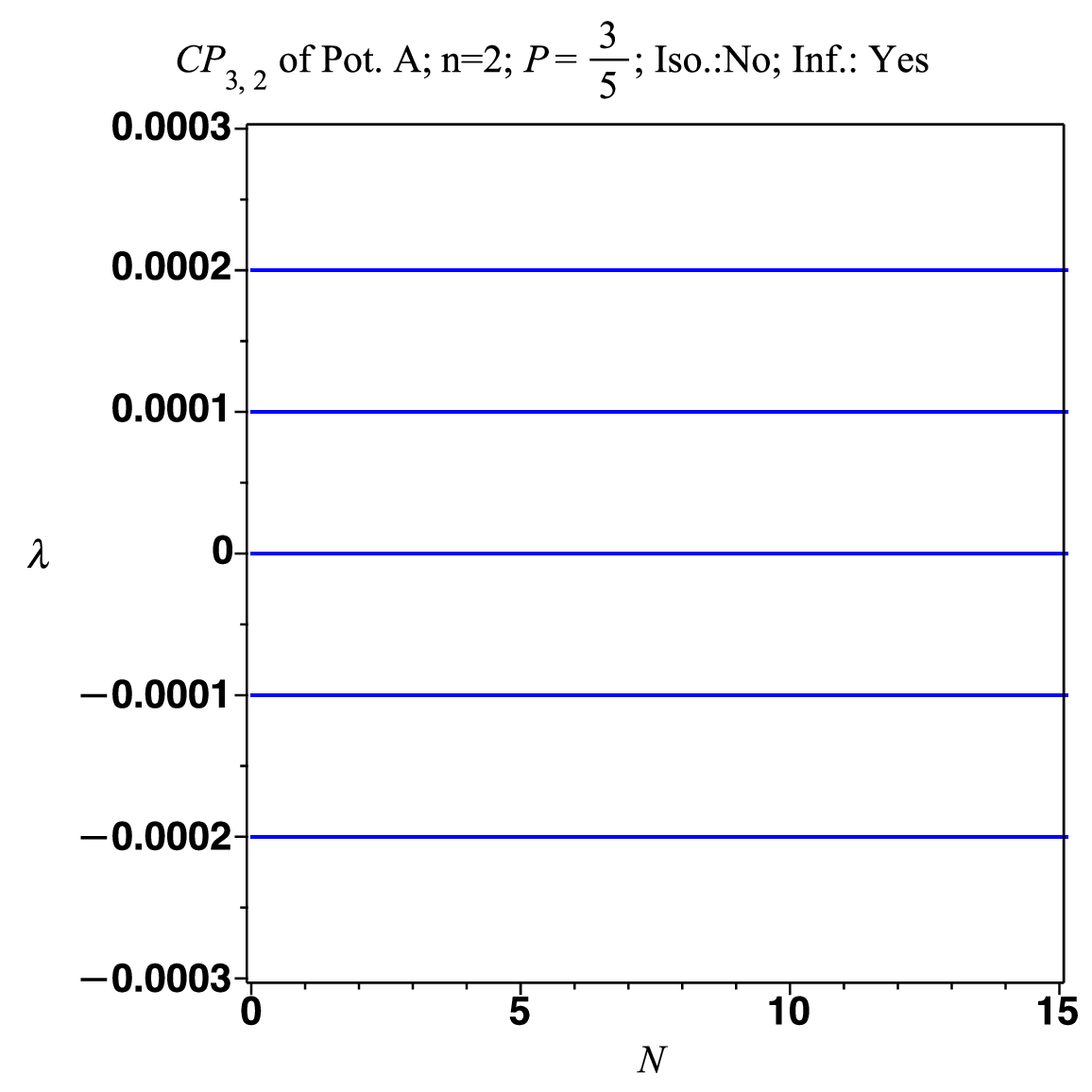}
		}
		\caption{
Perturbative behavior of potential $\mathrm{A}$ around the equilibrium point $\mathrm{CP}_{A,3-2}$ for $\lambda=1$, $n=2$, and $P=3/5$ (Inflation without Strong Isotropization).
		}
		\label{FigPotA32-inf}
	\end{figure}
	\begin{figure}
		\centering
		\subfloat[]{
			\includegraphics[width=1.9 in, height= 1.9 in]{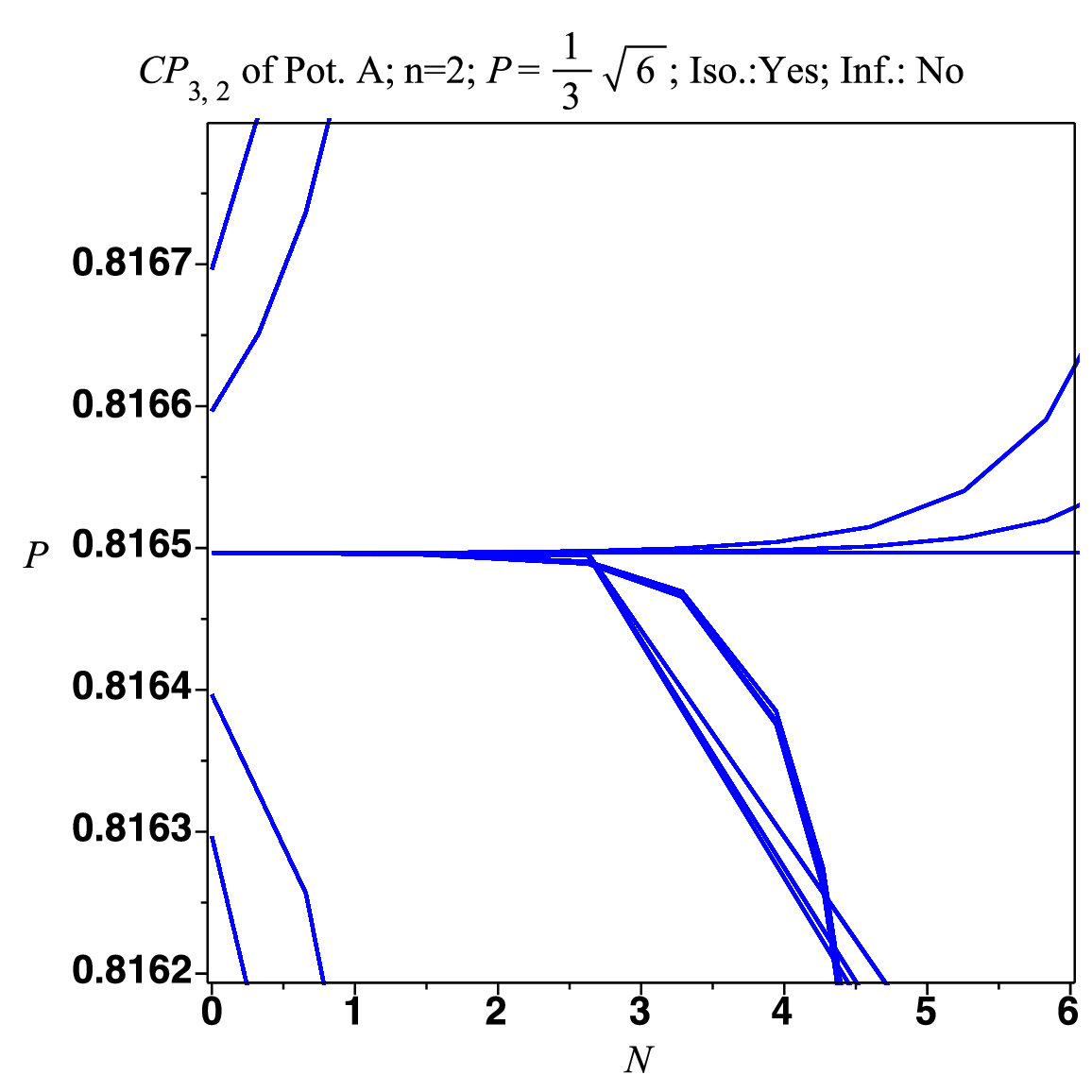}
		}
		\quad
		\subfloat[]{
			\includegraphics[width=1.9 in, height= 1.9 in]{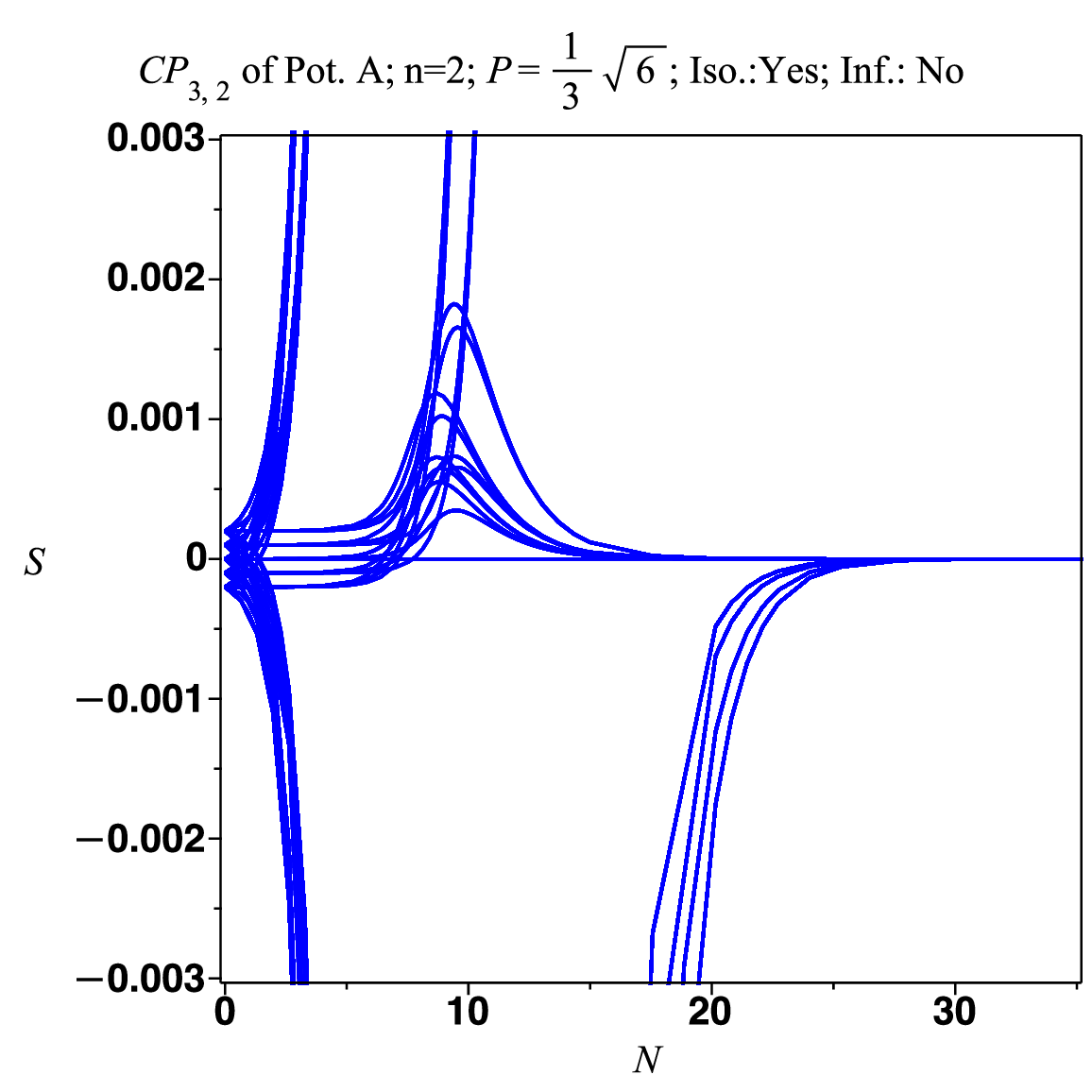}
		}
		\quad
		\subfloat[]{
			\includegraphics[width=1.9 in, height= 1.9 in]{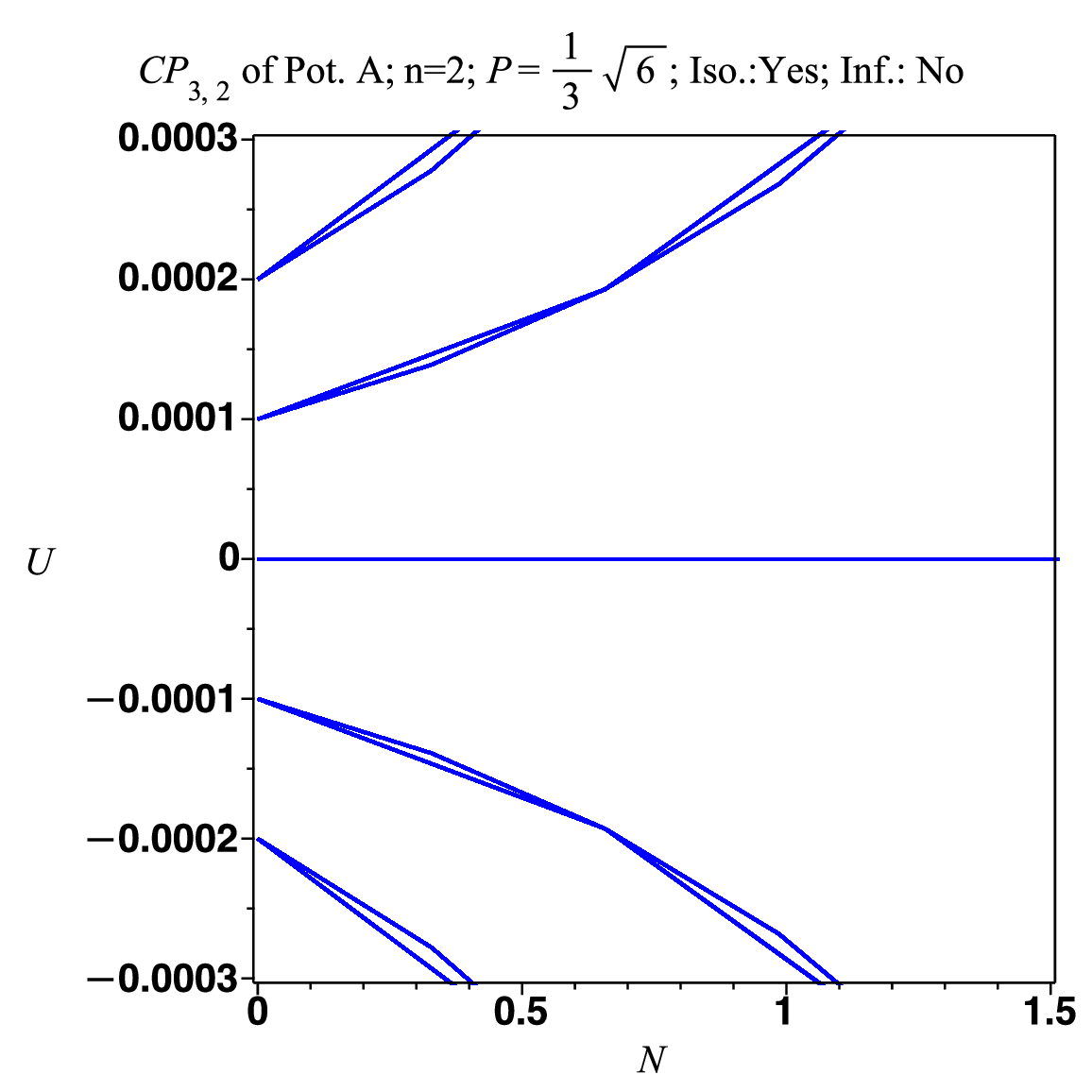}
		}
		\quad
		\subfloat[]{
			\includegraphics[width=1.9 in, height= 1.9 in]{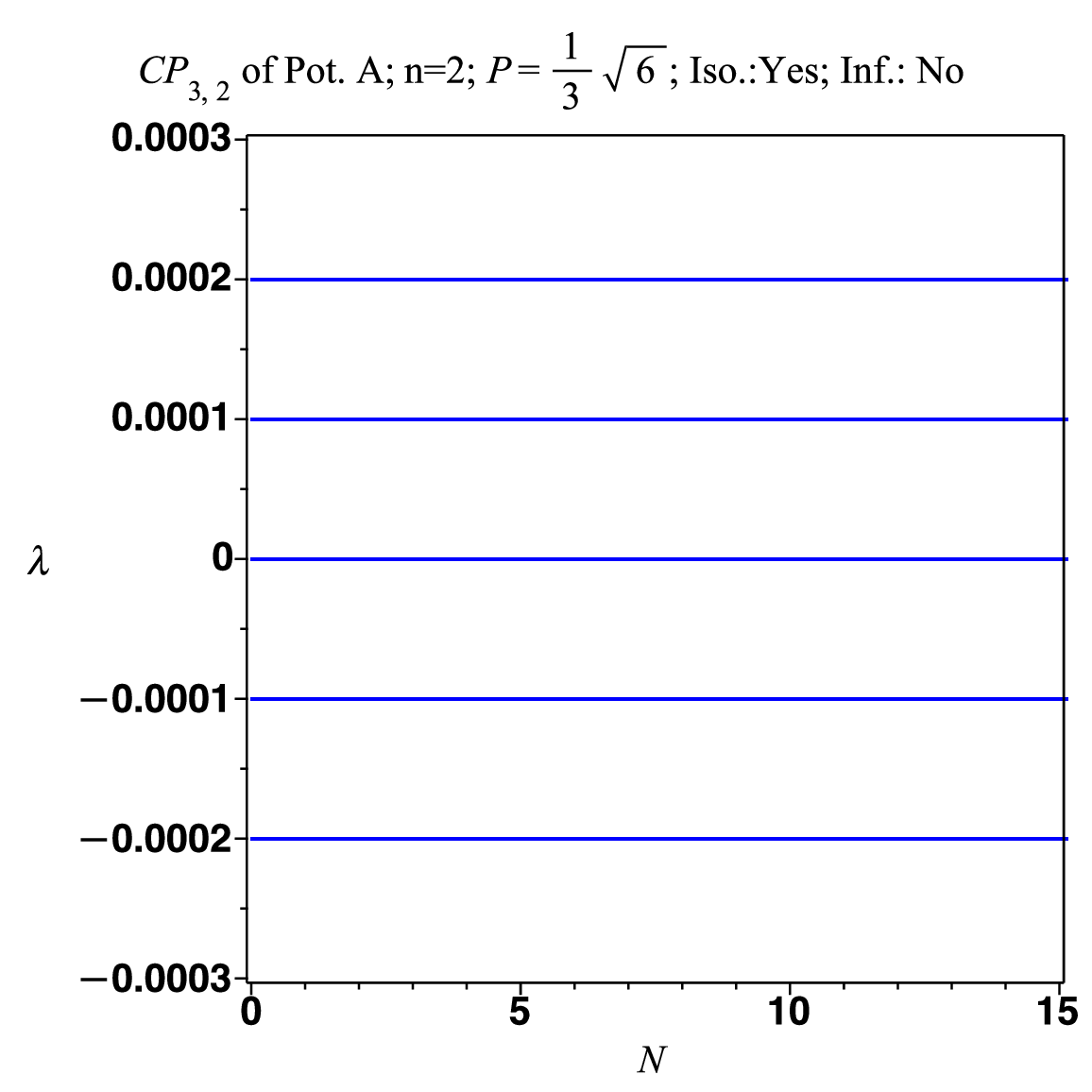}
		}
		\caption{
				Perturbative behavior of potential $\mathrm{A}$ around the equilibrium point $\mathrm{CP}_{A,3-2}$ for $\lambda=1$, $n=2$, and $\sqrt{6}/3$ (Strong Isotropization without Inflation).
		}
		\label{FigPotA32-iso}
	\end{figure}
\end{landscape}
	\begin{landscape}
	\begin{figure}
		\centering
		\subfloat[]{
			\includegraphics[width=1.9 in, height= 1.9 in]{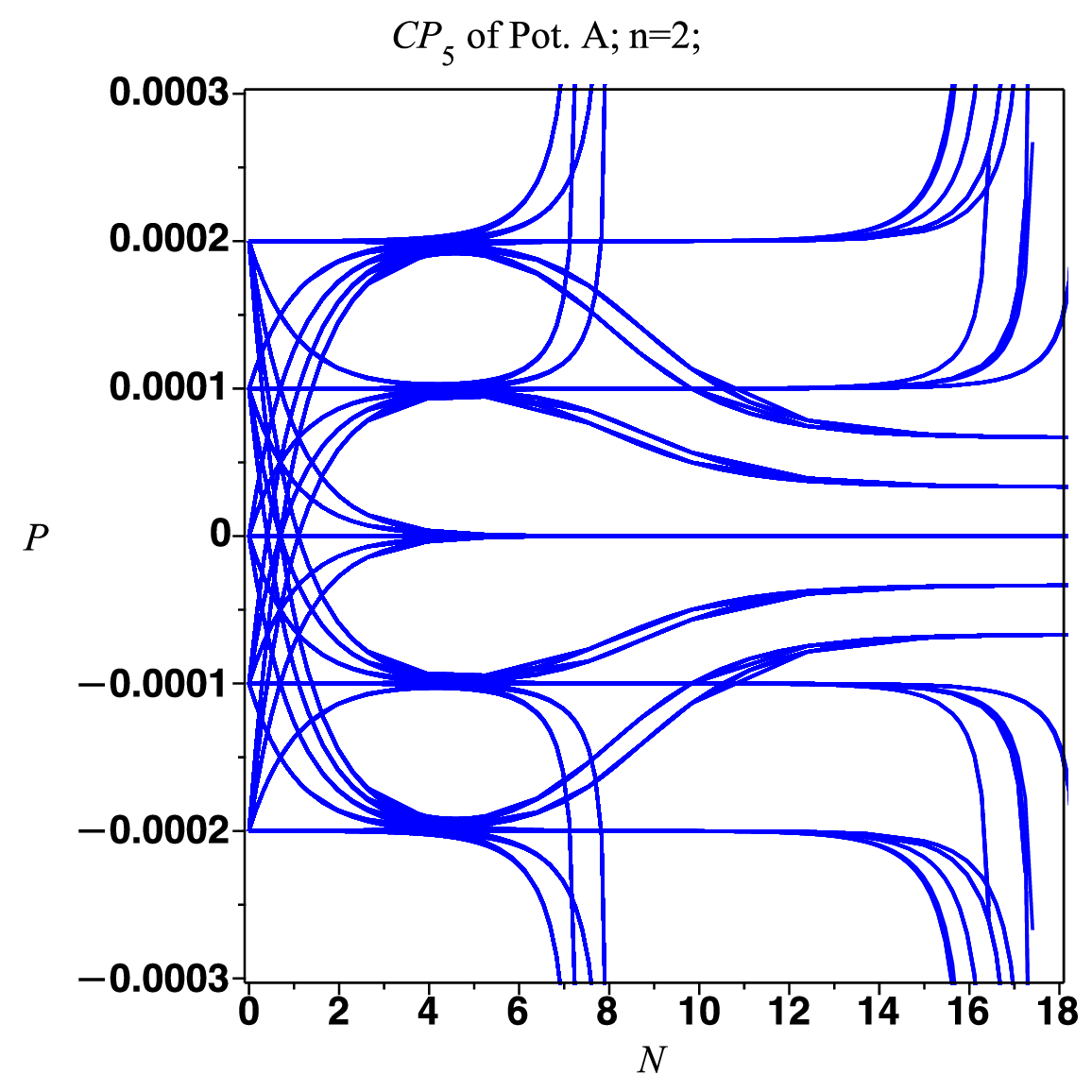}
		}
		\quad
		\subfloat[]{
			\includegraphics[width=1.9 in, height= 1.9 in]{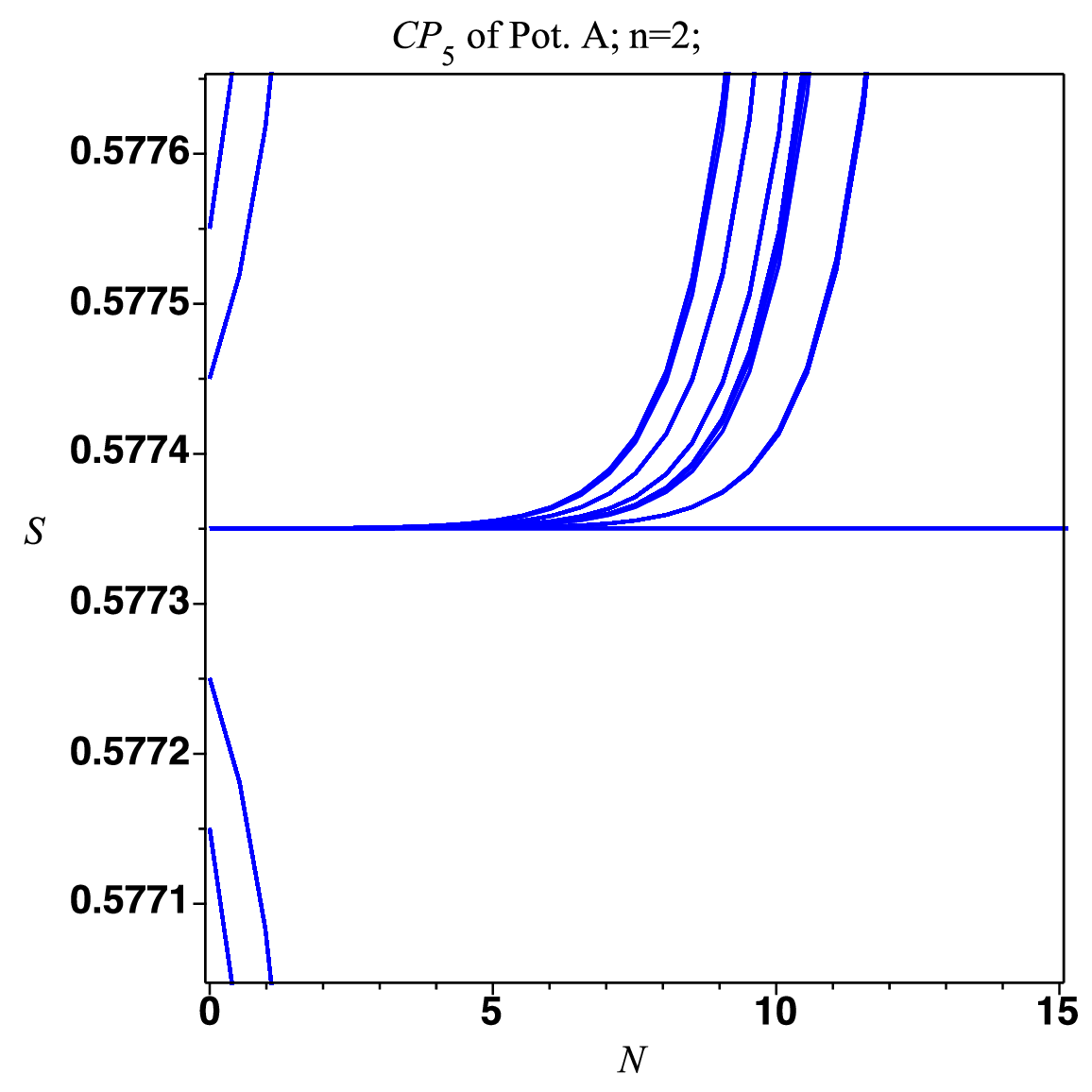}
		}
		\quad
		\subfloat[]{
			\includegraphics[width=1.9 in, height= 1.9 in]{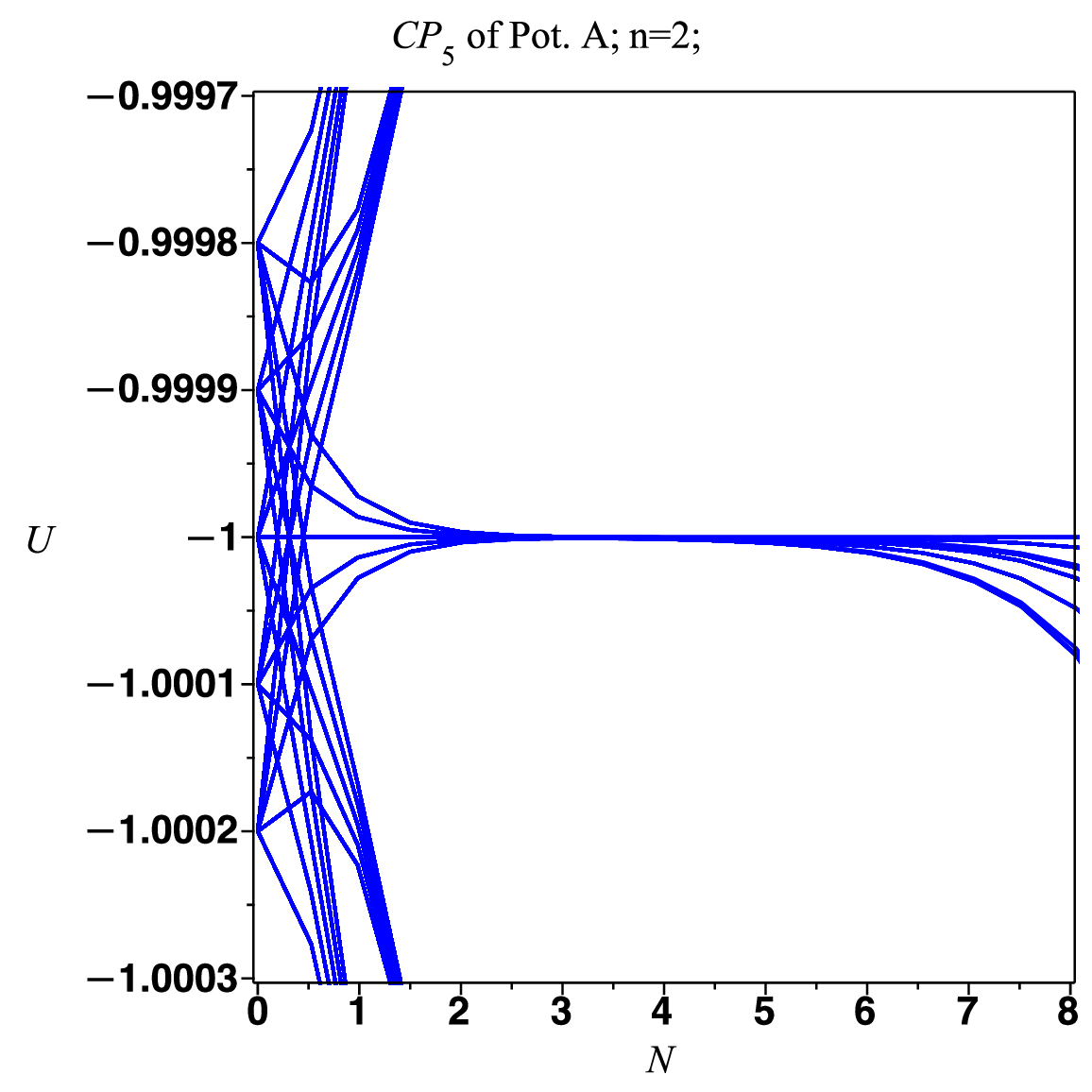}
		}
		\quad
		\subfloat[]{
			\includegraphics[width=1.9 in, height= 1.9 in]{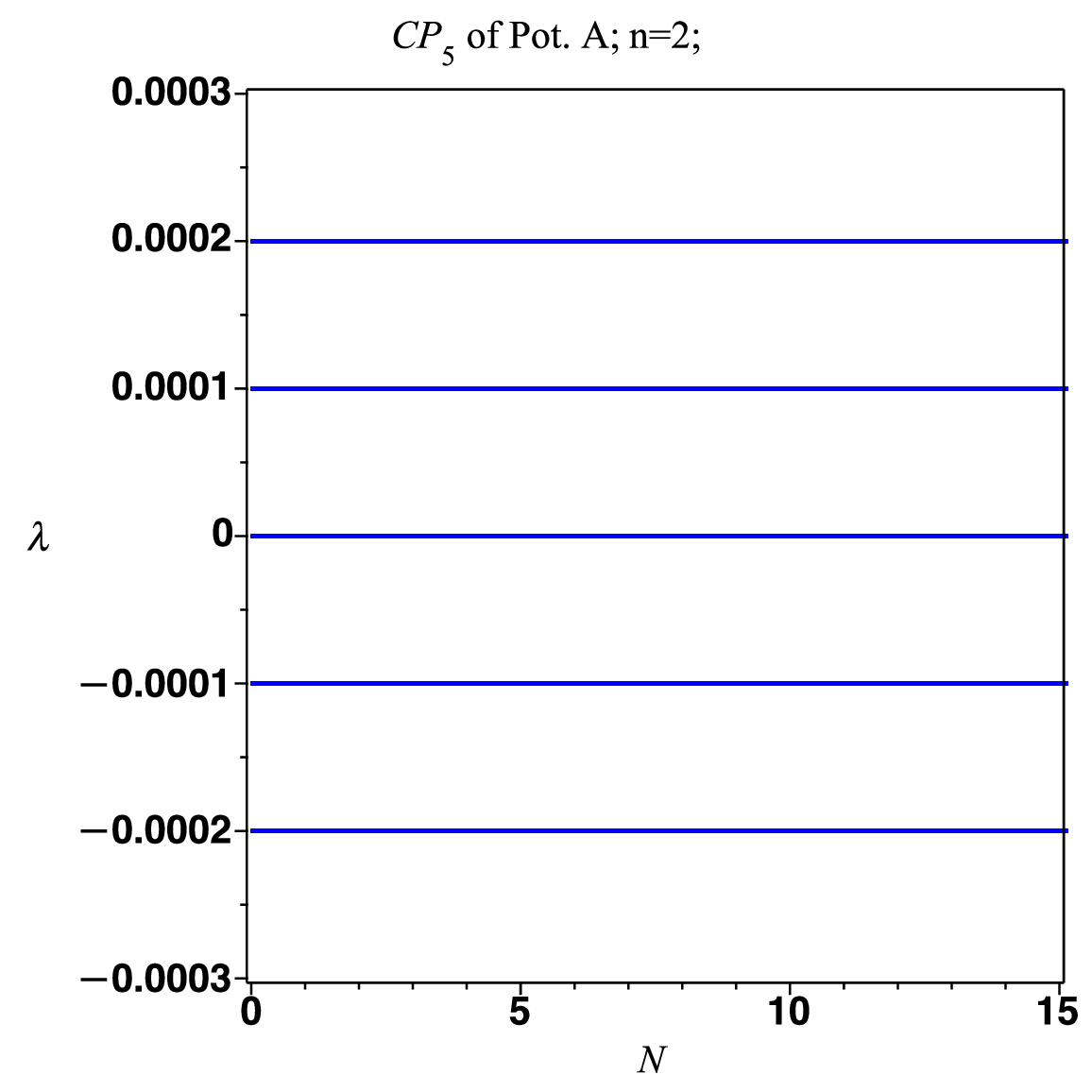}
		}
		\caption{
Perturbative behavior of potential $\mathrm{A}$ around the equilibrium point $\mathrm{CP}_{A,4}$ for $\lambda=1$ and $n=2$.
		}
		\label{FigPotA4}
	\end{figure}
	\begin{figure}
		\centering
		\subfloat[]{
			\includegraphics[width=1.9 in, height= 1.9 in]{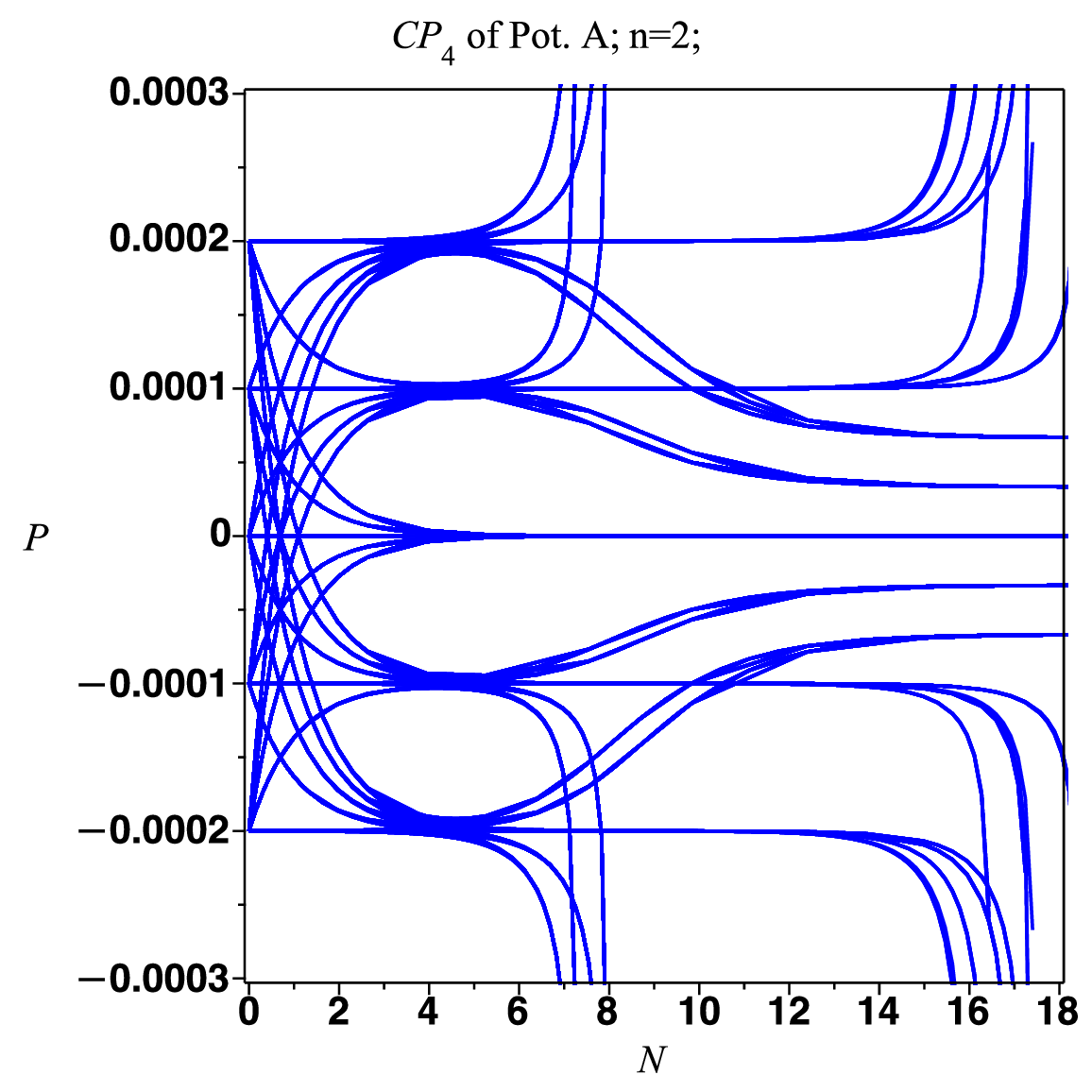}
		}
		\quad
		\subfloat[]{
			\includegraphics[width=1.9 in, height= 1.9 in]{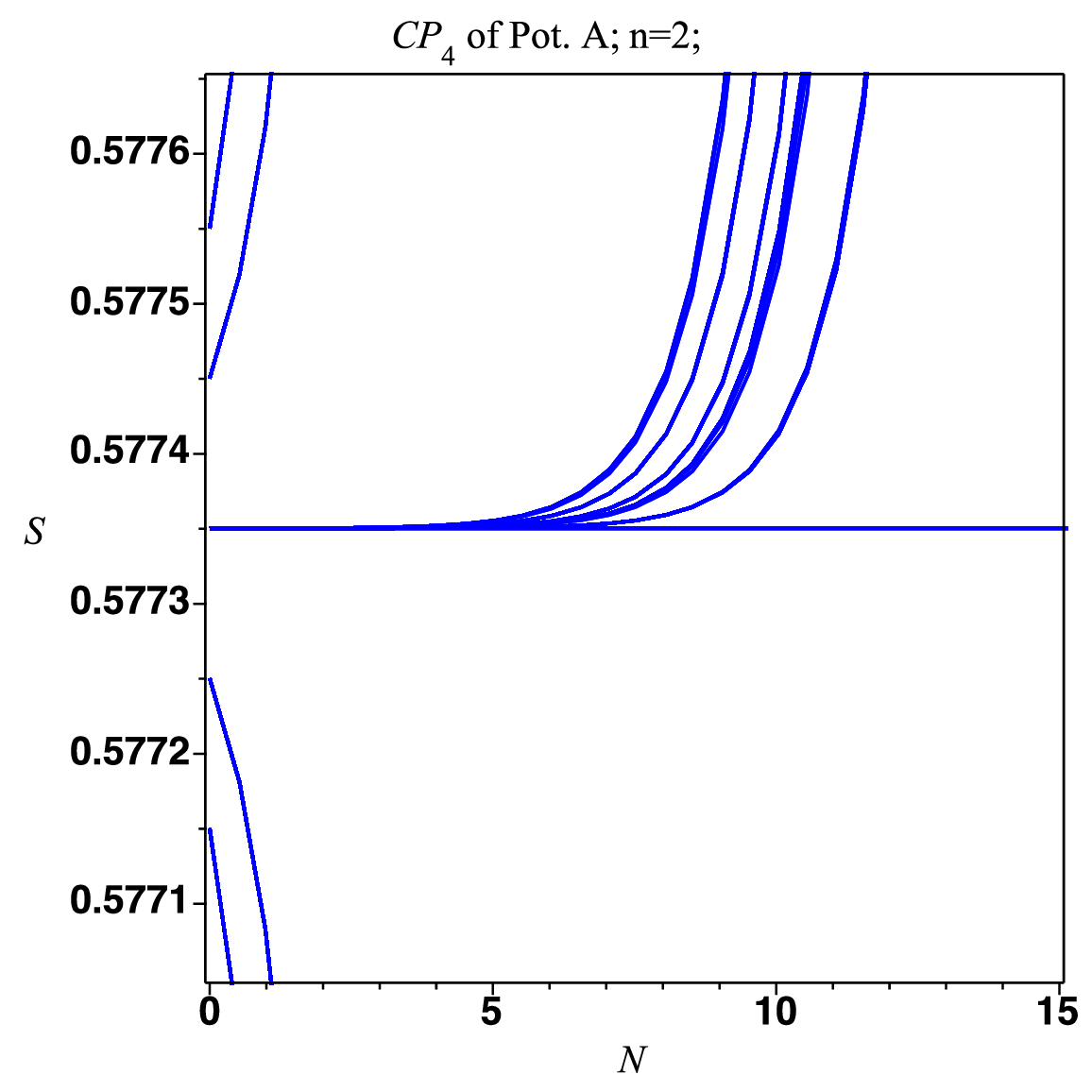}
		}
		\quad
		\subfloat[]{
			\includegraphics[width=1.9 in, height= 1.9 in]{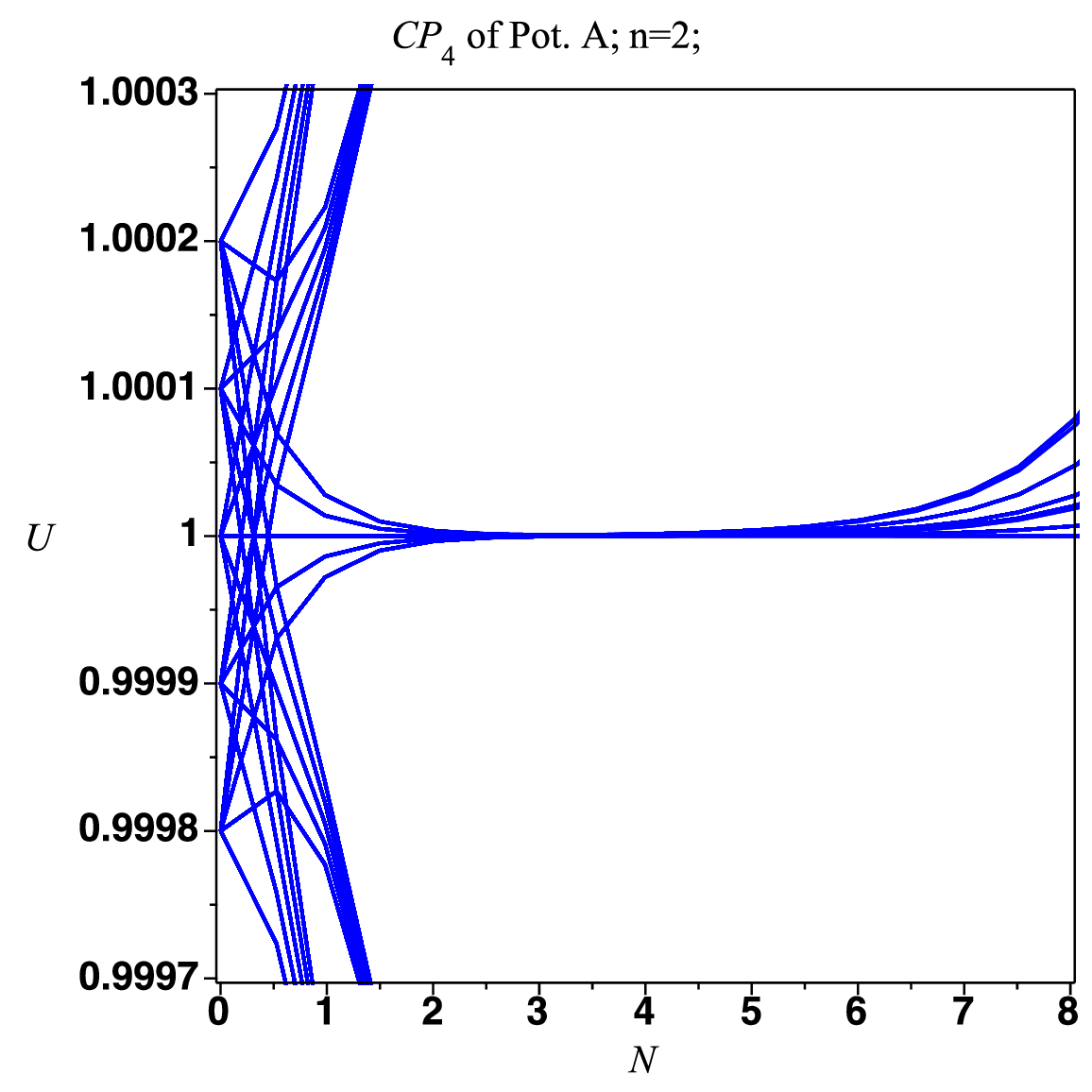}
		}
		\quad
		\subfloat[]{
			\includegraphics[width=1.9 in, height= 1.9 in]{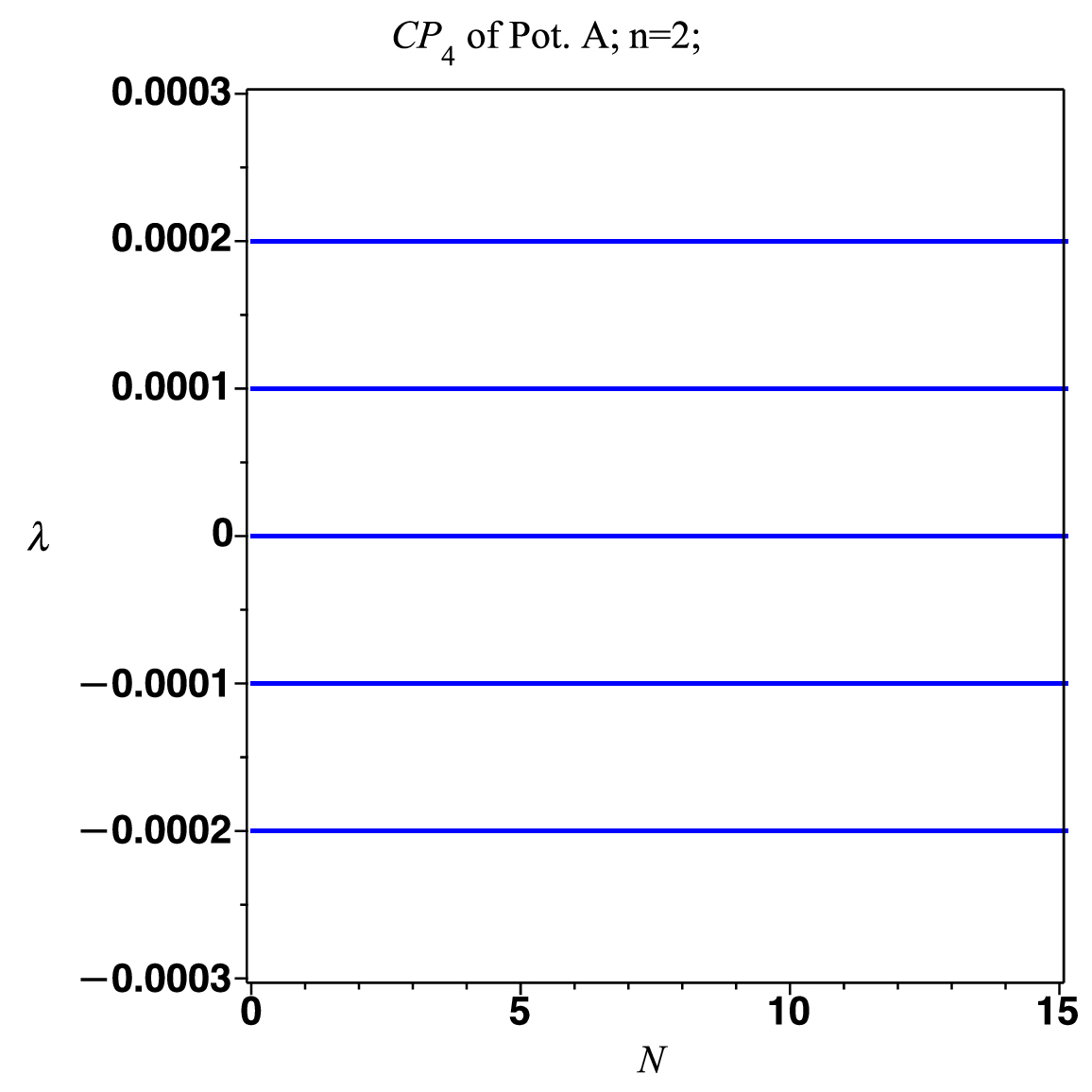}
		}
		\caption{
Perturbative behavior of potential $\mathrm{A}$ around the equilibrium point $\mathrm{CP}_{A,5}$ for $\lambda=1$ and $n=2$.
		}
		\label{FigPotA5}
	\end{figure}
\end{landscape}
	\begin{landscape}
	\begin{figure}
		\centering
		\subfloat[]{
			\includegraphics[width=1.9 in, height= 1.9 in]{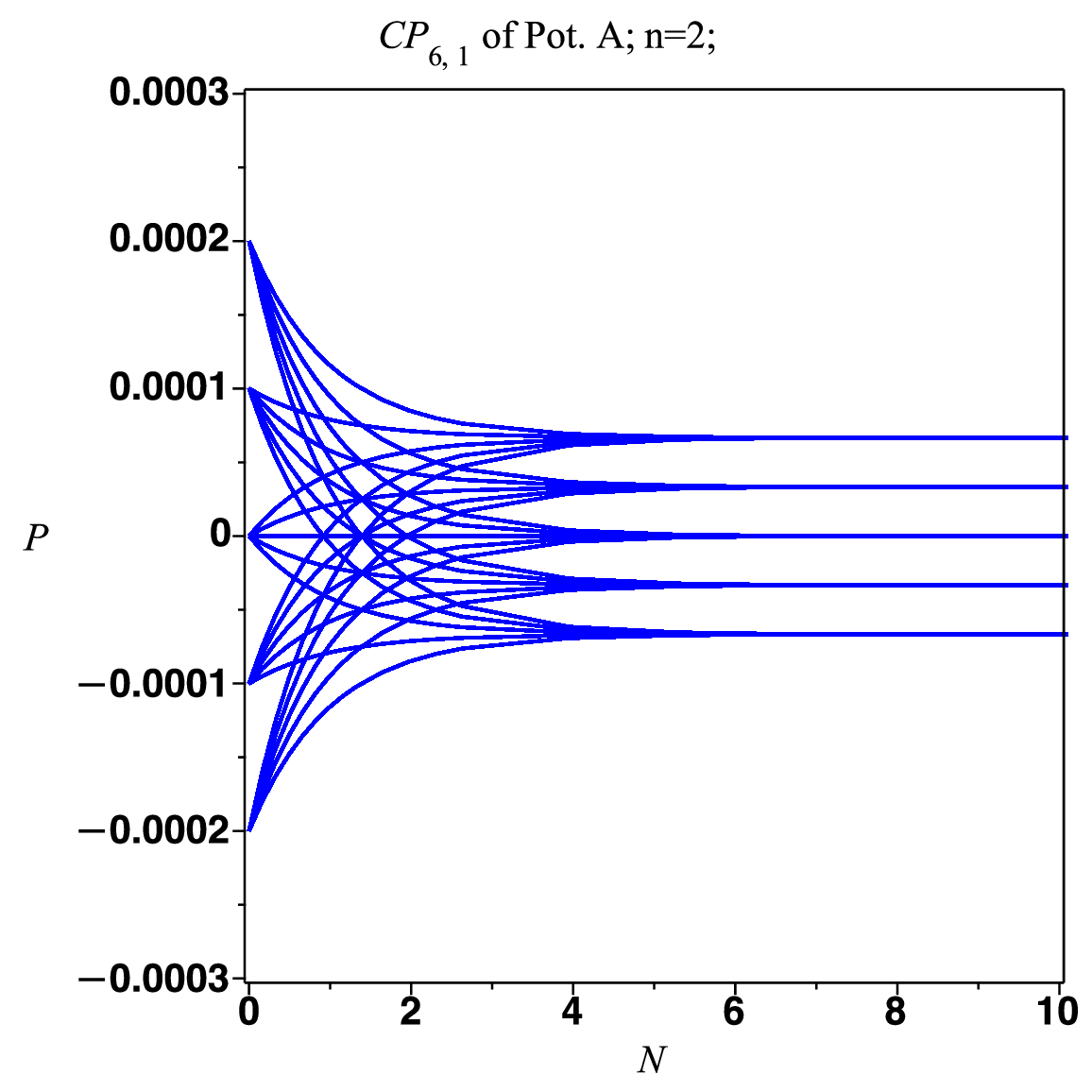}
		}
		\quad
		\subfloat[]{
			\includegraphics[width=1.9 in, height= 1.9 in]{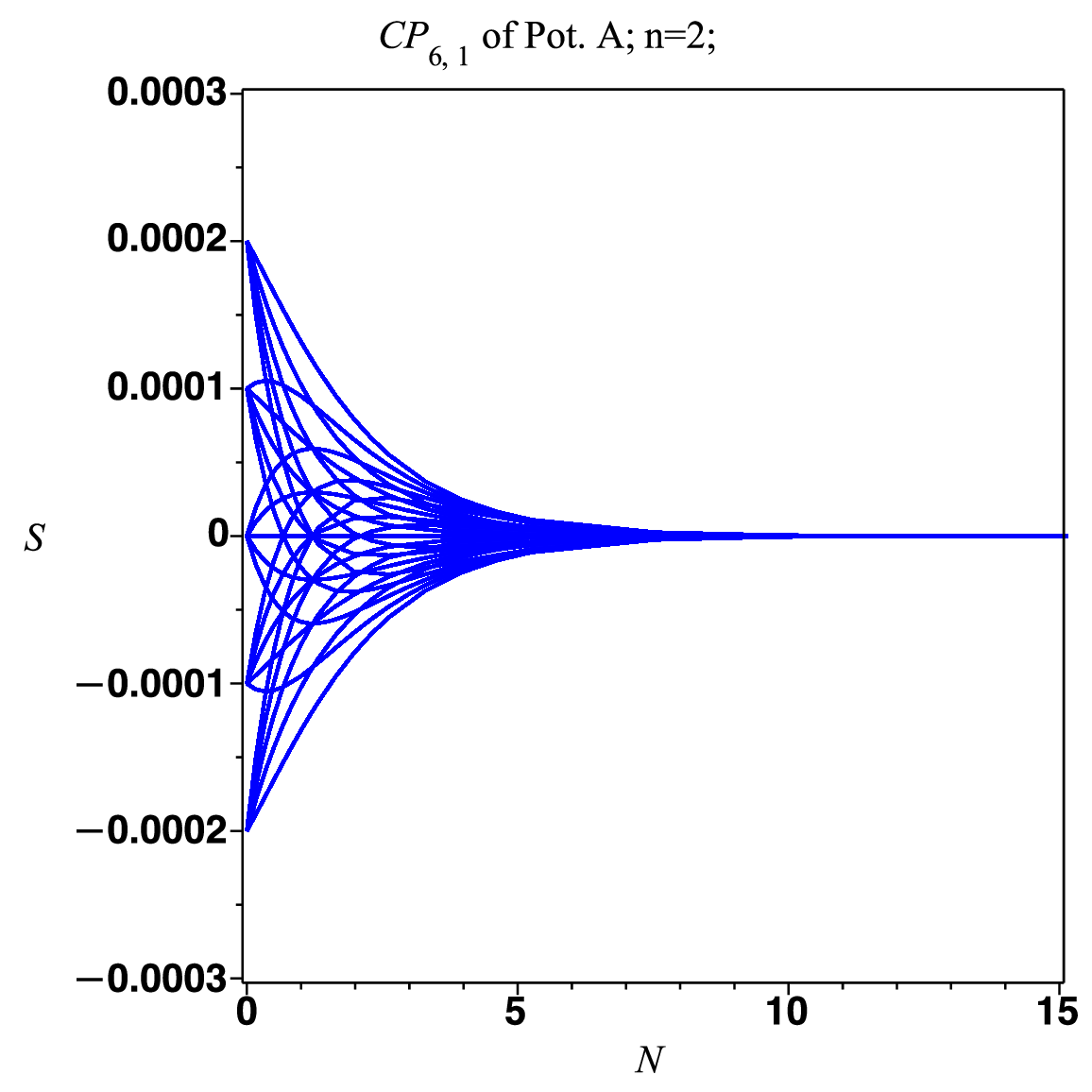}
		}
		\quad
		\subfloat[]{
			\includegraphics[width=1.9 in, height= 1.9 in]{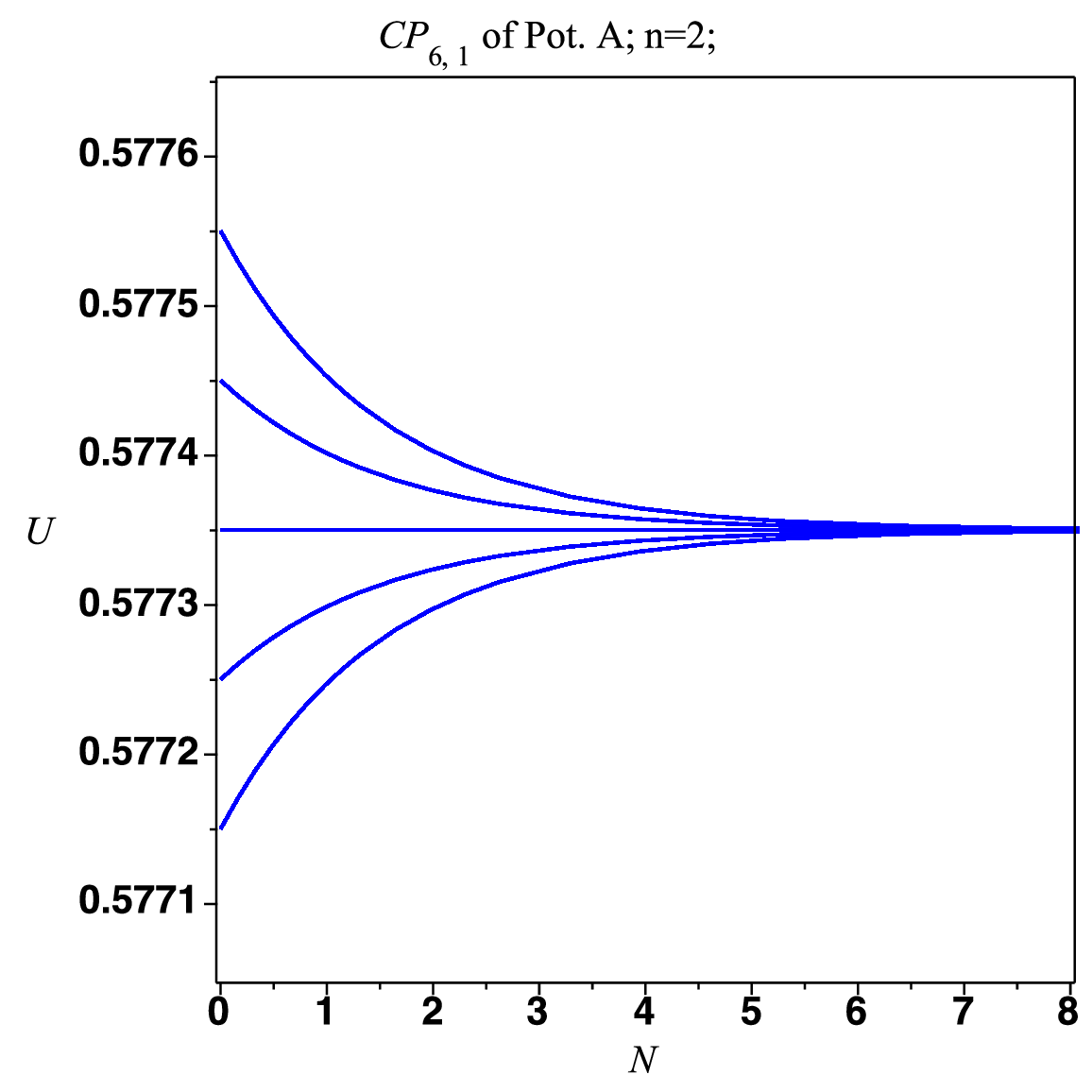}
		}
		\quad
		\subfloat[]{
			\includegraphics[width=1.9 in, height= 1.9 in]{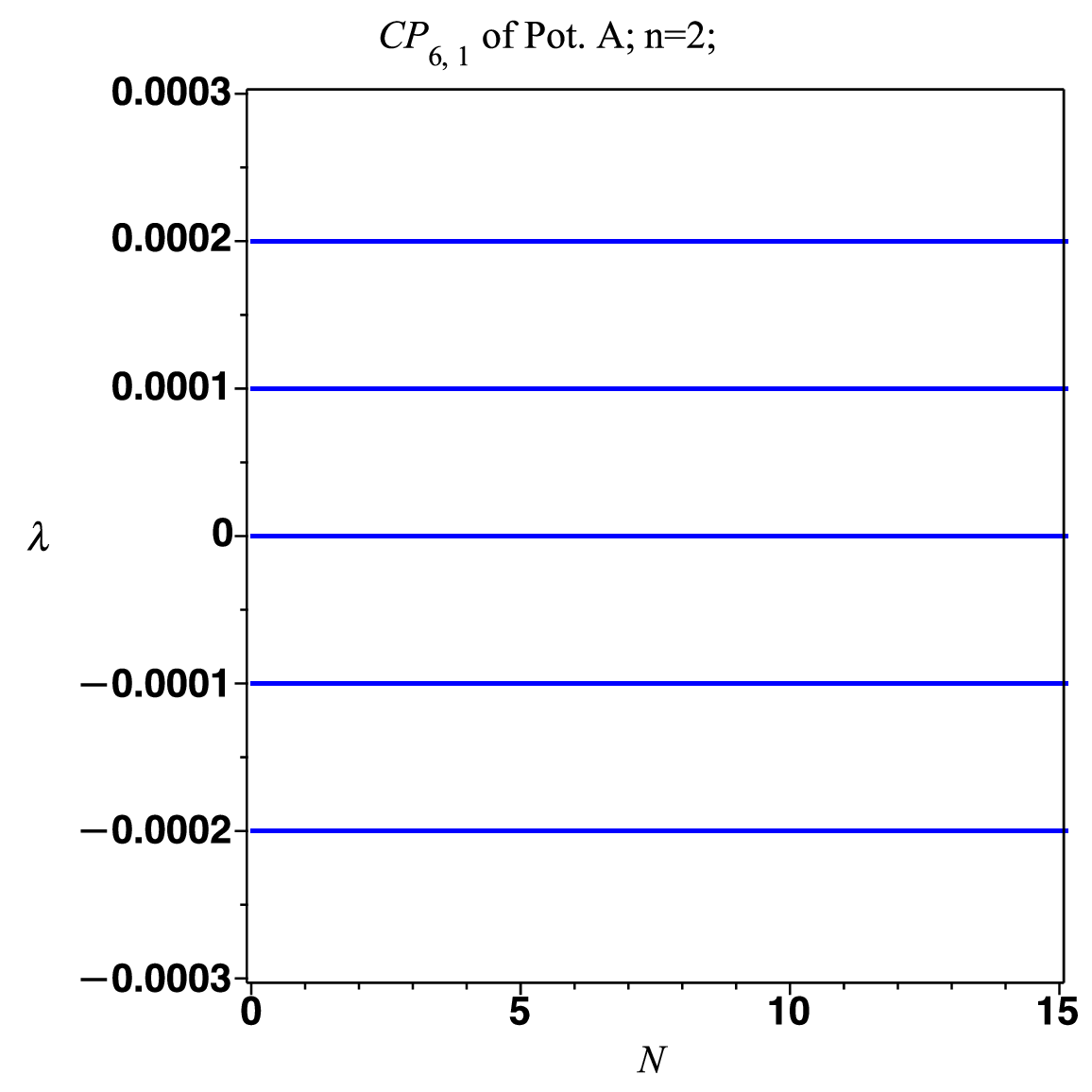}
		}
		\caption{
Perturbative behavior of potential $\mathrm{A}$ around the equilibrium point $\mathrm{CP}_{A,6-1}$ for $\lambda=1$ and $n=2$.
		}
		\label{FigPotA61}
	\end{figure}
	\begin{figure}
		\centering
		\subfloat[]{
			\includegraphics[width=1.9 in, height= 1.9 in]{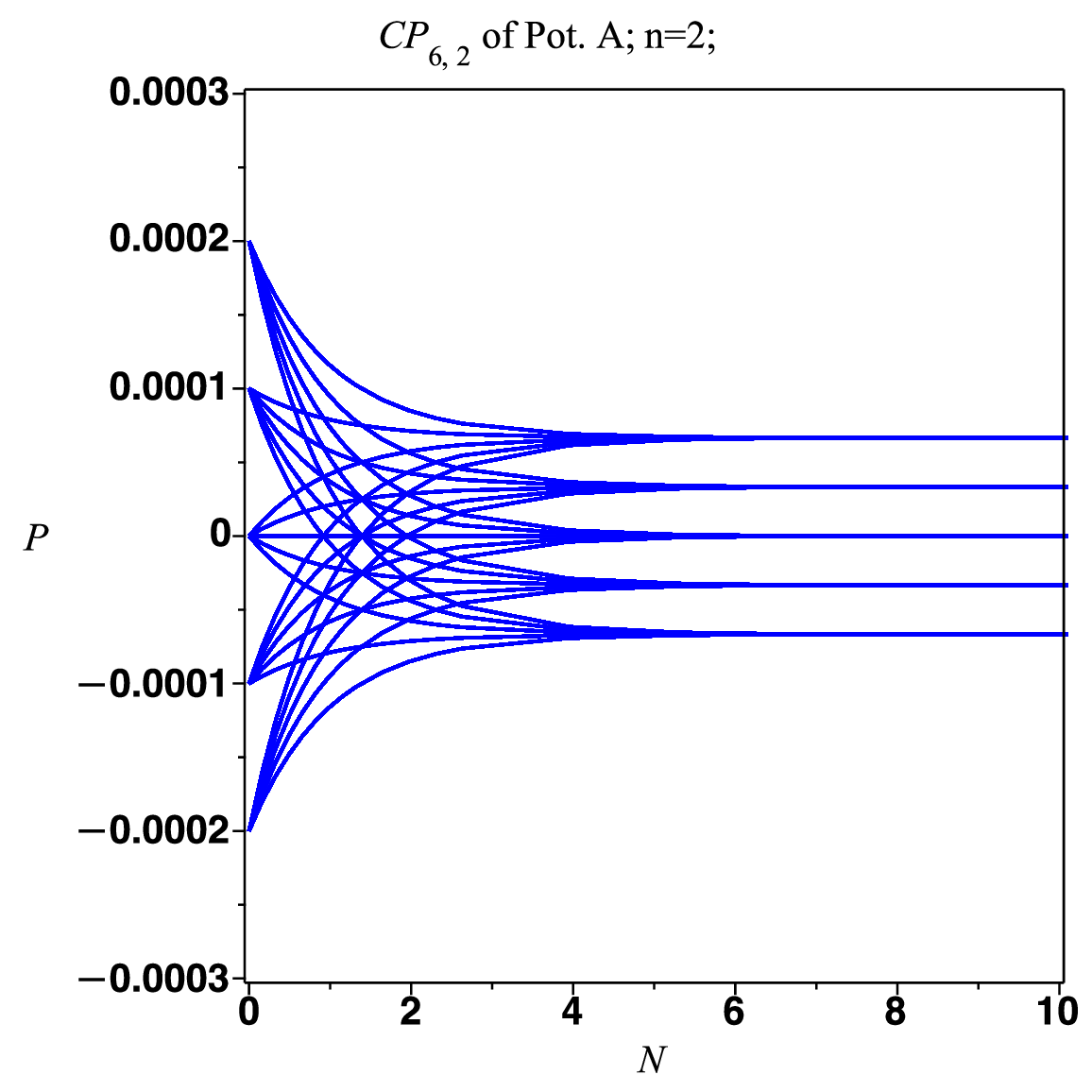}
		}
		\quad
		\subfloat[]{
			\includegraphics[width=1.9 in, height= 1.9 in]{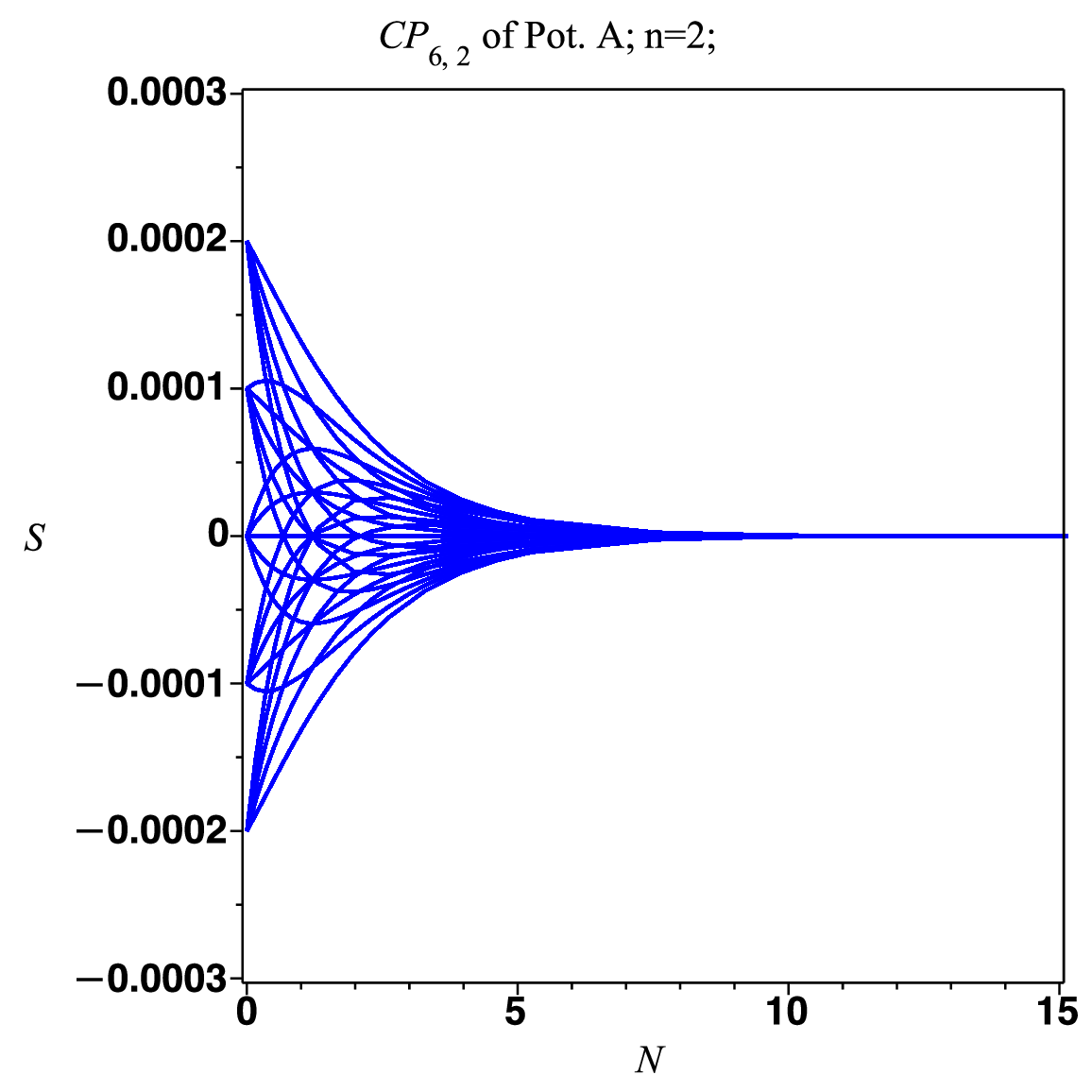}
		}
		\quad
		\subfloat[]{
			\includegraphics[width=1.9 in, height= 1.9 in]{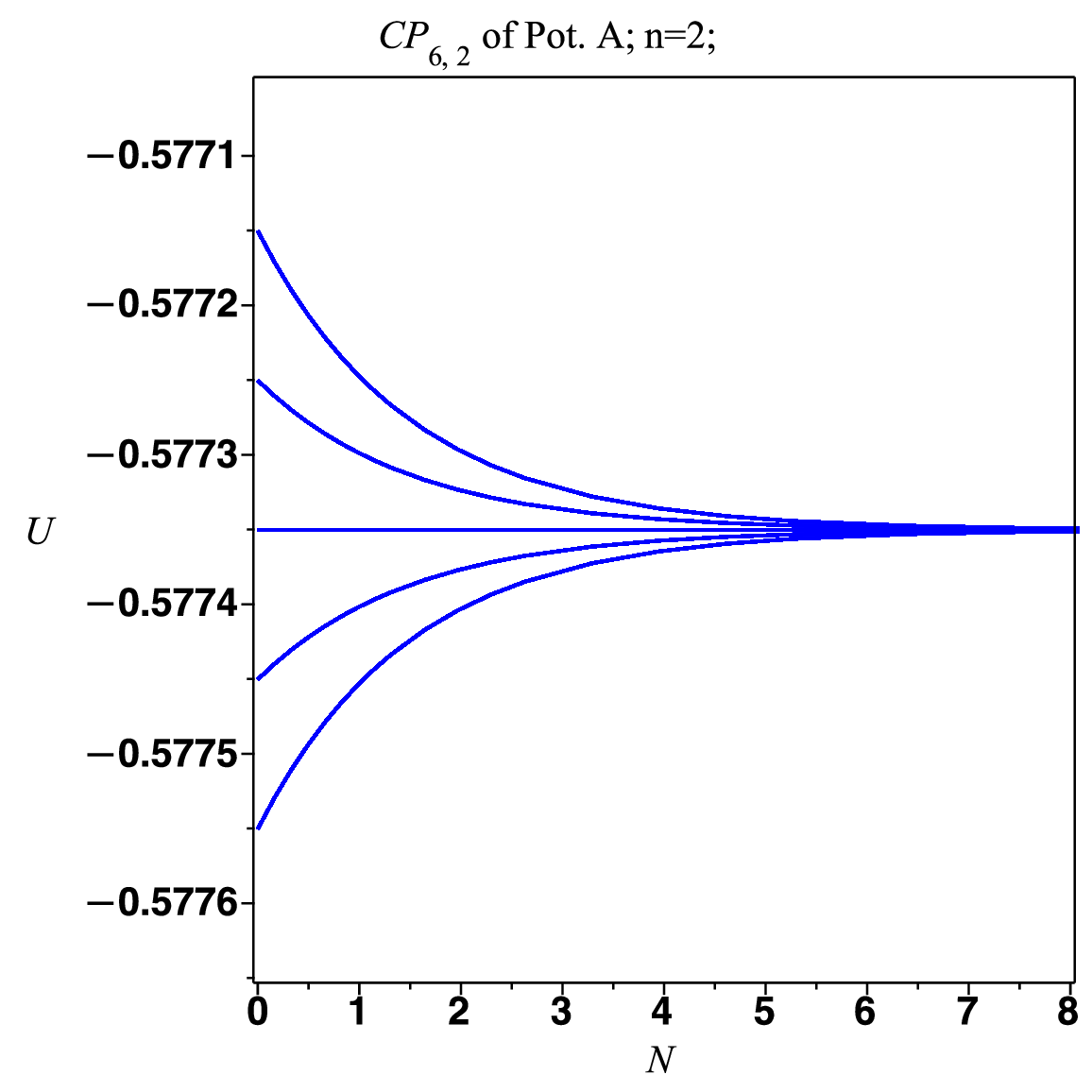}
		}
		\quad
		\subfloat[]{
			\includegraphics[width=1.9 in, height= 1.9 in]{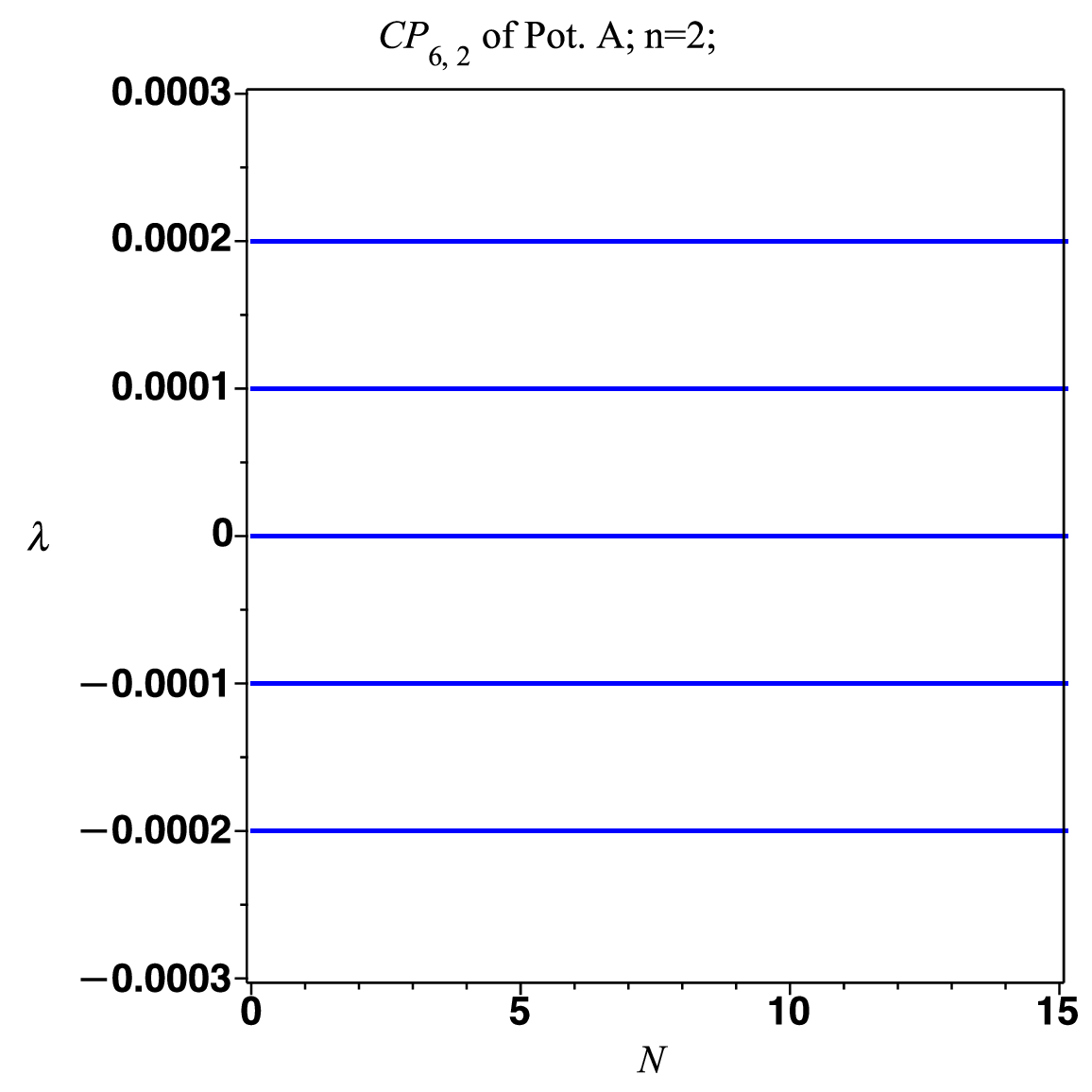}
		}
		\caption{
Perturbative behavior of potential $\mathrm{A}$ around the equilibrium point $\mathrm{CP}_{A,6-2}$ for $\lambda=1$ and $n=2$.
		}
		\label{FigPotA62}
	\end{figure}
\end{landscape}

\phantomsection
\section*{Acknowledgment}
\addcontentsline{toc}{section}{Acknowledgment}

This work has been supported financially by Iranian National Ellits Foundations.

\phantomsection

\end{document}